\documentclass[fleqn,usenatbib,useAMS]{mnras}

\usepackage{newtxtext,newtxmath}
\usepackage[T1]{fontenc}
\usepackage{ae,aecompl}

\usepackage{graphicx}	
\usepackage{amsmath}	
\usepackage{amssymb}	
\usepackage{commath}    

\newcommand{\ceeoh}[2]{CO(#1-#2)}
\newcommand{\icfost}{IC~5063}
\newcommand{\PFF}{P44}    
\newcommand{\PFV}{P45}    
\makeatletter
\newcommand*{\rom}[1]{\expandafter\@slowromancap\romannumeral #1@}
\makeatother
\newcommand{\halpha}{H$\alpha$}

\newcommand{\ergs}{\,\text{erg}\,\text{s}^{-1}}
\newcommand{\kms}{\,\text{km}\,\text{s}^{-1}}
\newcommand{\cc}{\,\text{cm}^{-3}}
\newcommand{\kelvin}{\,\text{K}}
\newcommand{\pc}{\,\text{pc}}
\newcommand{\kpc}{\,\text{kpc}}
\newcommand{\kyr}{\,\text{kyr}}
\newcommand{\Myr}{\,\text{Myr}}
\newcommand{\GHz}{\,\text{GHz}}
\newcommand{\WpHz}{\,\text{W}\,\text{Hz}^{-1}}
\newcommand{\Msun}{\,\text{M}_{\sun}}
\newcommand{\Lsun}{\,\text{L}_{\sun}}
\newcommand{\Msunyr}{\,\text{M}_{\sun}\,\text{yr}^{-1}}

\newcommand{\kB}{k_\text{B}}
\newcommand{\mmpp}{\bar{\mu}}
\newcommand{\amu}{\,\text{u}}

\newcommand{\Pjet}{P_\text{jet}}
\newcommand{\pjet}{p_\text{jet}}
\newcommand{\ejet}{\varepsilon_\text{jet}}
\newcommand{\Ajet}{A_\text{jet}}
\newcommand{\vjet}{\beta_\text{jet}}
\newcommand{\rhojet}{\rho_\text{jet}}
\newcommand{\rjet}{r_\text{jet}}
\newcommand{\thjet}{\theta_\text{jet}}
\newcommand{\tjet}{t_\text{jet}}
\newcommand{\Popfghz}{P_{1.4\GHz}}

\newcommand{\emiss}{\epsilon}
\newcommand{\vel}{\text{v}}
\newcommand{\critdens}{n_\text{crit}}
\newcommand{\fmol}{f_\text{mol}}
\newcommand{\trw}{\phi_{w}}
\newcommand{\vrad}{\vel_\text{rad}}
\newcommand{\vcyl}{\vel_\text{cyl}}
\newcommand{\tcool}{t_\text{cool}}
\newcommand{\tdyn}{t_\text{dyn}}

\newcommand{\LFIR}{L_\text{FIR}}
\newcommand{\mbh}{M_\text{bh}}

\title[The jet-ISM interactions in \icfost{}]{The jet-ISM interactions in \icfost{}}

\author[D. Mukherjee et al.]{%
Dipanjan Mukherjee$^{1,7}$\thanks{E-mail: dipanjan.mukherjee@unito.it},
Alexander Y. Wagner$^{2,6}$,
Geoffrey V. Bicknell,$^{1}$
\newauthor
Raffaella Morganti$^{3,4}$,
Tom Oosterloo$^{3,4}$,
Nicole Nesvadba$^{5}$,
Ralph S. Sutherland$^{1}$
\\
$^{1}$Research School of Astronomy \&{} Astrophysics, Mount Stromlo Observatory, Cotter Road, Weston Creek, ACT 2611, Australia \\
$^{2}$Center for Computational Sciences, University of Tsukuba, 1-1-1 Tennodai, Tsukuba, Ibaraki 305-8577 Japan\\
$^{3}$ASTRON, the Netherlands Institute for Radio Astronomy, Postbus 2, 7990 AA, Dwingeloo, The Netherlands\\
$^{4}$Kapteyn Astronomical Institute, University of Groningen, PO Box 800, 9700 AV Groningen, The Netherlands\\
$^{5}$Institut d'Astrophysique Spatiale, CNRS, Centre Universitaire d'Orsay, Bat. 120$-$121, 91405 Orsay, France\\
$^{6}$Institut d'Astrophysique de Paris, 98 bis bd Arago, F-75014 Paris, France \\
$^{7}$Dipartimento di Fisica, Universit\`a degli Studi di Torino, Via Pietro Giuria 1, 10125 Torino, Italy
}

\date{Accepted XXX. Received YYY; in original form ZZZ}
\pubyear{2017}

\begin{document}
\label{firstpage}
\pagerange{\pageref{firstpage}--\pageref{lastpage}}
\maketitle

\begin{abstract}
The interstellar medium of the radio galaxy \icfost{} is highly perturbed by an AGN jet expanding in the gaseous disc of the galaxy. We model this interaction with relativistic hydrodynamic simulations and multiphase initial conditions for the interstellar medium and compare the results with recent observations. As the jets flood through the inter-cloud channels of the disc, they ablate, accelerate, and disperse clouds to velocities exceeding $400\kms$. Clouds are also destroyed or displaced in bulk from the central regions of the galaxy. Our models with jet powers of $10^{44}\ergs$ and $10^{45}\ergs$ are capable of reproducing many of the features seen in the position-velocity diagram published in Morganti et al. (2015) and confirm the notion that the jet is responsible for the strongly perturbed gas dynamics seen in the ionized, neutral, and molecular gas phases. 
In our simulations, we also see strong venting of the jet plasma perpendicular to the disc, which entrains clumps and diffuse filaments into the halo of the galaxy. Our simulations are the first 3D hydrodynamic simulations of the jet and ISM of \icfost{}.
\end{abstract}




\section{Introduction}

Studying the interaction of AGN jets with the interstellar matter (ISM) of gas-rich galaxies can shed light on how galaxies and black holes co-evolve \citep{Wagner2011a}. Many low and high-redshift radio-galaxies show highly disturbed gas kinematics and outflows \citep{Morganti2005a, Garcia-Burillo2014a, Collet2015a,Nesvadba2017a}. The outflows are sometimes found to be dominated in mass by the molecular phase \citep{dasyra12a}. When resolved, these outflows often appear aligned with the radio jet \citep{Mahony2013a, Nesvadba2008a}, which suggests that the jet is directly responsible for the turbulent dispersion of the ISM and acceleration of outflows. 

\icfost{} is Type 2 Seyfert in a nearby elliptical galaxy (z=0.0113) with powerful ($P_{1.4 \mbox{\small  GHz}}=3\times10^{23} \mbox{W Hz}^{-1}$) extended radio structures, which exhibit strong interactions with the ISM \citep{Morganti1998a,Morganti2007a, Tadhunter2014a}. The galaxy contains a gaseous disk with large scale dust lanes \citep[see Fig.~1 of ][]{Oosterloo2017a}, possibly resulting from a merger with a gas rich galaxy \citep{morganti98a} in the past. The galaxy is seen almost edge on in the sky and the jet appears to be aligned parallel to the disc. This is not an uncommon feature of Seyfert galaxies, in which kiloparsec-scale radio structures are found to be randomly oriented with respect to the galaxy's major axis \citep{Gallimore2006a} or misaligned with the central gas disc \citep{cecil02a,Dopita2015a}.  

\icfost{} contains a reservoir of molecular gas exceeding $10^9\Msun$ \citep{Morganti2015a}. Within the region containing the strong radio emission, \icfost{} exhibits kinematically perturbed gas in the warm ionized phase \citep{Morganti2007a, Dasyra2015a}, neutral (HI) phase \citep{Morganti1998a, Oosterloo2000a}, warm molecular phase \citep{Tadhunter2014a}, and cold molecular phase recently observed with ALMA by \cite[][M15 henceforth]{Morganti2015a} and \cite{Dasyra2016a}. On the whole, the strong velocity dispersions of $\gtrsim500\kms$ seen in the various phases appear to be caused by the jet and the energy bubble it drives, with a particularly strong interaction with the ISM occurring in the western hotspot, about $500 \pc$ from the core.\footnote{We use $1\arcsec{} = 232\pc$, as in \citet{Oosterloo2017a}.}

The outflowing molecular gas was found to be optically thin, and the mass estimate of outflowing molecular gas is at least $1.3\times10^6\Msun$ \citep{Oosterloo2017a}. In comparison, the mass of the neutral HI outflow associated with the western hot spot is approximately $3.6\times10^6\Msun$ \citep{Morganti2007a}. The mass outflow rate at the western hotspot in molecular gas is estimated to be at least $4\Msunyr$ \citep{Oosterloo2017a}.

It is not clear whether the outflowing molecular gas is mostly driven from the disc \citep{Scannapieco2017a}, or whether a substantial fraction of the molecular emission in outflow comes from molecules formed in outflowing diffuse gas, which is cooling rapidly \citep{Richings2017a}. Observations of the near-infrared H$_2$ lines imply an outflowing H$_2$ mass at the western hotspot of only approximately $8\times10^2\Msun$, suggesting that the warm molecular gas component is only a transitory phase in the process of gas cooling behind strong radiative shocks driven into the clouds accelerated by the jets \citep{Tadhunter2014a}. The detailed mechanism of molecule formation in outflows is, however, not well understood.

In this paper, we present idealized relativistic hydrodynamic simulations of the inner kilo-parsec region of \icfost{}, concentrating on the kinematic signatures resulting from the interaction of the relativistic jets with the inhomogeneous ISM. The parameters of the simulations  are chosen based on observational constraints outlined in M15.

In \S~\ref{sec:pvintro}, we summarize the main features and interpretations of the kinematic signatures presented through the position-velocity (PV) diagram in Fig.~3 in M15. We describe the equations that are solved, the numerical methods and set up in \S~\ref{sec:eqns} and the parameters of the simulations in \S~\ref{sec:parameters}. The results are presented in \S~\ref{sec:results}, in which we first describe the evolution of the jet and the ISM and then focus on a comparison of the observed PV diagram (M15) with synthetic PV diagrams computed from our simulations. Additional results from our simulations are noted in a third subsection. We provide a summary and short discussion in \S~\ref{sec:summary}.

\section{Kinematic signatures in \ceeoh{2}{1}}\label{sec:pvintro}
The principal aim of the work presented in this paper is to investigate whether jet-ISM interactions are responsible for the enhanced velocity dispersion observed in the ISM of \icfost{}. To this end, we pay particular attention to the position-velocity (PV) diagram presented in Figure 3 of M15. The PV diagram show several characteristics indicative of strong jet-ISM interactions:
\begin{enumerate}
\item The dispersion in velocity space in the regions impacted by the jets ($\pm 2.5$\arcsec{}) are broadened to a few $100\kms$. The unperturbed gas at the outer edges has a much lower dispersion ($\sim 40\kms$).
\item The broadening across the region impacted by the jet is not smooth. A number of spiky features are seen. A particularly strong feature is the dispersion in the western part that coincides with the radio hotspot.  
\item The broadening across the region impacted by the jet is not symmetric about the center of the galaxy. The western side appears to be more strongly dispersed than the eastern side.
\item There is substantial emission in the regions away from the mean rotation curve (forbidden quadrants) implying strong gas motions in directions against the rotation of the disc.  
\item The gradient of the rotation profile in the region impacted by the jet appears to be shallower than the gradient expected from the gravitational potential (white line in Fig 3 of M15).
\end{enumerate}

M15 outlined a model, sketched in their Fig.~7, to explain some of the features described above. In their model, the jet-ISM interactions displaced clouds that were located near the center of the galaxy to larger disc radii, leading to clouds rotating more slowly with respect to the unperturbed gas. This model explained the first and last points in the above list, but could not directly explain the other points.

The asymmetry of the gas kinematics and the asymmetric radio morphology hint at a clumpy gas distribution in the disc \citep{Gaibler2011a}. Theoretical papers investigating the interactions of outflows with a turbulent gas distribution \citep{Sutherland2007a, Wagner2011a, Gaibler2011a, Wagner2012a, wagner13a, mukherjee16a} have shown that modelling the ISM as an inhomogeneous medium is crucial for correctly modelling observational characteristics and calculating feedback efficiencies.

\section{Numerical set up}\label{sec:eqns}
We use the open-source software PLUTO \citep{Mignone2007a} version 4.1 to numerically solve the equations of relativistic hydrodynamics, using an Eulerian finite-volume Godunov-type scheme as described in \citet{mignone05a}. We also solve two additional advection equations, which trace jet plasma and dense clouds, respectively, in our simulations. The code and integration schemes used are identical to those used by \citet{mukherjee16a}. We employ the third order Runge-Kutta time-stepping scheme together with the piecewise parabolic method of spatial interpolation. 

The temperature $T$ of the gas is given by the ideal gas law, $p=\rho \kB T / \mmpp$, where $\kB$ is Boltzmann's constant, and $\mmpp$ is the mean mass per particle. Atomic cooling is implemented using a tabulated non-equilibrium radiative cooling function $\Lambda(T)$ for an optically thin plasma generated by the MAPPINGS~\rom{5} code \citep{Sutherland2017} for solar metallicity gas. The cooling is dominated by emission from collisionally ionized atoms. 

Between the cooling floor of $100\kelvin$ and the peak of the cooling curve at just above $10^5\kelvin$, the mean mass per particle decreases monotonically from $\mmpp\approx1.24\amu$ to $\mmpp\approx0.604\amu$. The value of $\mmpp$ for a given temperature is also obtained from the precomputed Mappings~\rom{5} cooling function, in which $\mu$ is tabulated as a function of $p/\rho$. At relativistic pressures, a relativistic correction to Bremsstrahlung  as implemented by \citet{Krause2007a} is employed.  

Molecular cooling is not included in this work, and the cooling below $10^4 \kelvin$ down to the cooling floor of 100 K is by metal line cooling alone. In our simulations, the gas in the dense cores is quickly driven to the cooling floor even with purely atomic cooling, so any additional effects due to cooling by molecules does not influence the dynamics of the gas in this work.

\section{Simulation setup and choice of parameters}\label{sec:parameters}

\begin{table*}
\centering
\caption{Simulation parameters}
\label{tab:parameters}
\begin{tabular}[ht!]{ c l c c }
\hline
Parameter & Description & \multicolumn{2}{c}{Value} \\
\hline
\textbf{Jet Parameters} &&& \\
$\Pjet$ & Jet power & $10^{44}\ergs$ (\PFF) & $10^{45}\ergs$ (\PFV) \\
$\Gamma$ & Jet bulk Lorentz factor at nozzle & 4 & 6\\
$\chi$ & Jet proper density parameter$^1$ & 4.8 & 173 \\
$\rjet$ & Jet radius & \multicolumn{2}{c}{$25\pc$} \\
$\thjet$ & Jet opening angle & \multicolumn{2}{c}{$20^\circ$} \\
\hline
\textbf{Potential \& Hot Halo} &&& \\
$M$ & Total mass in Hernquist profile & \multicolumn{2}{c}{$1.67\times10^{11}\Msun$} \\
$a$ & Scale length of Hernquist profile & \multicolumn{2}{c}{$2.8\kpc$} \\
$\rho_h (0)$ & Central density of hot atmosphere & \multicolumn{2}{c}{$0.1\cc$} \\
$T_h$ & Temperature of hot atmosphere & \multicolumn{2}{c}{$6\times10^6\kelvin$} \\
\hline
\textbf{Clouds and disc} &&& \\
$T_c$ & Critical temperature for warm phase & \multicolumn{2}{c}{$3\times10^4\kelvin$} \\
$\bar\rho(0,0)$ & Mean central density of clouds & \multicolumn{2}{c}{$200\cc$} \\
$\sigma$ & Combined turbulent and thermal dispersion & \multicolumn{2}{c}{$40 \> \rm km \> s^{-1}$} \\
$e_{\rm K}$ & Ratio of azimuthal speed to Keplerian speed & \multicolumn{2}{c}{0.9} \\
\hline
\end{tabular}\\
$^1$ Ratio of jet rest mass energy to enthalpy (eq.~\ref{eqn:jet_chi}).
\end{table*}

The simulations of the jet-ISM interactions in \icfost{} are idealized isolated galaxy simulations. The focus of the simulations is on the interaction of the jet plasma with the hot, warm, and cold gas phases in the ISM. 

A cubical region of $2\kpc$ width around the central black hole is initialized with a jet nozzle at the center and a two-phase ISM representing the hot phase and warm to cold phase in \icfost{}. We use a Cartesian coordinate system ($X$--$Y$--$Z$) in which the galaxy disc is co-planar with the $X$--$Y$ plane; the $Z$--axis is the symmetry axis of the disc. The jet is aligned in the plane of the disc along the $X$--axis.
The computational domain is divided into a grid of $512^3$ uniform cells, providing a spatial resolution of approximately $4\pc$ per cell.

The duration of the simulations is set by time required for the jet head to advance to at least $0.5\kpc$ from the center of the galaxy. This timescale is of order $1\Myr$ (see Appendix~\ref{sec:tjet}). We can therefore ignore the self-gravity of the warm phase gas and the effects of star-formation and supernova feedback in these simulations. We also ignore the effects of magnetic fields, which are expected to make only a small difference in terms of the energy transfer efficiency from jets to the ISM \citep{Asahina2017a}.

The aim of the simulations is to provide a fully hydrodynamic model that qualitatively, and to some extent quantitatively, explains the observations of \icfost{}. We have therefore not performed a costly parameter space study, but have chosen parameters informed from observations. We present two models with differing jet powers. The parameters for the simulations are listed in Table~\ref{tab:parameters}. In the following section we explain in more detail the setup of the simulations and the parameters chosen for the gravitational potential, the different phases of the ISM, and the jet.

\subsection{Gravitational potential and initial gas disc}\label{sec:potential}

In modelling the gravitational potential of the galaxy, we do not need to take the gravitational field of the black hole into account since, for a mass $\mbh\approx 2.8 \times 10^8 \Msun$ \citep{Nicastro2003a} and stellar velocity dispersion $\sigma \approx 160 \kms$ \citep{woo02a} the radius of influence of the black hole is $\sim G  \mbh / \sigma^2 \approx 46 \pc $, comparable to the radius of injection of the jet ($\sim 40$ pc) in the simulation. For comparison with the jet powers employed in the simulations, the Eddington luminosity of the black hole \emph{is} important and this is approximately $3.5 \times 10^{46} \ergs$.

We use the analytic Hernquist potential \citep{hernquist90a} to model the gravitational potential of \icfost{}.  The parameters of this potential are the total mass, $M$ and the scale length, $a$, where the mass , $M(r)$,  and the gravitational potential, $\Phi(r)$, as functions of spherical radius, $r$, are given by: 
\begin{align}
M(r) &= M \frac{r^2}{(r + a)^2} \\
\Phi(r) &= -\frac {GM}{r + a} \;.
\end{align}
We use the inferred effective radius $R_{\rm e} = 21.5\arcsec = 5139\pc$ \citep[M15, ][]{kulkarni98a} and the central velocity dispersion $160\kms$ to evaluate the parameters $M$ and $a$, as follows. Since the gravitational potential is most likely dominated by baryonic matter within the effective radius we use the \citet{hernquist90a} relation between half-light radius and scale length, viz, $R_{\rm e}  = 1.8153 \, a$, to estimate $a \approx 11.8\arcsec \approx 2.8\kpc$.

In the Hernquist model, the projected isotropic velocity dispersion rises rapidly to a maximum $\approx 0.319 \, (GM/a)^{1/2}$ at a projected radius $x \approx 0.224 \, a \approx 2\arcsec \approx 477\pc$ for \icfost{}. We equate the maximum velocity dispersion with the observed central value of $160\kms$ to obtain a mass $M \approx 1.67 \times 10^{11}\Msun$. 

These estimates can be checked for consistency using the observed major axis HI rotational velocity beyond the region of influence of the jet. Let $\theta \approx 74^\circ$ be the angle between the disc normal and the line of sight (M15). Then, in the Hernquist model, the observed rotational velocity is given by:
\begin{equation}
v_{\rm rot} = \pm \left( \frac {GM}{a} \right)^{1/2} \, \frac {(r/a)^{1/2}}{1+ r/a} \, \sin \theta \;.
\label{eqn:vcirc}
\end{equation}
The region just beyond the influence of the jet is at an angular distance of 4\arcsec, implying an observed rotational velocity $\approx \pm 209\kms$. This compares favourably with the observed speed of $\pm 190\kms$ M15.

\begin{figure*}
\centering
\includegraphics[height=5.1cm]{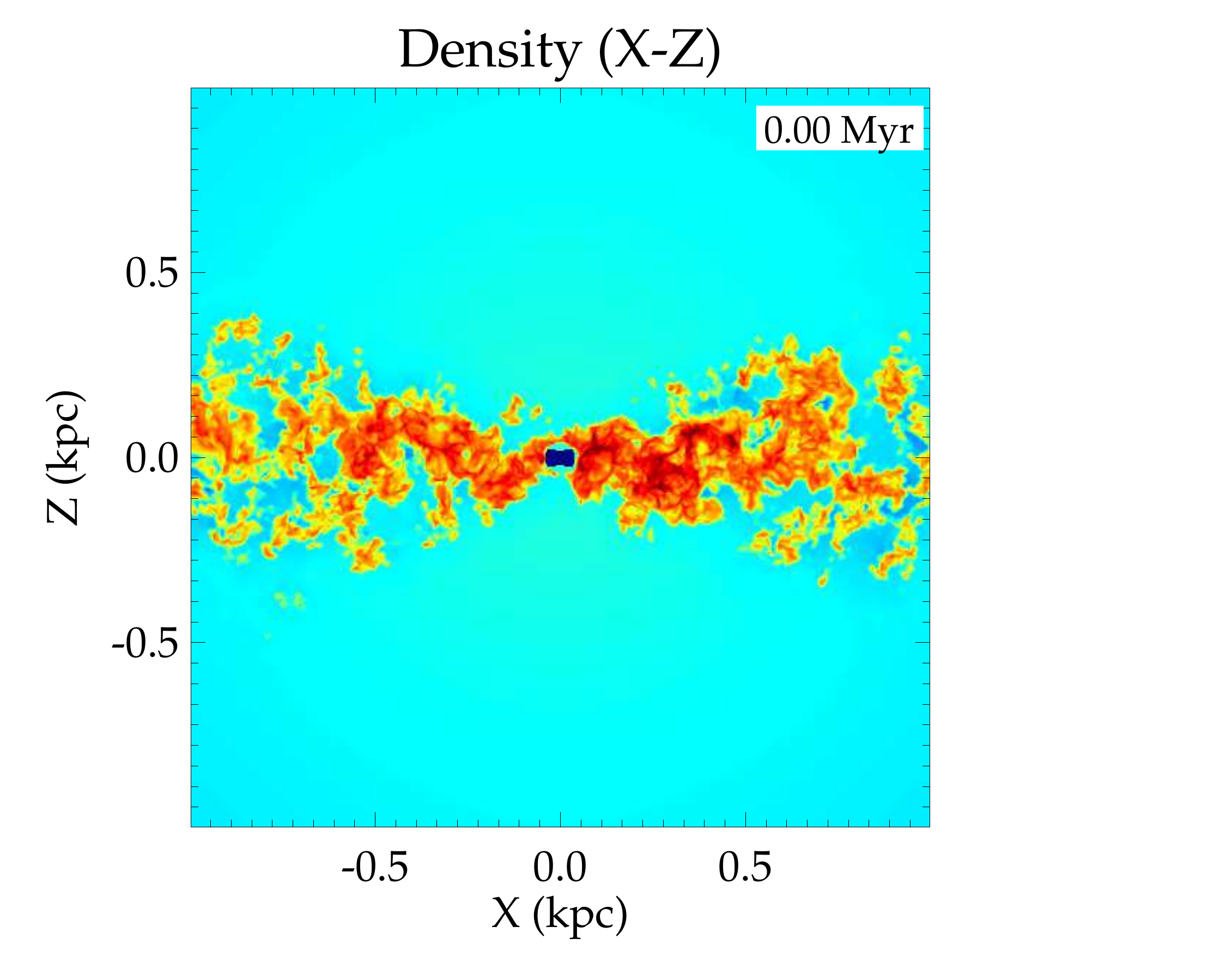}\hspace{-1.5cm}
\includegraphics[height=5.1cm]{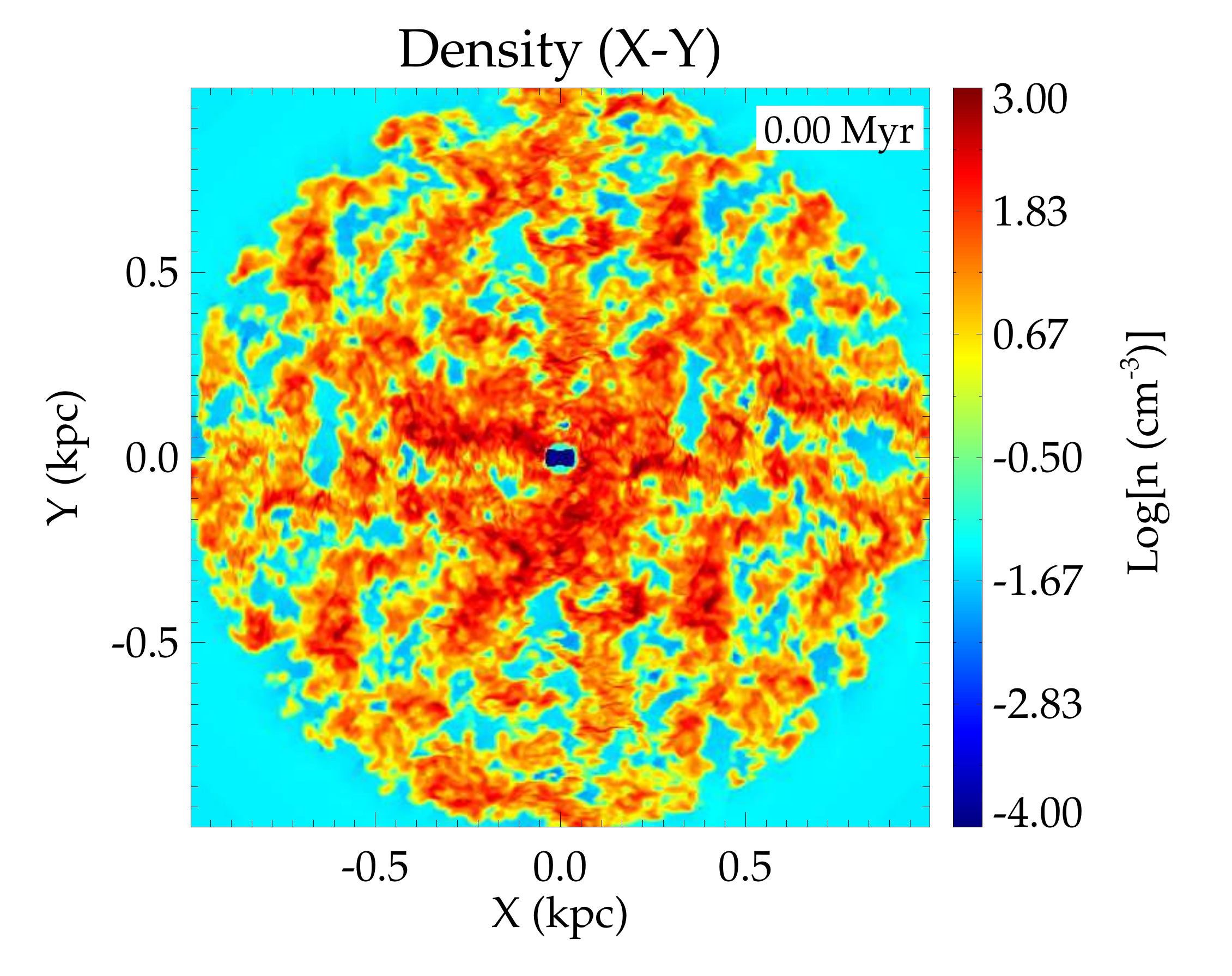}
\includegraphics[height=5.1cm]{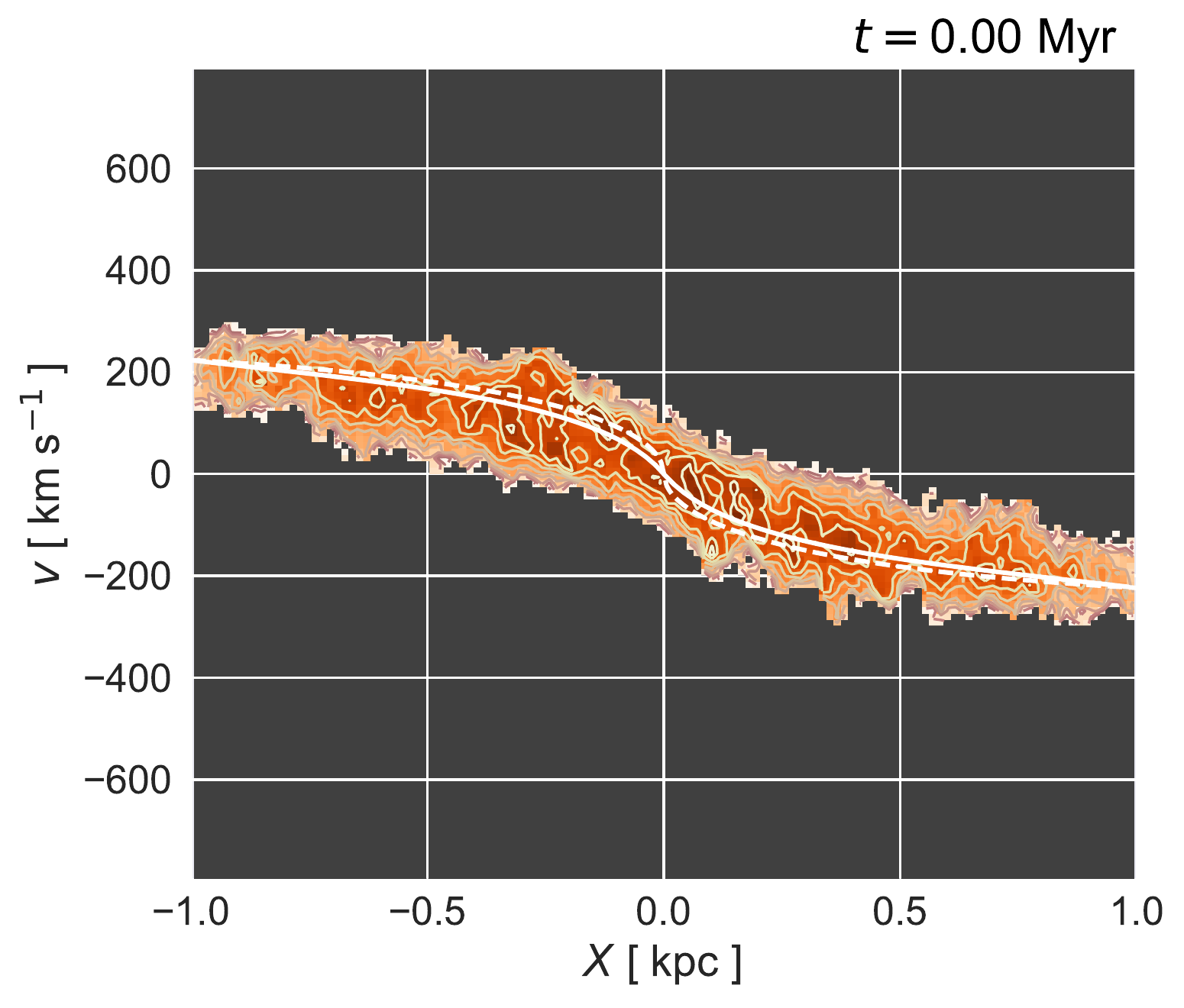}
\caption{\small Density ($\log(n [\cc])$) in the $X-Z$ (left) and $X-Y$ (center) planes at t=0. Right: Synthetic PV diagram based on \ceeoh{2}{1} emission at $t=0$ (based on Eq.~\ref{eqn:emiss}). The dashed and solid lines trace the circular velocity according to Eq.~\eqref{eqn:vcirc} and the projected circular velocity weighted by the mean density profile of the turbulent disc given by Eq.~\eqref{eqn:disc}}
\label{fig:ic}
\end{figure*}
The initial distribution of gas in the galaxy is defined by a hot isothermal atmosphere in hydrostatic equilibrium. Assuming the central density at $r=0$ to be $\rho_h(0)$, for gas with temperature $T_h$, mean molecular weight $\mu=0.6$ and atomic mass unit $m_a$, the halo density profile is given by 
\begin{equation}
    \rho _h=\rho_h(0) \left[\exp \left(\frac{-\phi(r)}{\kB T_h/(\mu m_a)}\right)\right]\;.
\end{equation}
The temperature of the hot gas is initialised to $T_h=6 \times 10^6$ K (as in Table~\ref{tab:parameters}).
 Superimposed on the halo is a turbulent disc prescribed using the approach described in \citet{Sutherland2007a}, which is summarized below. A number of cells in the computational volume are replaced by denser, cooler gas using the following approach. 
 
Let the cool gas density in cylindrical coordinates be $\rho(R,z)$, with $\rho(0,0)$ being the value at the center of the galaxy, $\sigma_t$ the turbulent velocity dispersion and $T$ the mean temperature. The azimuthal velocity of the disc is defined to be a constant fraction, $e_K$ of the Keplerian velocity. Then the \emph{mean} density of the disc is given by:
\begin{equation}
\frac {\bar \rho(R,z)}{\bar \rho(0,0)} = 
\exp - \left[ \frac {\Phi(R,z) - e_K^2 \Phi(R,0) - (1-e_K^2) \Phi(0,0)}{\sigma^2} \right]
\label{eqn:disc}
\end{equation}   
where
\begin{equation}
\sigma^2 =  \sigma_t^2 + \frac {k \tilde T}{\mmpp} \;.
\label{eqn:sigma}
\end{equation}
The distribution of dense gas is constructed by multiplying a fractal cube by the mean density described by Eq.~\eqref{eqn:disc}. The fractal cube consists of a log-normal density distribution with unit mean and a power-law spectrum in Fourier space. The temperature of the gas is defined by pressure equilibrium with the hot atmosphere. This gas distribution then replaces a number of cells in the hot atmosphere, except when the temperature in a cell exceeds a critical value $T_{\rm c}$, corresponding to a low density $\rho_c = \mmpp  \, p_{\rm hot}/kT_{\rm c}$. While the gas temperature, in principle, contributes to the overall dispersion parameter $\sigma$, in these simulations the contribution is very small since $\tilde T \sim 100\kelvin$ because of the cooling floor implemented. Hence, we define the cool gas density by neglecting the temperature contribution and assigning the parameter $\sigma_t$ alone.   

A Gaussian turbulent velocity field, with velocity dispersion $\sigma$ is imposed on the dense gas. This is an addition to the \citet{Sutherland2007a} Ansatz and was first implemented in \citet{mukherjee16a}. The parameters of the distribution are summarized in Table~\ref{tab:parameters} together with the other parameters of the simulation. The initial conditions are shown by two orthogonal mid-plane density slices through the simulation box at time $t=0$ and the PV diagram of the rotating disc (line of sight along $Y$) in Fig.~\ref{fig:ic}.

\subsection{The relativistic jet}\label{sec:jet}

We present two models which have identical setups and parameters, except for the jet power. The jet power in model \PFF{} is $10^{44}\ergs$, and that in model \PFV{} is $10^{45}\ergs$. The Eddington ratios of these jets are approximately 0.015 and 0.15, respectively. 

The choice of jet powers is partly motivated by our previous studies of jet-ISM interactions \citep{Wagner2011a}, in which only jets with powers exceeding $10^{44}\ergs$ were found to be capable of powering velocity dispersions greater than $\sim 400 \kms$. Using the scaling relation by \citet{cavagnolo10a} with $\Popfghz=3\times10^{23}\WpHz$ gives a jet power of $9\times10^{43}\ergs$; the scaling relation by \citet{willott99a} yields $5\times10^{43}\ergs$ (M15). We discuss these estimates further in \S~\ref{sec:summary}.

The relations between the jet power and proper density to the jet pressure, $\pjet$, internal energy, $\ejet$, density, $\rhojet$, bulk Lorentz factor, $\Gamma$ and velocity, $\vjet$ are:
\begin{align}
\Pjet &= \frac{\gamma}{\gamma - 1} \pjet \Gamma^2 c \vjet \Ajet \left(1 + \frac{\Gamma - 1}{\Gamma} \chi \right) \label{eqn:jet_power} \\
\chi &=  \frac{\rhojet c^2}{\ejet + \pjet} = \frac{(\gamma - 1) \rhojet c^2}{\gamma \pjet} \;. \label{eqn:jet_chi} 
\end{align}
The adiabatic index is assumed to be $\gamma=5/3$ for the jet plasma, corresponding to that of an ideal equation of state. Such an assumption is appropriate for $\chi > 1$ \citep{mignone07a}, as used in our simulation. An ideal equation of state is also a better description for the dense thermal gas included in the non-homogeneous ISM. More on this has been discussed in section 2.3 of \citet{mukherjee16a}. The parameters for the jet are included in Table~\ref{tab:parameters}.

A small spherical region with a radius of $40\pc$ around the central black hole is clear of any warm phase gas and is set as a fixed internal boundary from which the jet emerges conically and bidirectionally with an opening angle of $20^\circ$. The jet radius at the inlet is $25\pc$. These initial conditions are prescribed to represent a freely expanding conical jet

The orientation of the jet with respect to the line of sight is not well constrained observationally, but the jet axis is thought to lie close to the plane of the disc and perpendicular to the line of sight. For simplicity, we orient the jet exactly along the disc axis. The evolution of the jet is dominated by its interaction with the warm phase of the ISM and the random and fractal clumpiness of the warm phase ensures that this choice of jet orientation does not lead to a vastly different solution than cases where the jet is slightly inclined with respect to the disc.

\section{Results}\label{sec:results}
We first describe the general features of the evolution of the jet and the ISM in the simulations. We then compare the kinematics of the jet-perturbed gas in the disc as seen in our simulations with the \ceeoh{2}{1} observations of M15.
\subsection{Evolution of jet and ISM}\label{sec:evolution}
\begin{figure*}
\centering
\includegraphics[width=6.2cm, keepaspectratio]{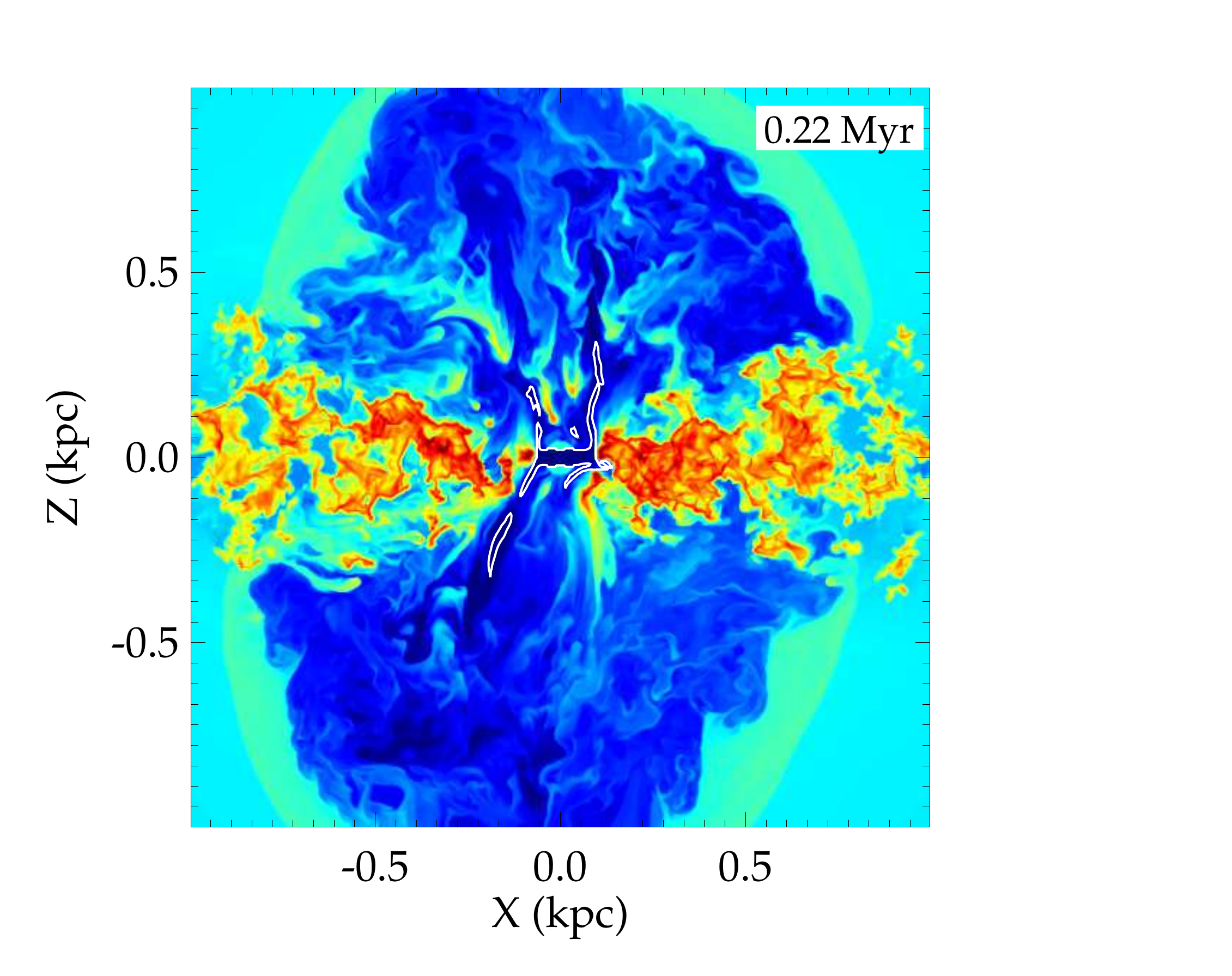} \vspace{-0.1cm}\hspace{-2.5cm}
\includegraphics[width=6.2cm, keepaspectratio]{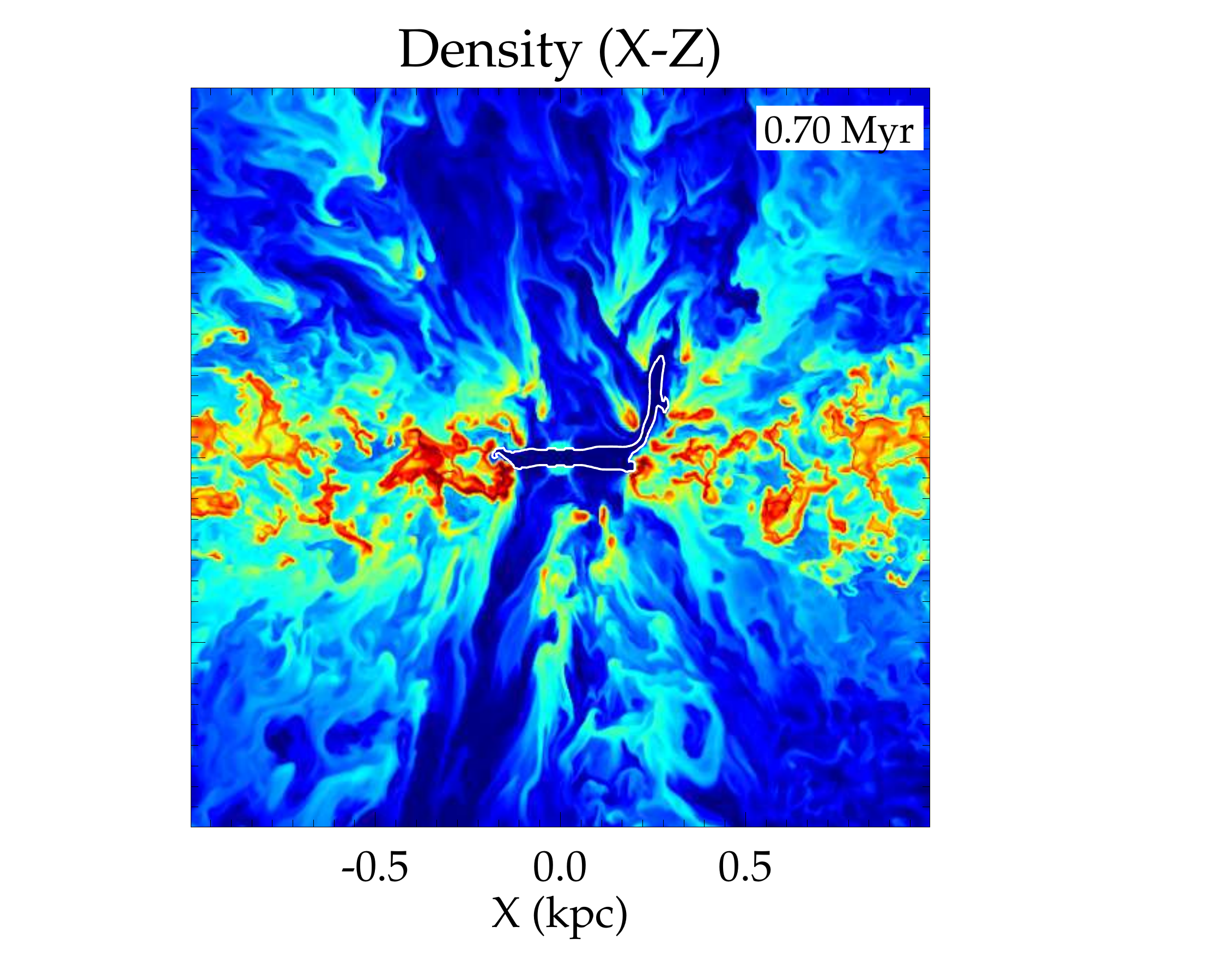} \vspace{-0.1cm}\hspace{-2.5cm}
\includegraphics[width=6.2cm, keepaspectratio]{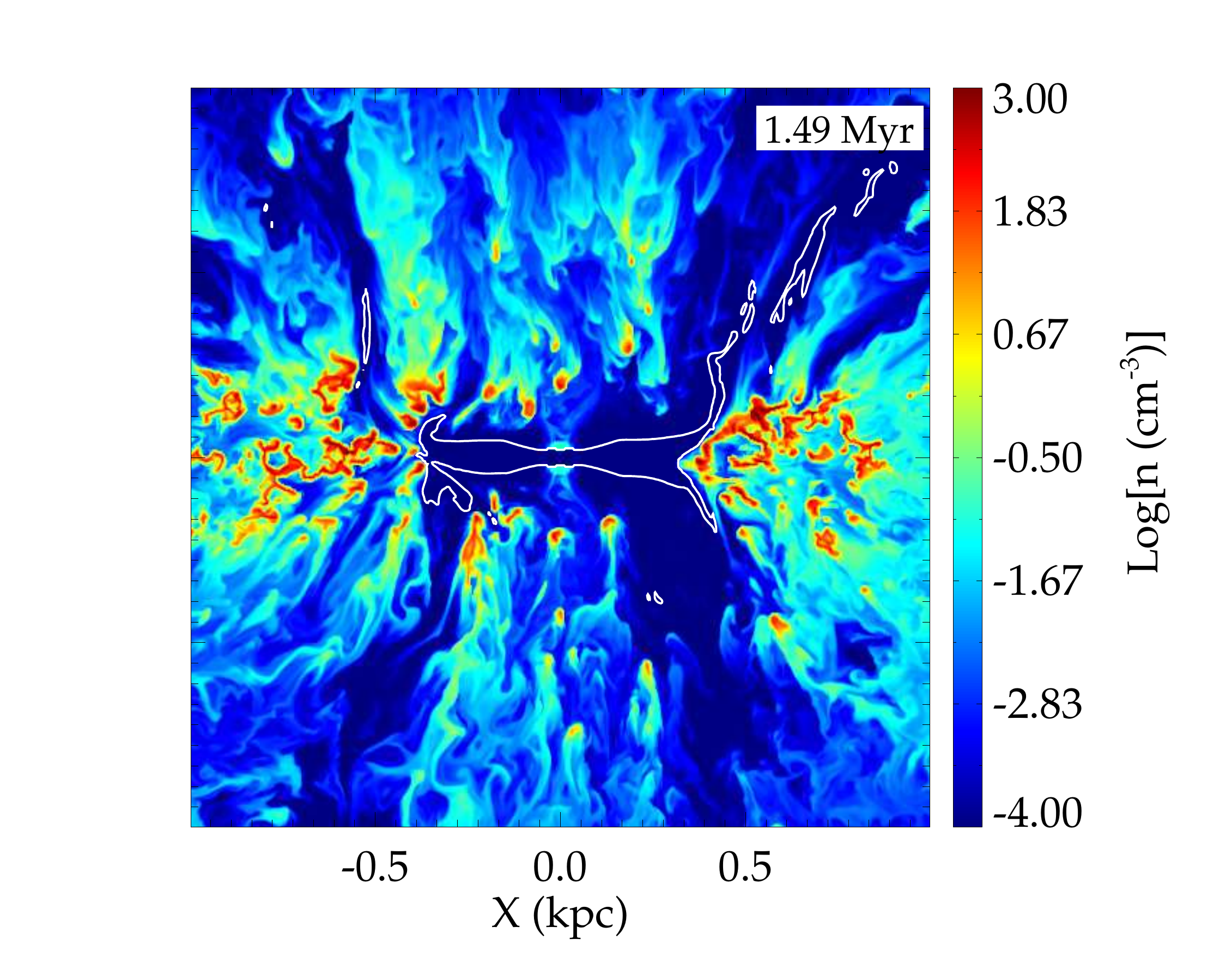} \vspace{-0.1cm}\linebreak
\includegraphics[width=6.2cm, keepaspectratio]{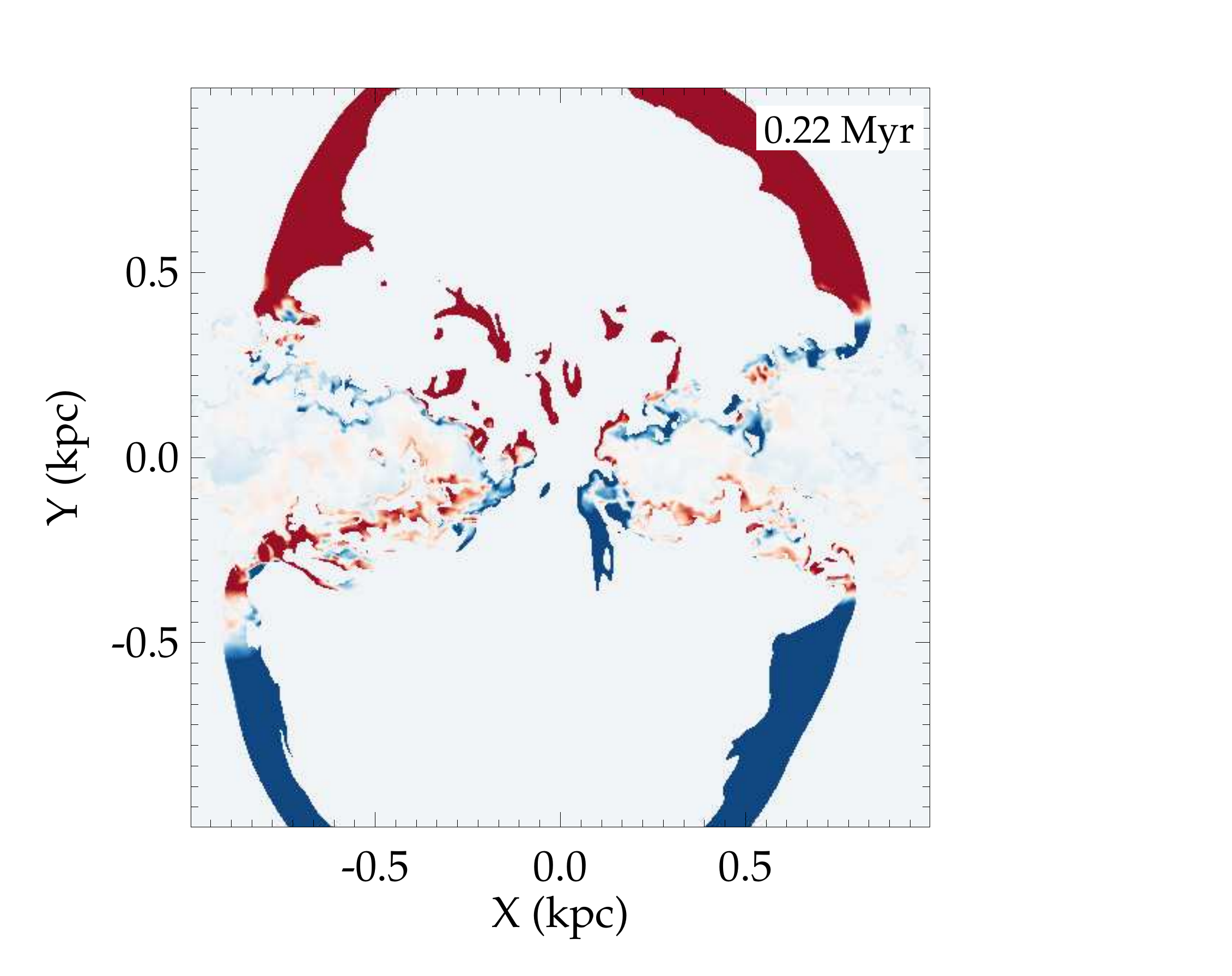} \vspace{-0.1cm}\hspace{-2.5cm}
\includegraphics[width=6.2cm, keepaspectratio]{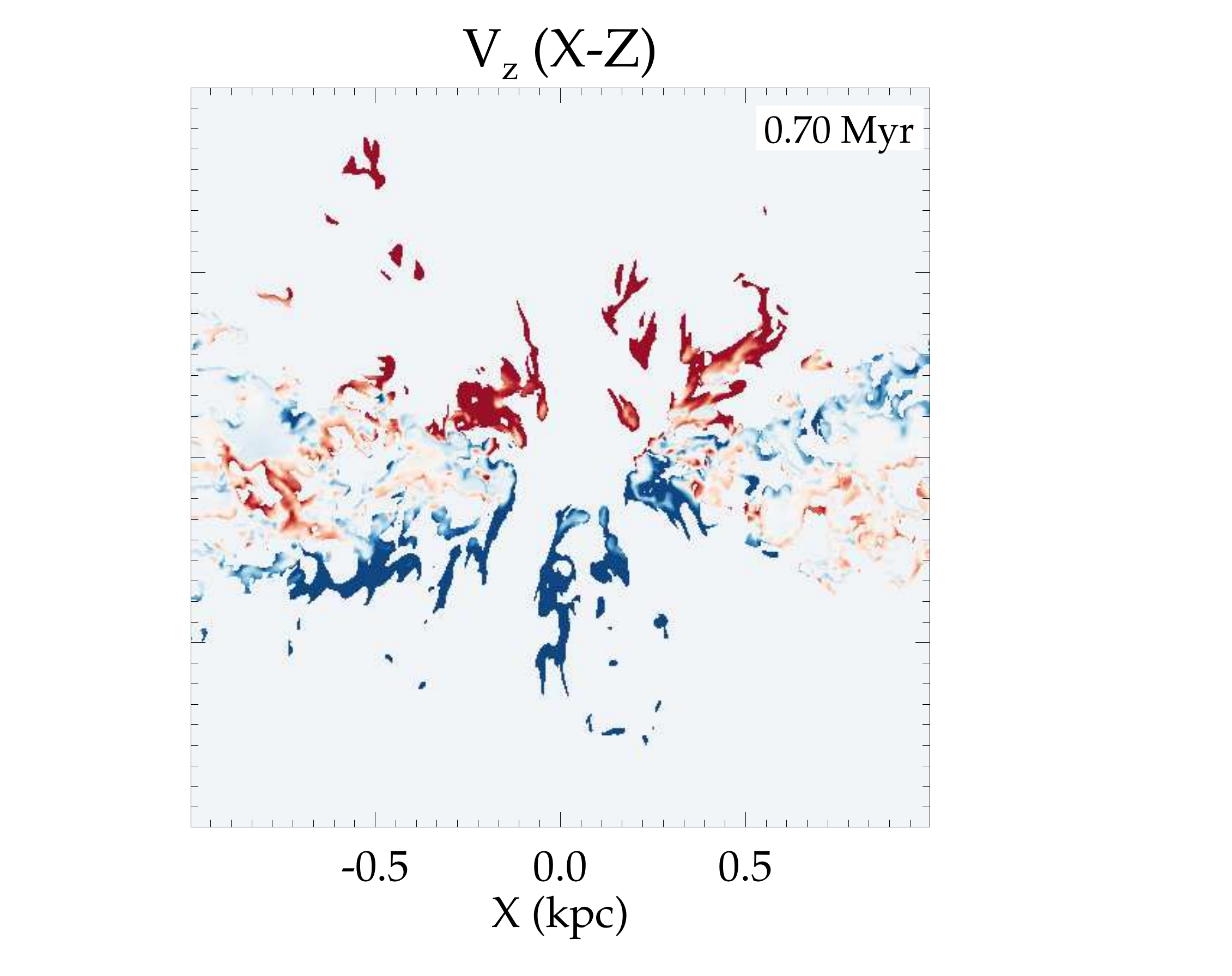} \vspace{-0.1cm}\hspace{-2.5cm}
\includegraphics[width=6.2cm, keepaspectratio]{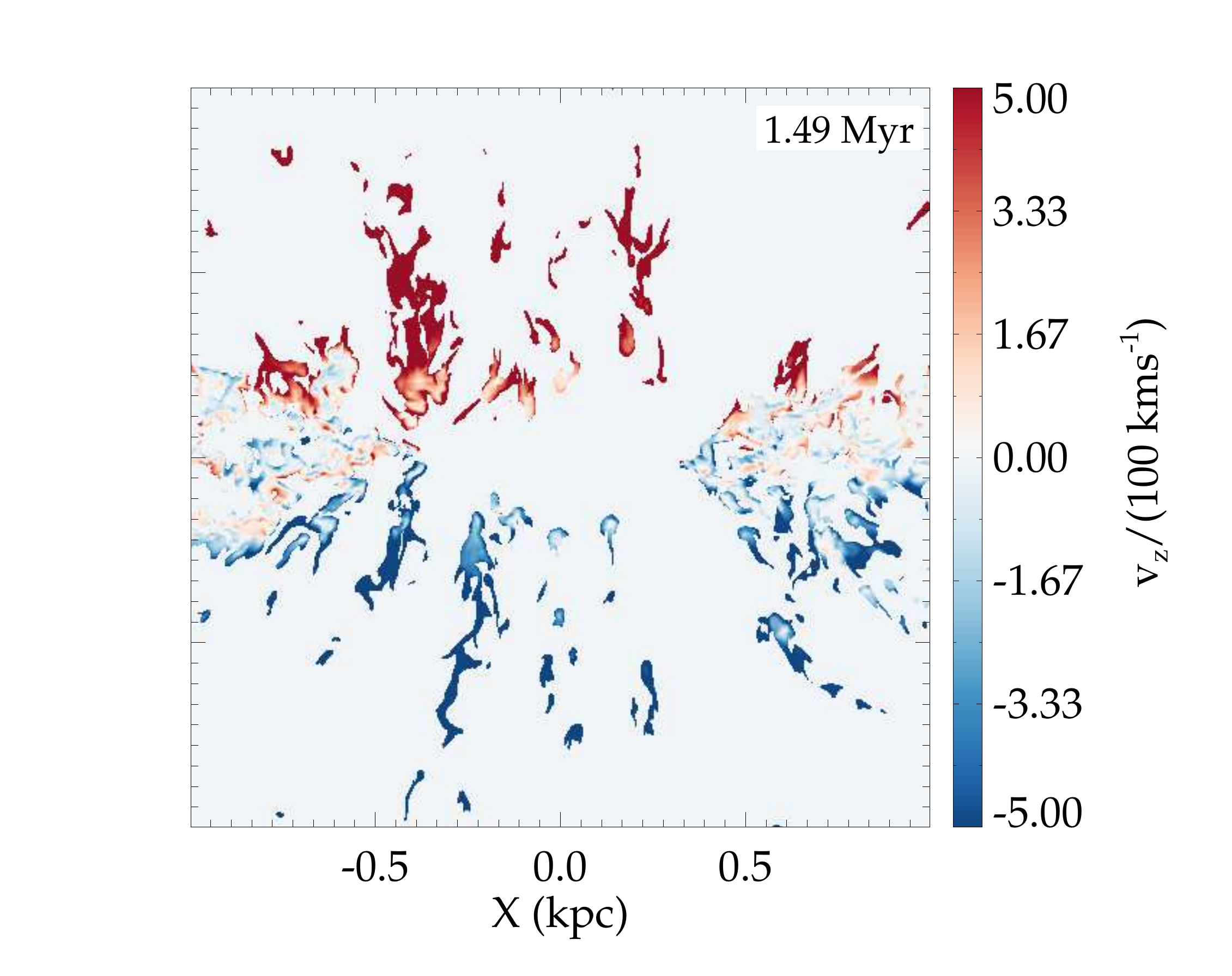} 
\includegraphics[width=6.2cm, keepaspectratio]{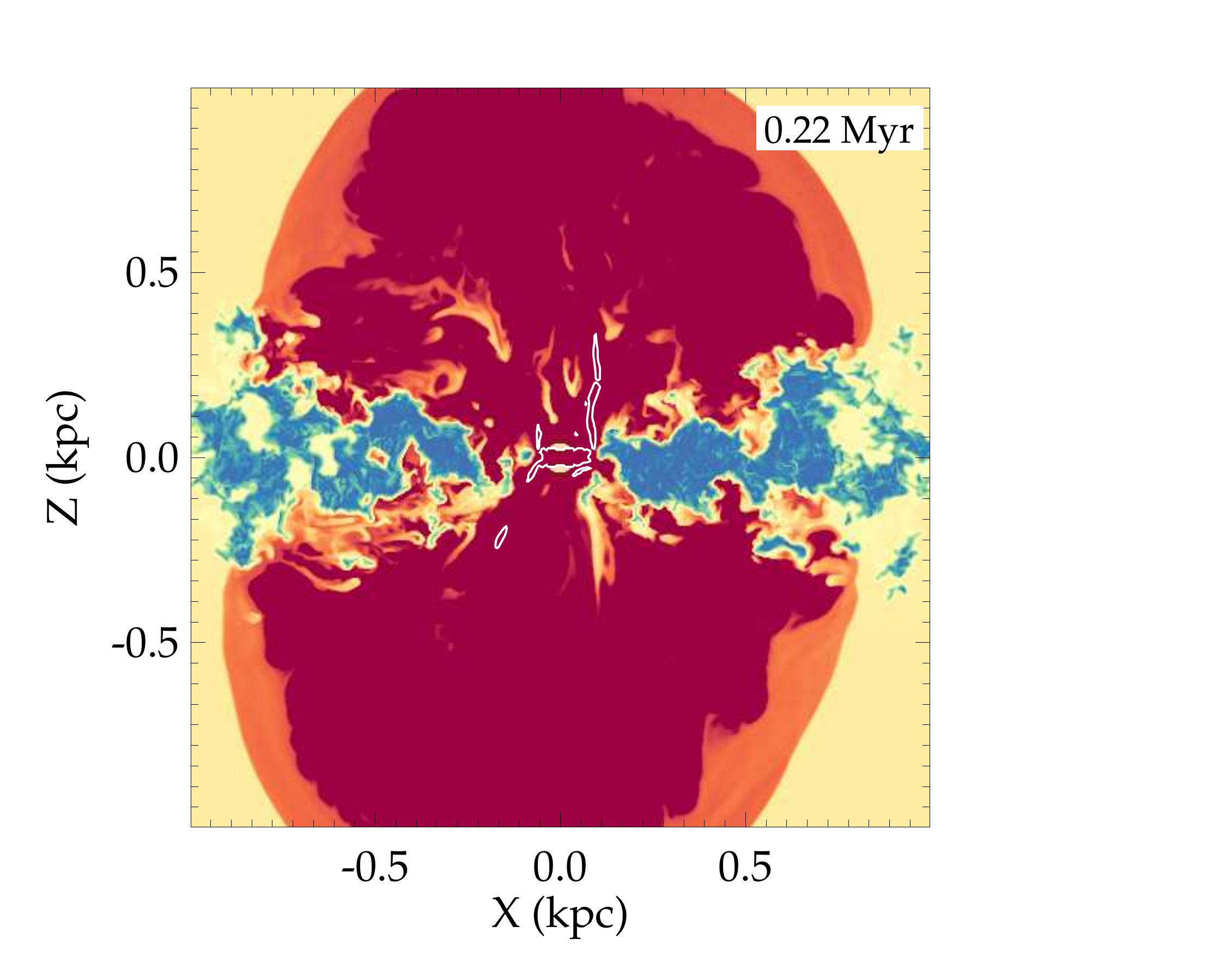} \vspace{-0.1cm}\hspace{-2.5cm}
\includegraphics[width=6.2cm, keepaspectratio]{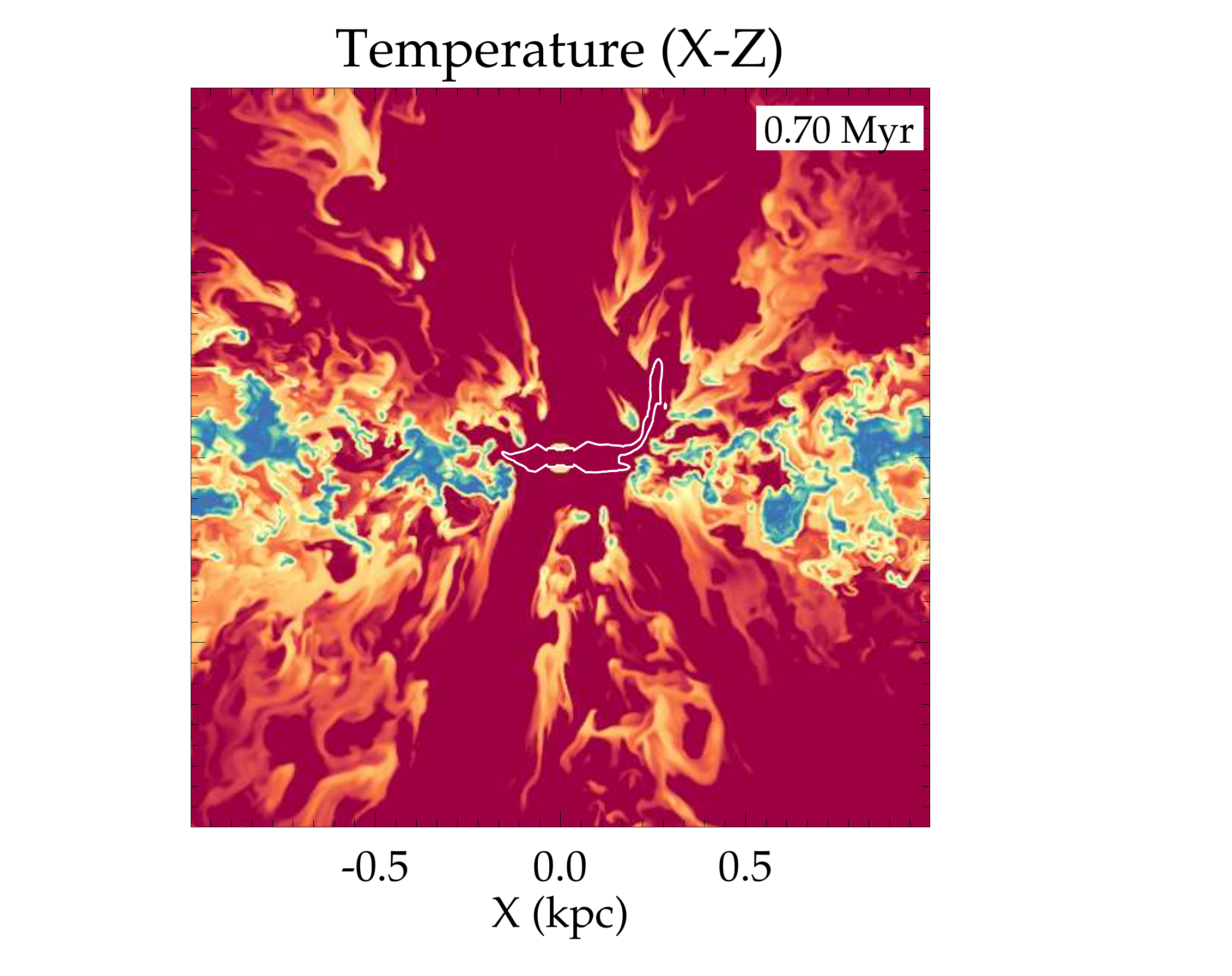} \vspace{-0.1cm}\hspace{-2.5cm}
\includegraphics[width=6.2cm, keepaspectratio]{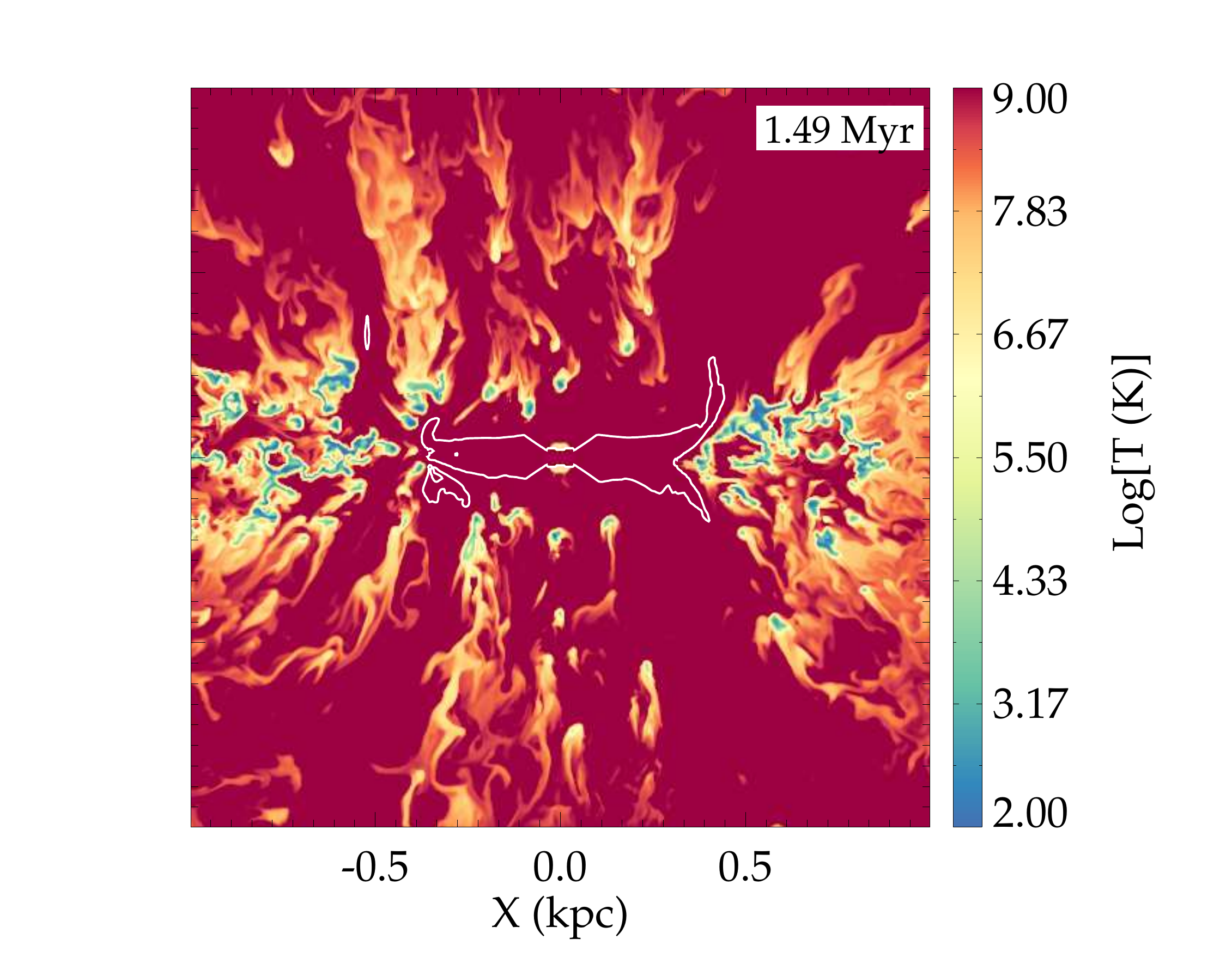} 
\caption{\small Mid-plane slices in the $X-Z$ plane ($Y=0$) of the evolution of simulation \PFF{}, for which the jet power is $\Pjet=10^{44}\ergs$. The last column of panels with $t=1.49\Myr$ corresponds roughly to the current extent of the jets in \icfost{}. The top row of panels shows the density ($\log(n [\cc])$), the middle row the vertical component of the velocity ($\vel_z$) normalised to $100 \kms$, for gas with density $n > 0.1\cc$, and the bottom row shows the temperature. The white line corresponds to the contours of jet tracer with value 0.1 (maximum being 1), demarcating the region with non-thermal plasma.}
\label{fig:jet44xz}
\end{figure*}
\begin{figure*}
\centering
\includegraphics[width=6.2cm, keepaspectratio]{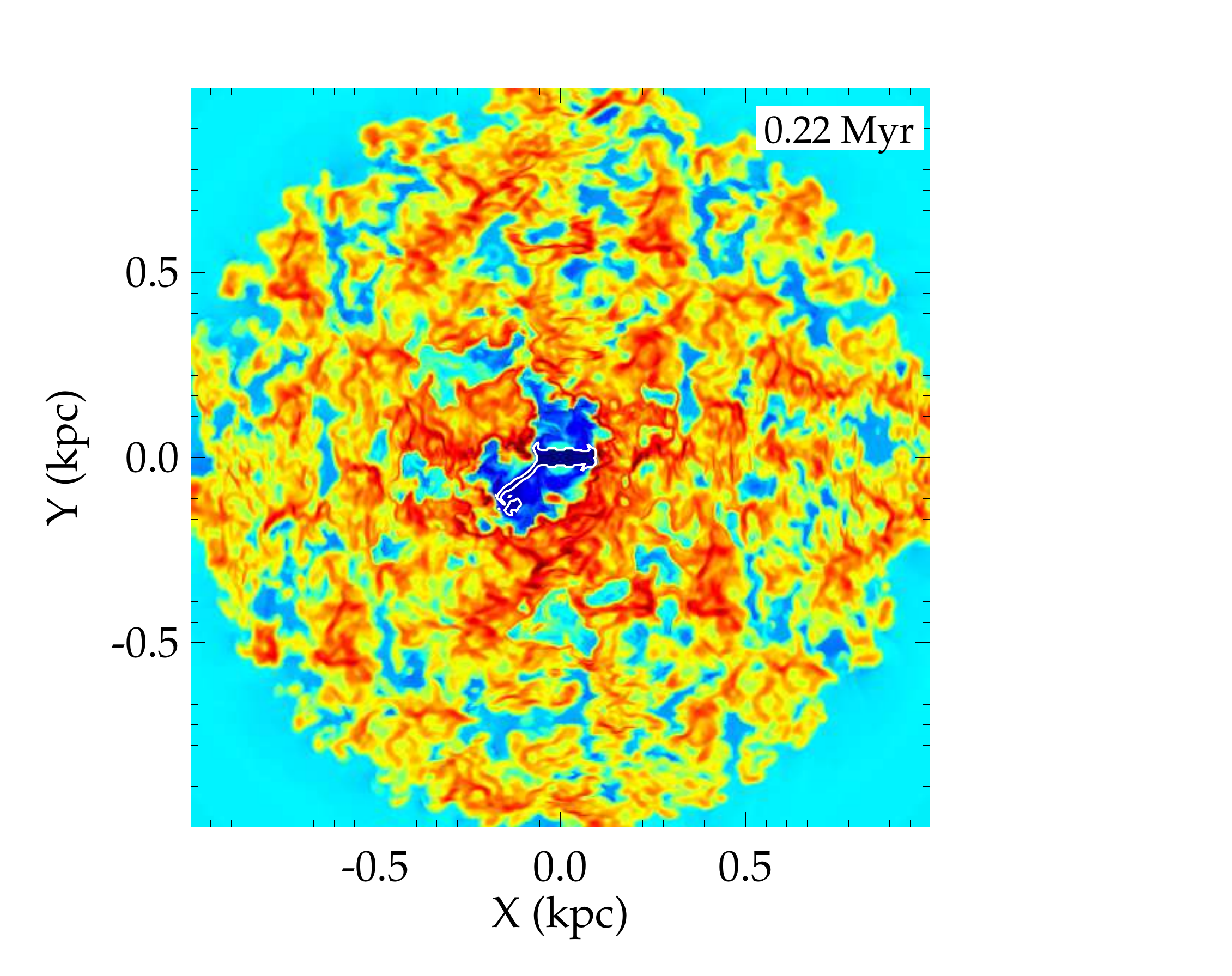} \vspace{-0.1cm}\hspace{-2.5cm}
\includegraphics[width=6.2cm, keepaspectratio]{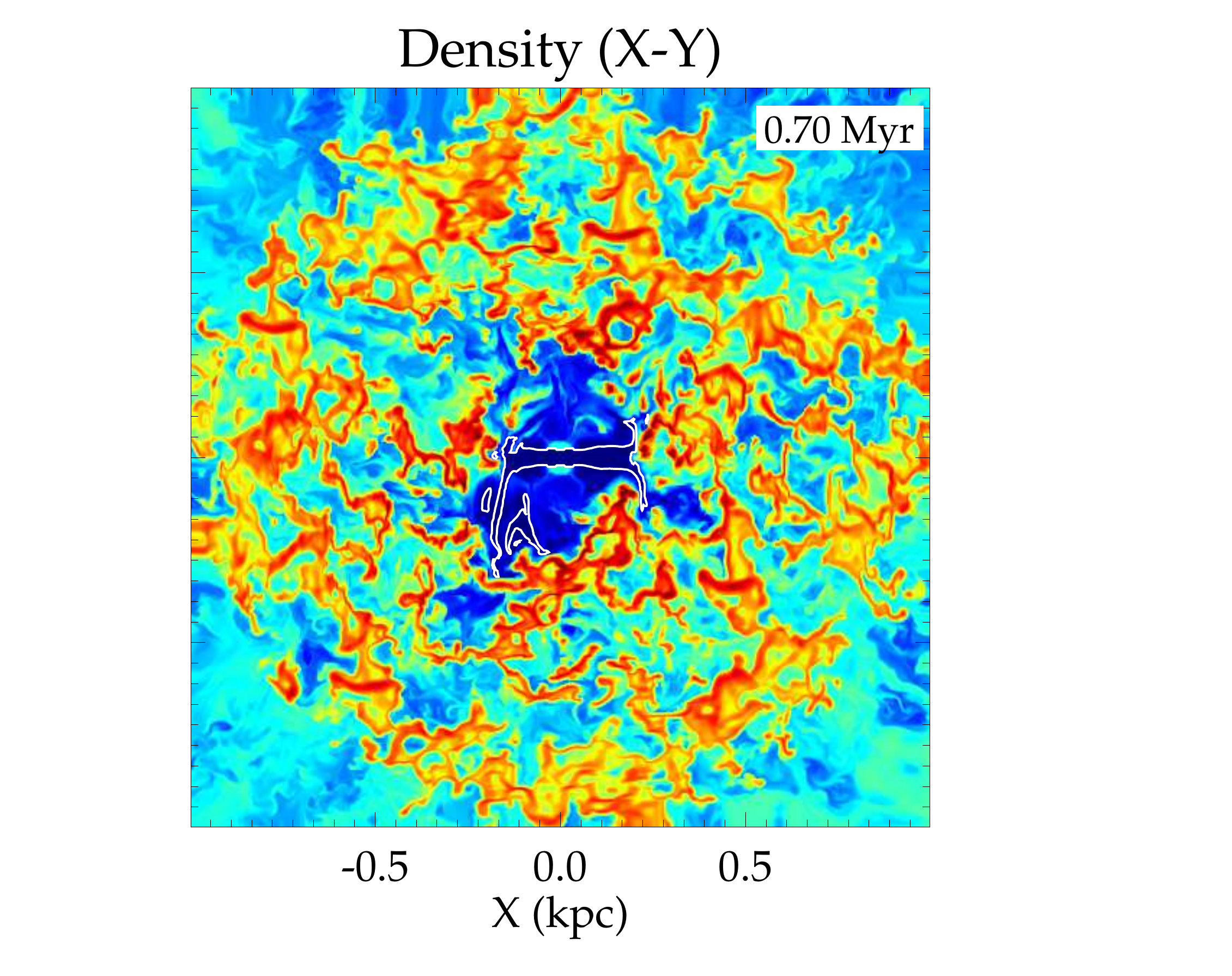} \vspace{-0.1cm}\hspace{-2.5cm}
\includegraphics[width=6.2cm, keepaspectratio]{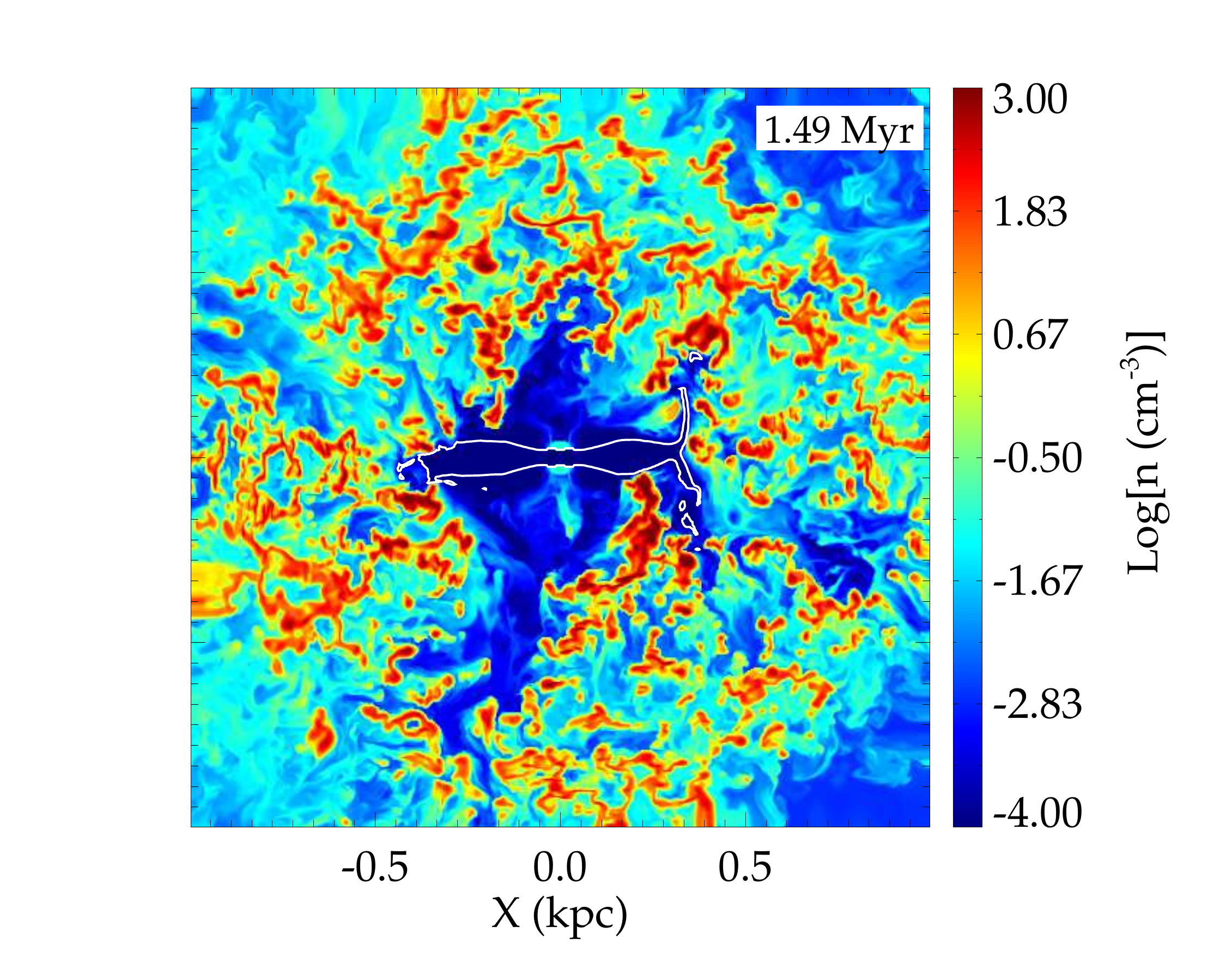} \vspace{-0.1cm}\linebreak
\includegraphics[width=6.2cm, keepaspectratio]{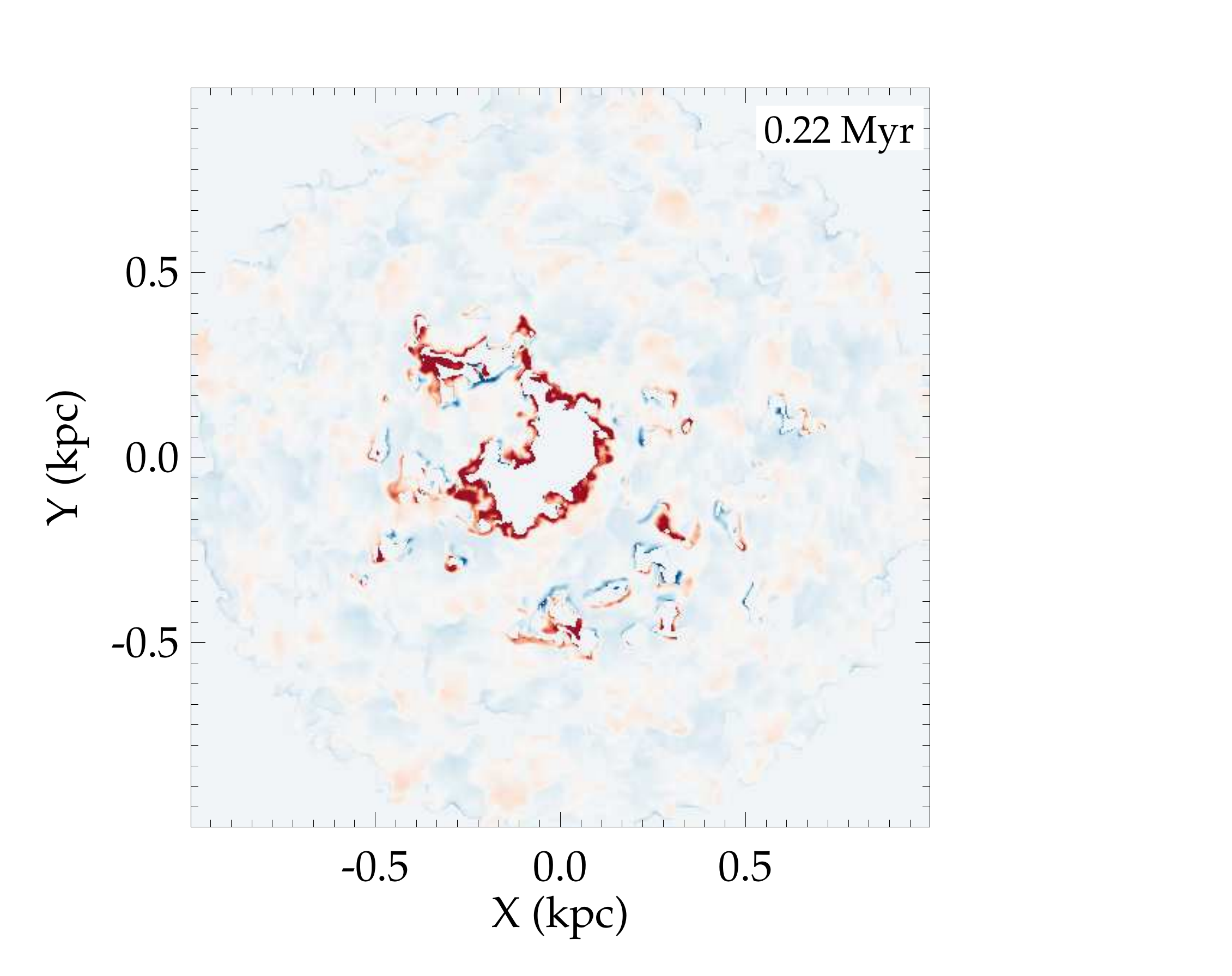} \vspace{-0.1cm}\hspace{-2.5cm}
\includegraphics[width=6.2cm, keepaspectratio]{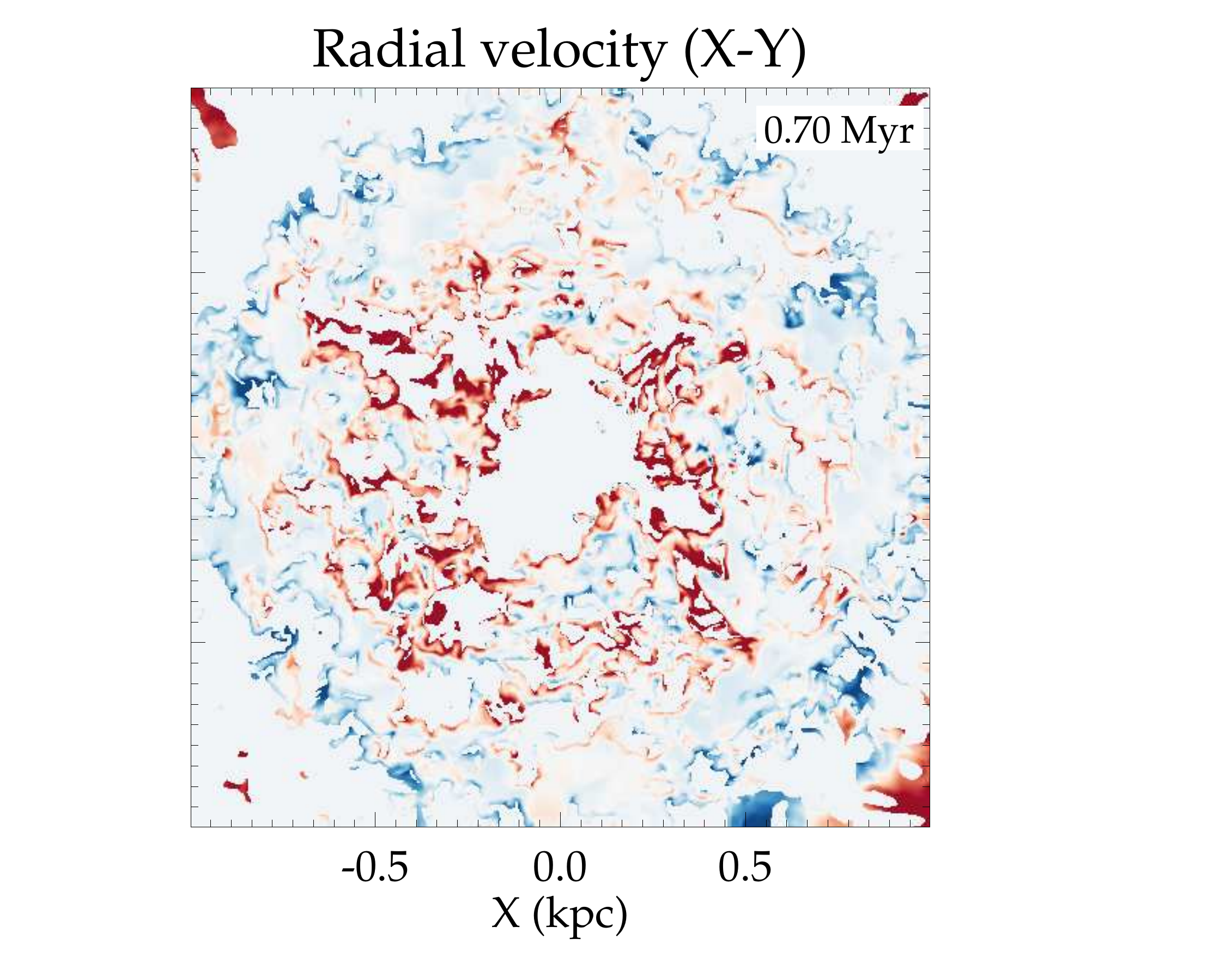} \vspace{-0.1cm}\hspace{-2.5cm}
\includegraphics[width=6.2cm, keepaspectratio]{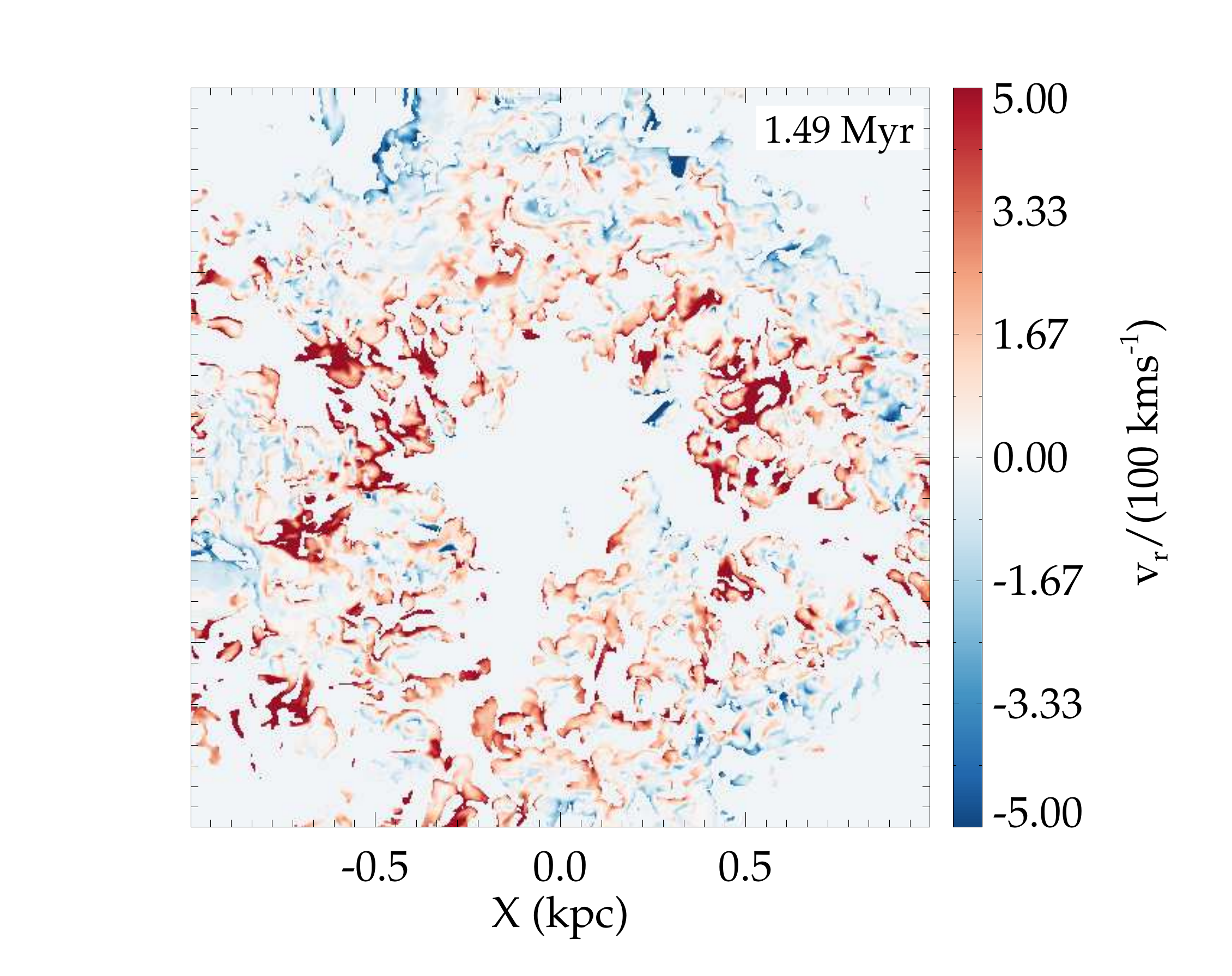} 
\caption{\small Mid-plane slices in the $X-Y$ plane ($Z=0$) of the evolution of simulation \PFF{}, for which the jet power is $\Pjet=10^{44}\ergs$. The last column of panels for which $t=1.49\Myr$ corresponds approximately to the current extent of the jets in \icfost{}. The top row shows the density ($\log(n [\cc])$) and the bottom row shows the cylindrically radial component of the velocity ($\vcyl=\sqrt{V_x^2+V_y^2}$) normalised to $100 \kms$, for gas with density $n>0.1\cc$.}
\label{fig:jet44xy}
\end{figure*}
\begin{figure*}
\centering
\includegraphics[width=6.2cm, keepaspectratio]{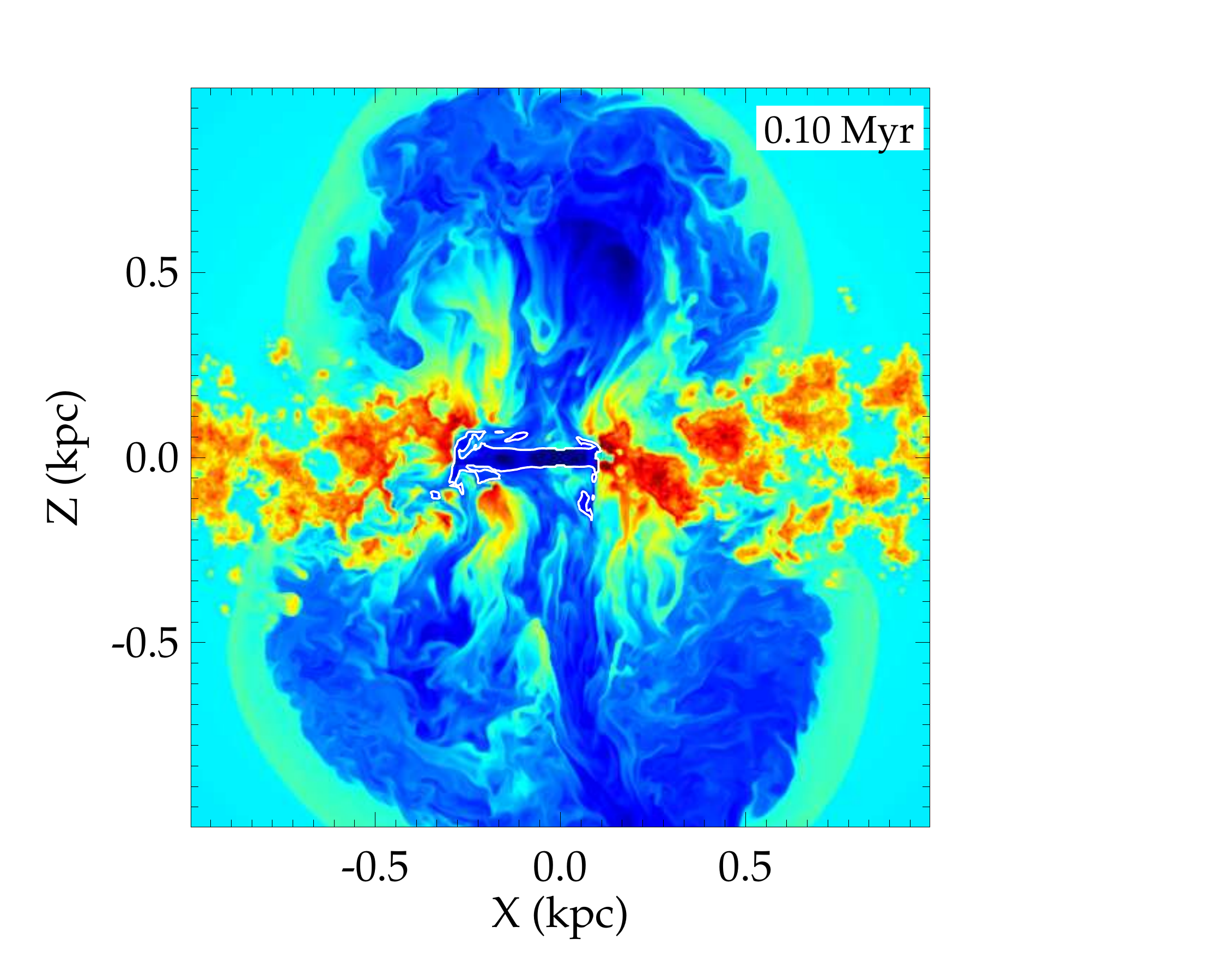} \vspace{-0.1cm}\hspace{-2.5cm}
\includegraphics[width=6.2cm, keepaspectratio]{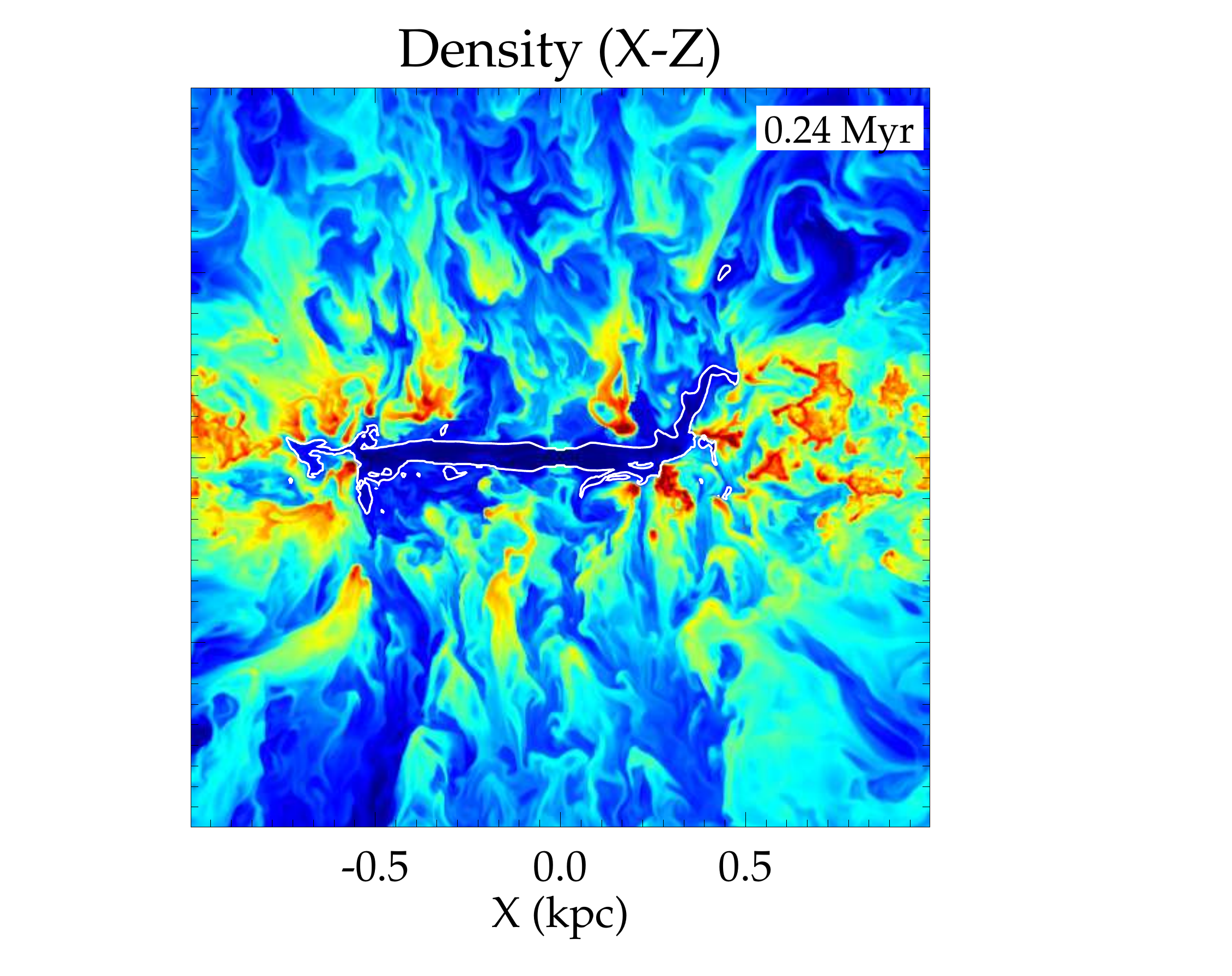} \vspace{-0.1cm}\hspace{-2.5cm}
\includegraphics[width=6.2cm, keepaspectratio]{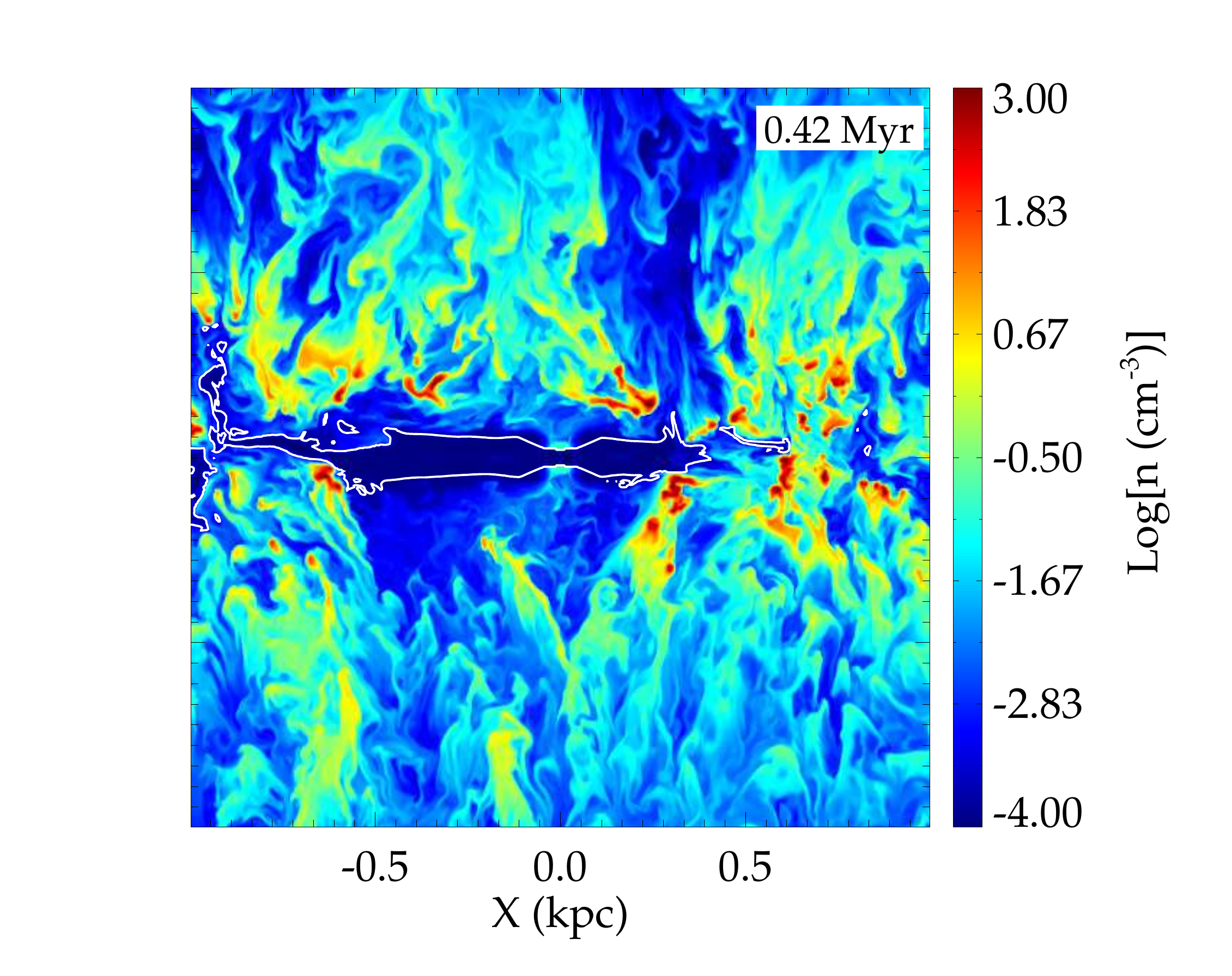} \vspace{-0.1cm}\linebreak
\includegraphics[width=6.2cm, keepaspectratio]{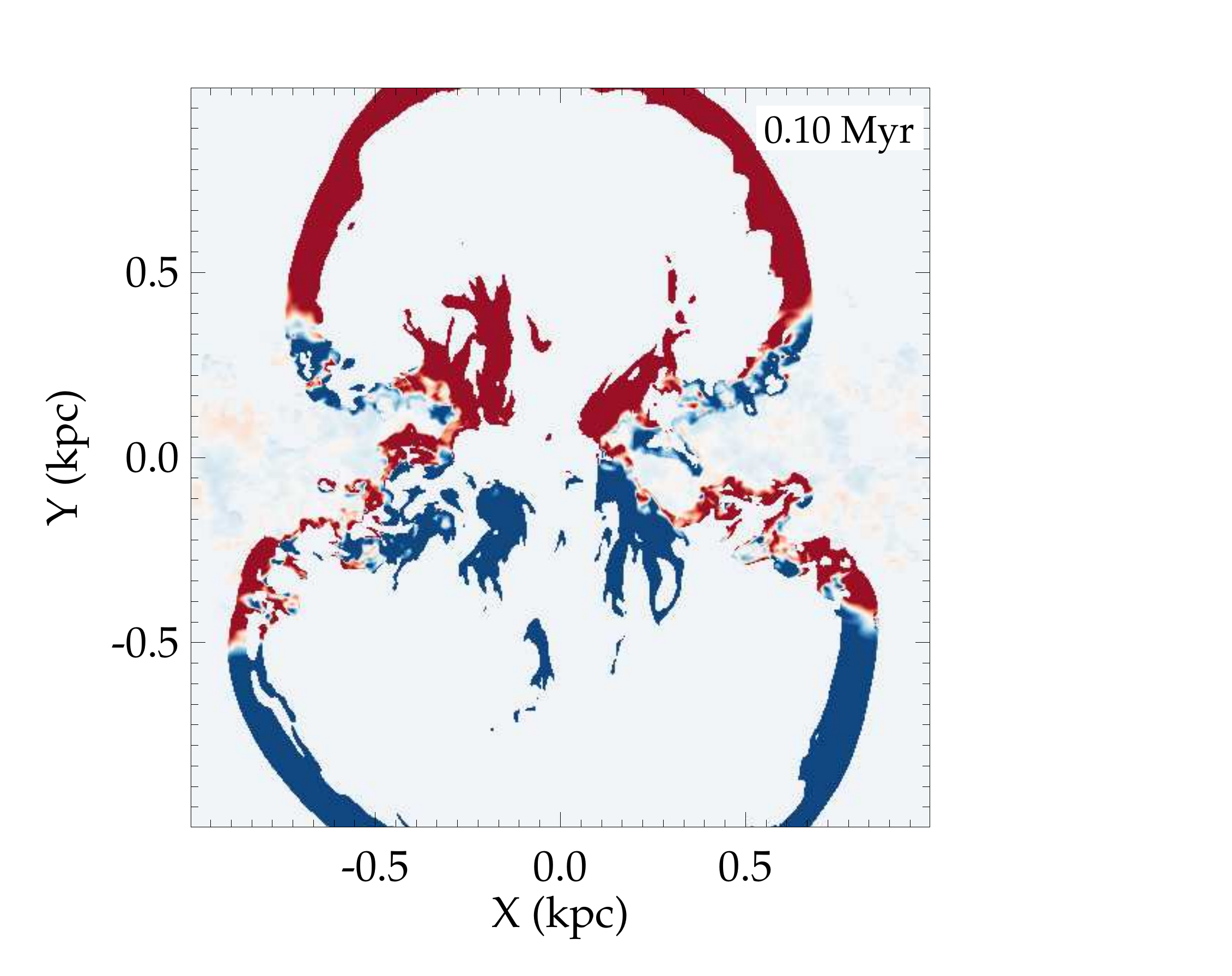} \vspace{-0.1cm}\hspace{-2.5cm}
\includegraphics[width=6.2cm, keepaspectratio]{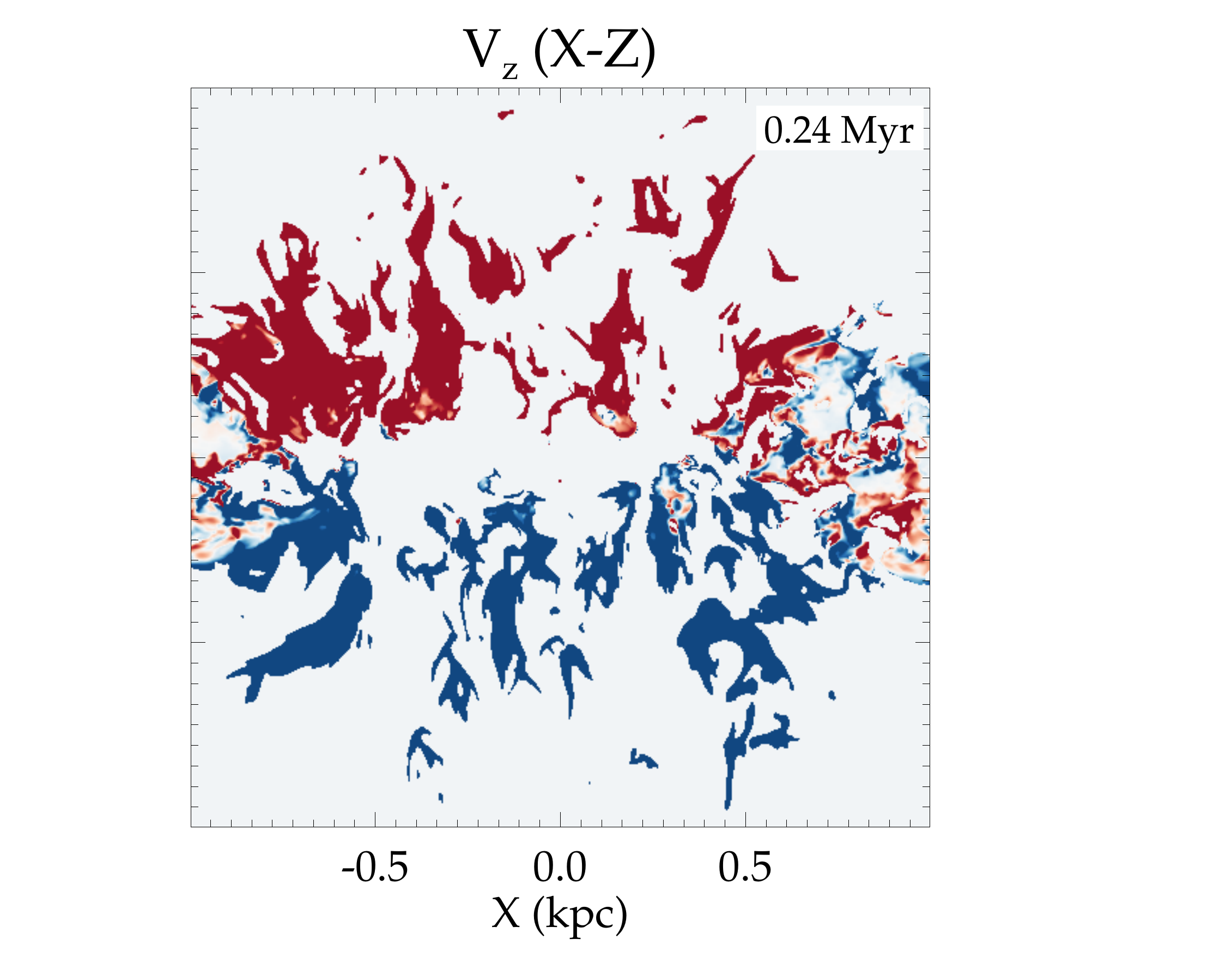} \vspace{-0.1cm}\hspace{-2.5cm}
\includegraphics[width=6.2cm, keepaspectratio]{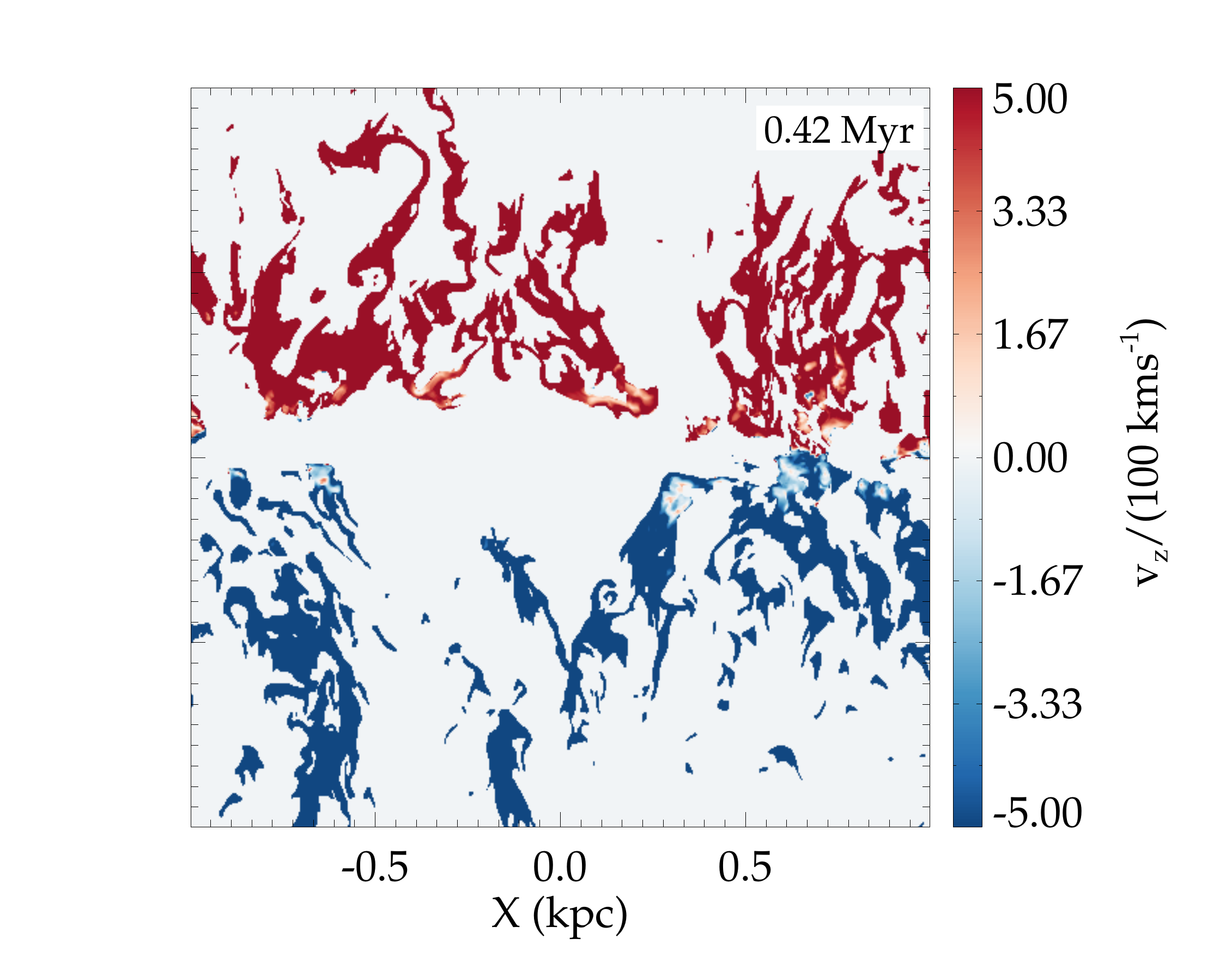} 
\includegraphics[width=6.2cm, keepaspectratio]{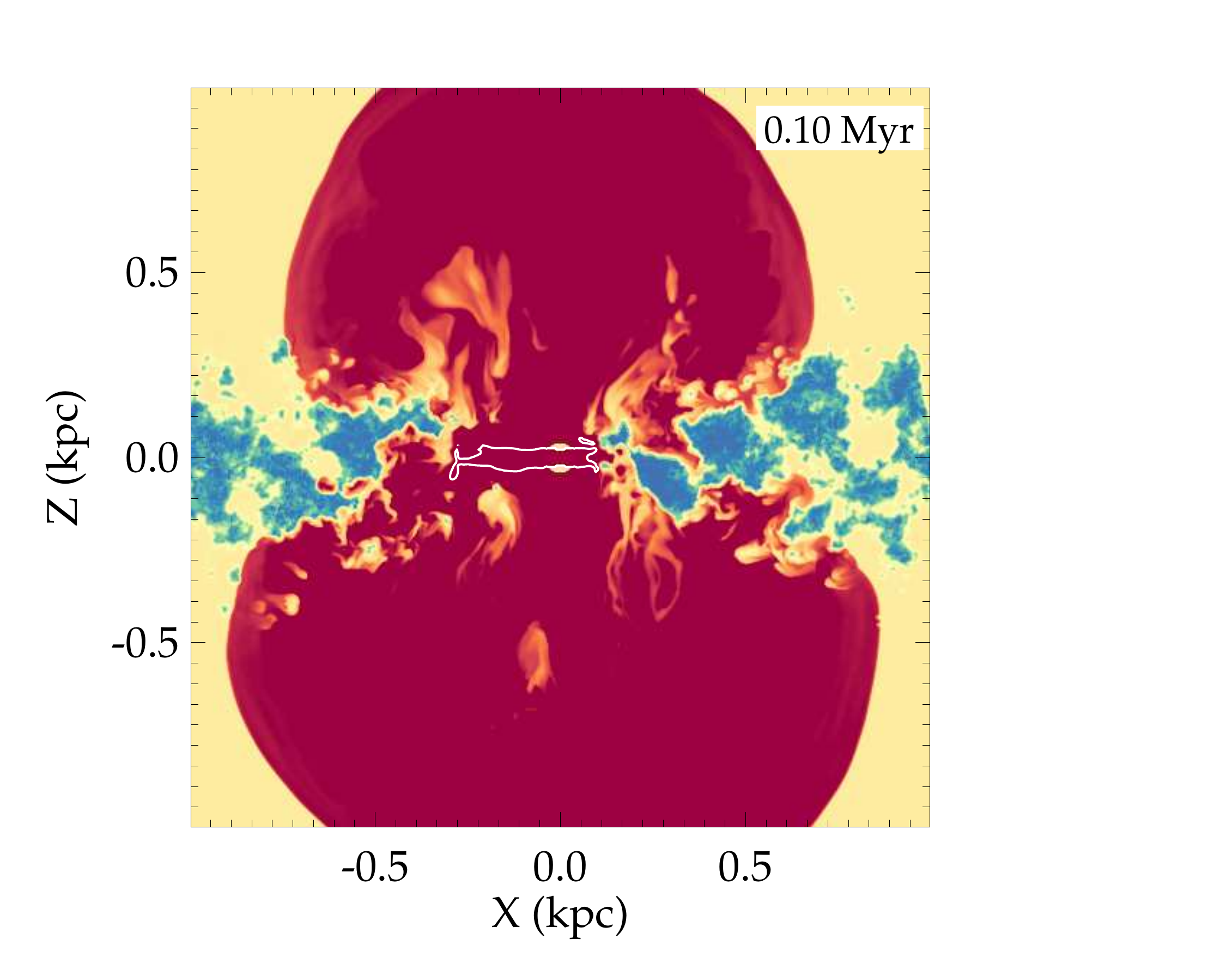} \vspace{-0.1cm}\hspace{-2.5cm}
\includegraphics[width=6.2cm, keepaspectratio]{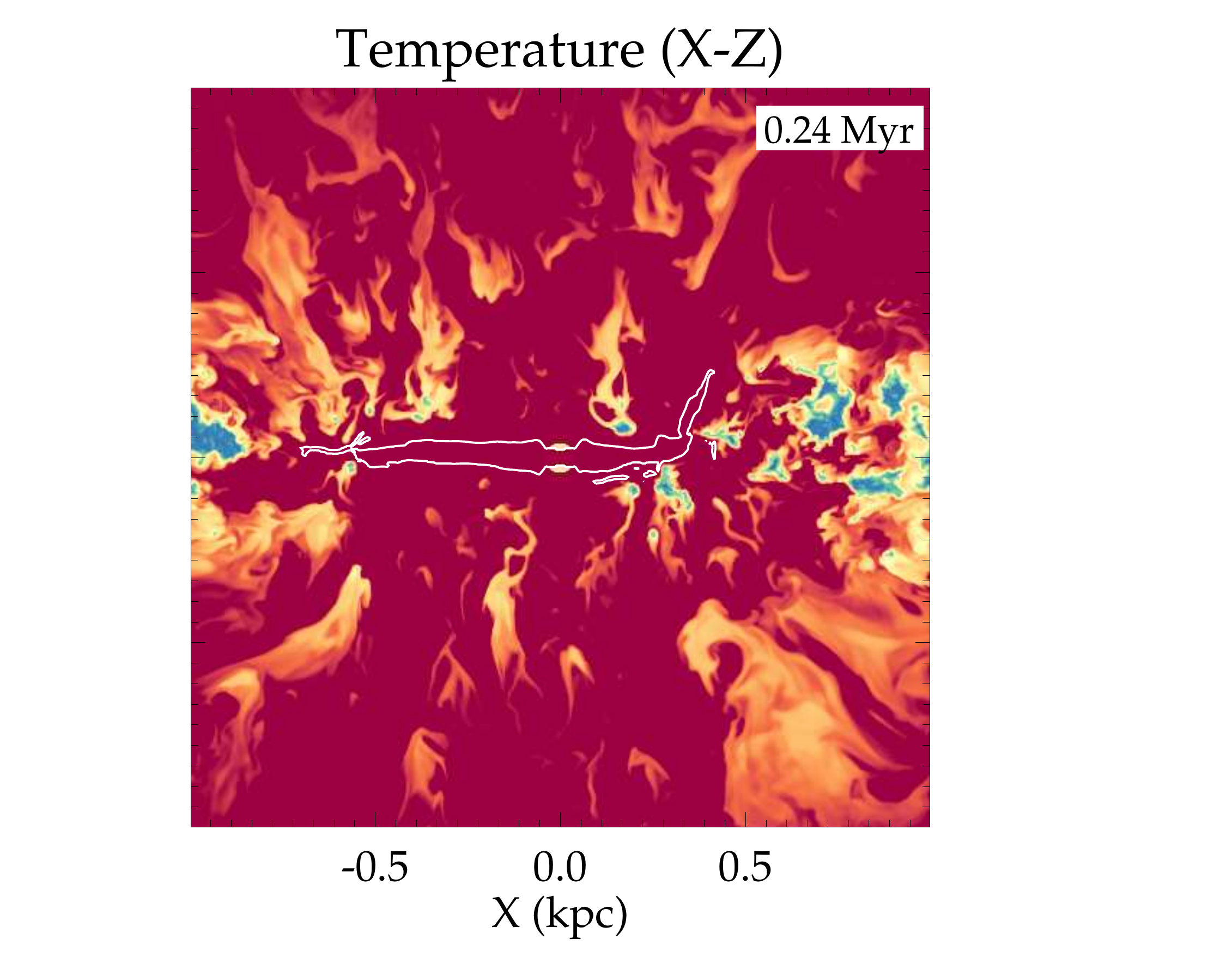} \vspace{-0.1cm}\hspace{-2.5cm}
\includegraphics[width=6.2cm, keepaspectratio]{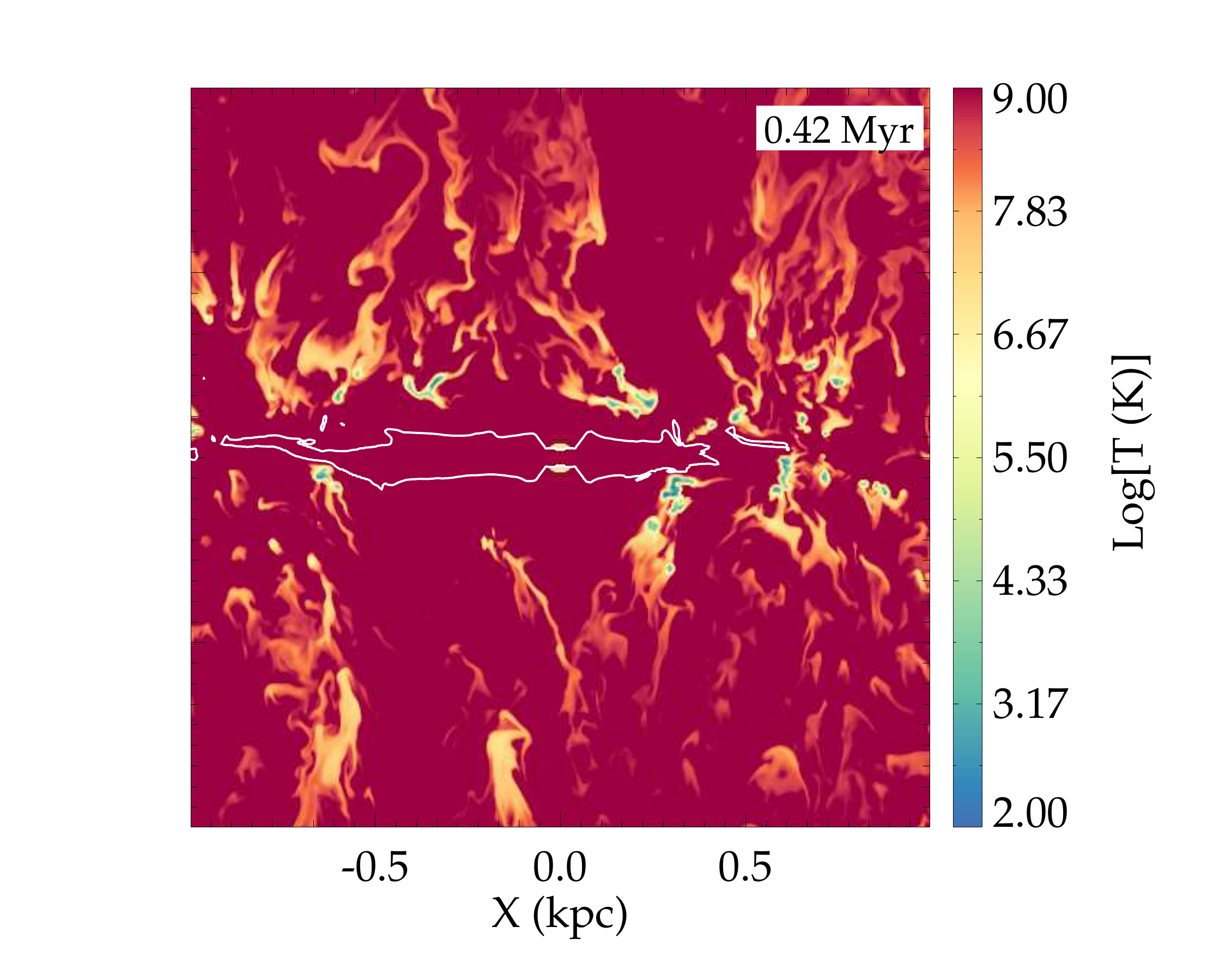} 
\caption{\small Same as Fig.~\ref{fig:jet44xz} but for simulation \PFV{}, for which the jet power is $\Pjet=10^{45}\ergs$. The central column of panels for which $t=0.24\Myr$ corresponds approximately  to the current extent of the jets in \icfost{}}
\label{fig:jet45xz}
\end{figure*}
\begin{figure*}
\centering
\includegraphics[width=6.2cm, keepaspectratio]{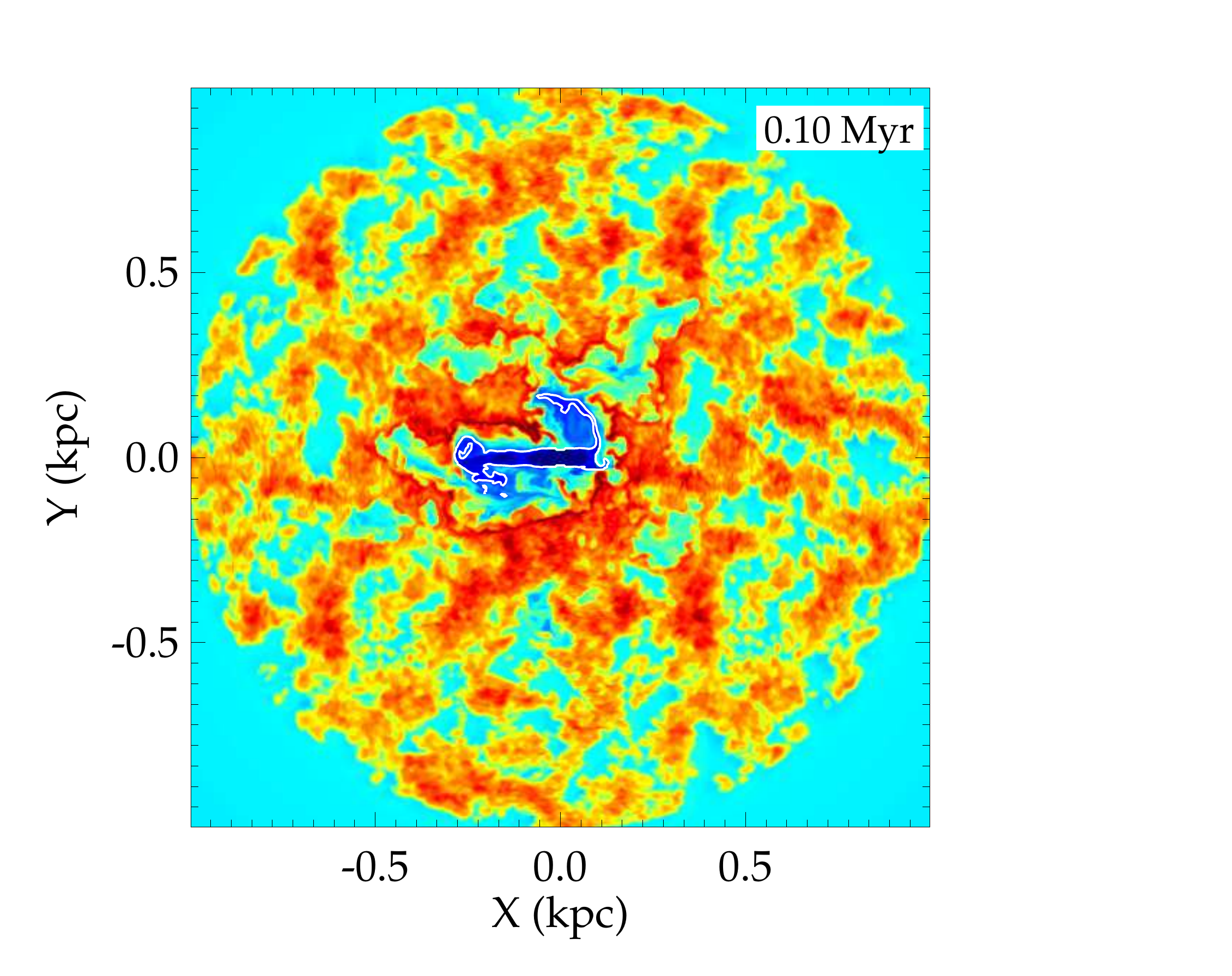} \vspace{-0.1cm}\hspace{-2.5cm}
\includegraphics[width=6.2cm, keepaspectratio]{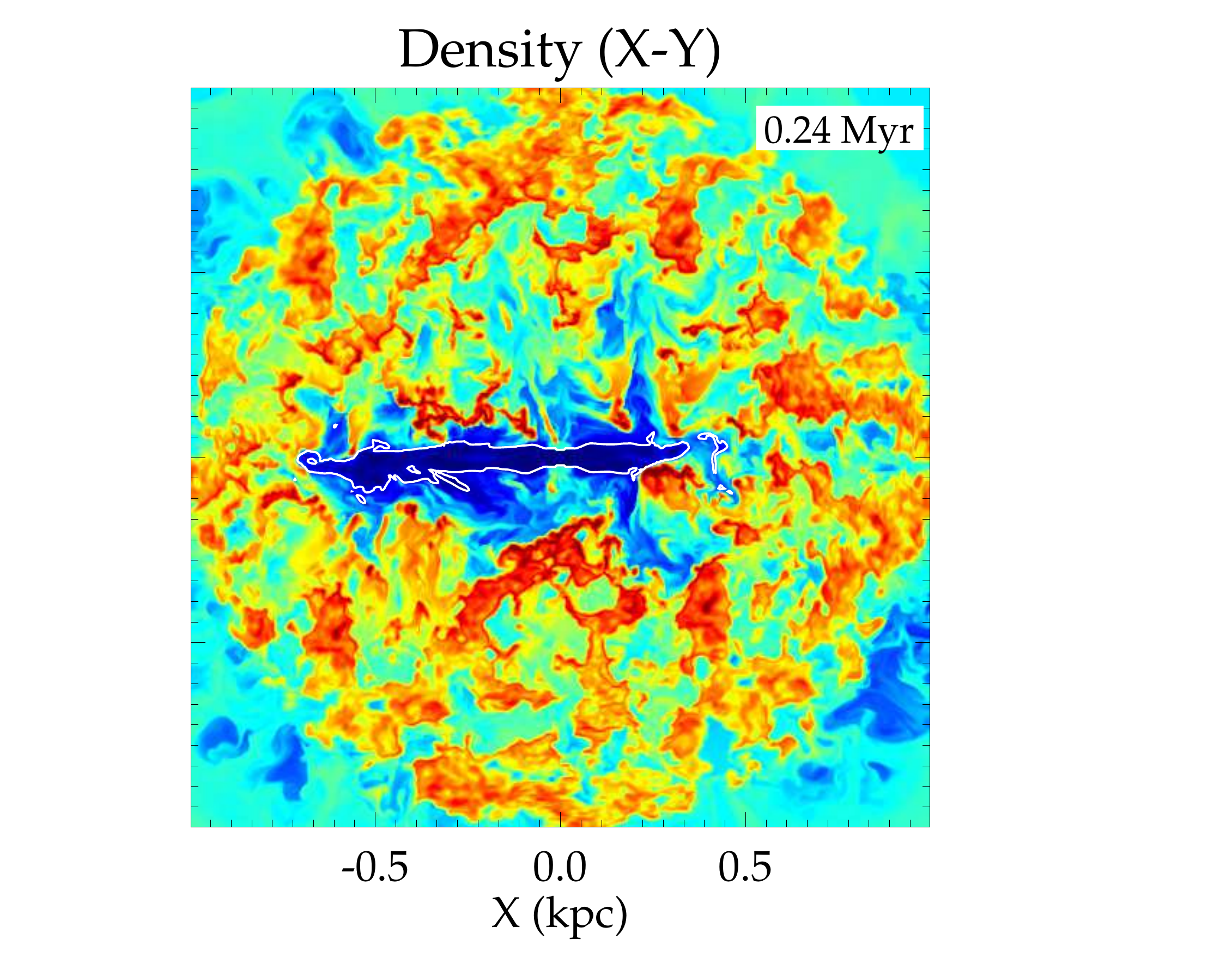} \vspace{-0.1cm}\hspace{-2.5cm}
\includegraphics[width=6.2cm, keepaspectratio]{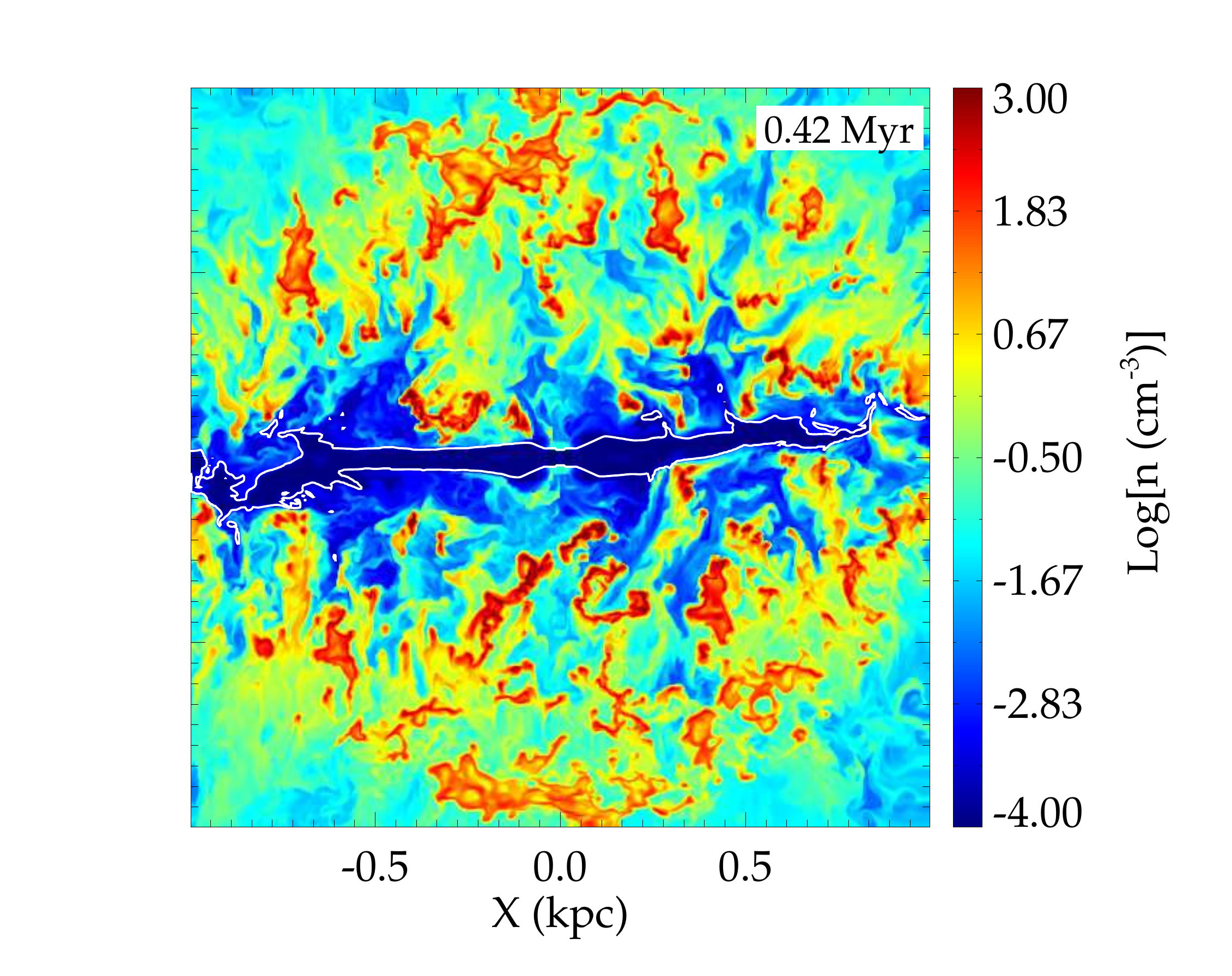} \vspace{-0.1cm}\linebreak
\includegraphics[width=6.2cm, keepaspectratio]{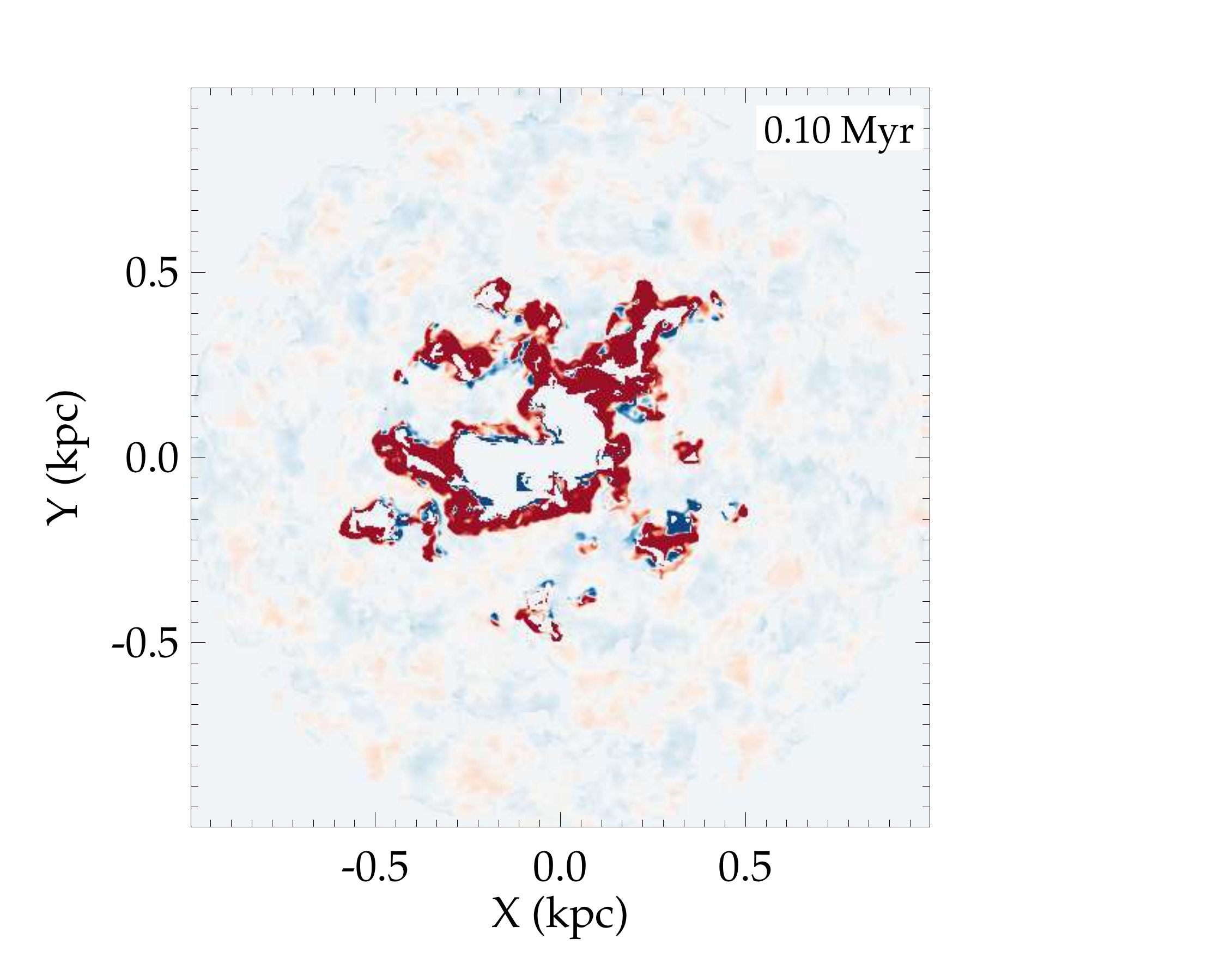} \vspace{-0.1cm}\hspace{-2.5cm}
\includegraphics[width=6.2cm, keepaspectratio]{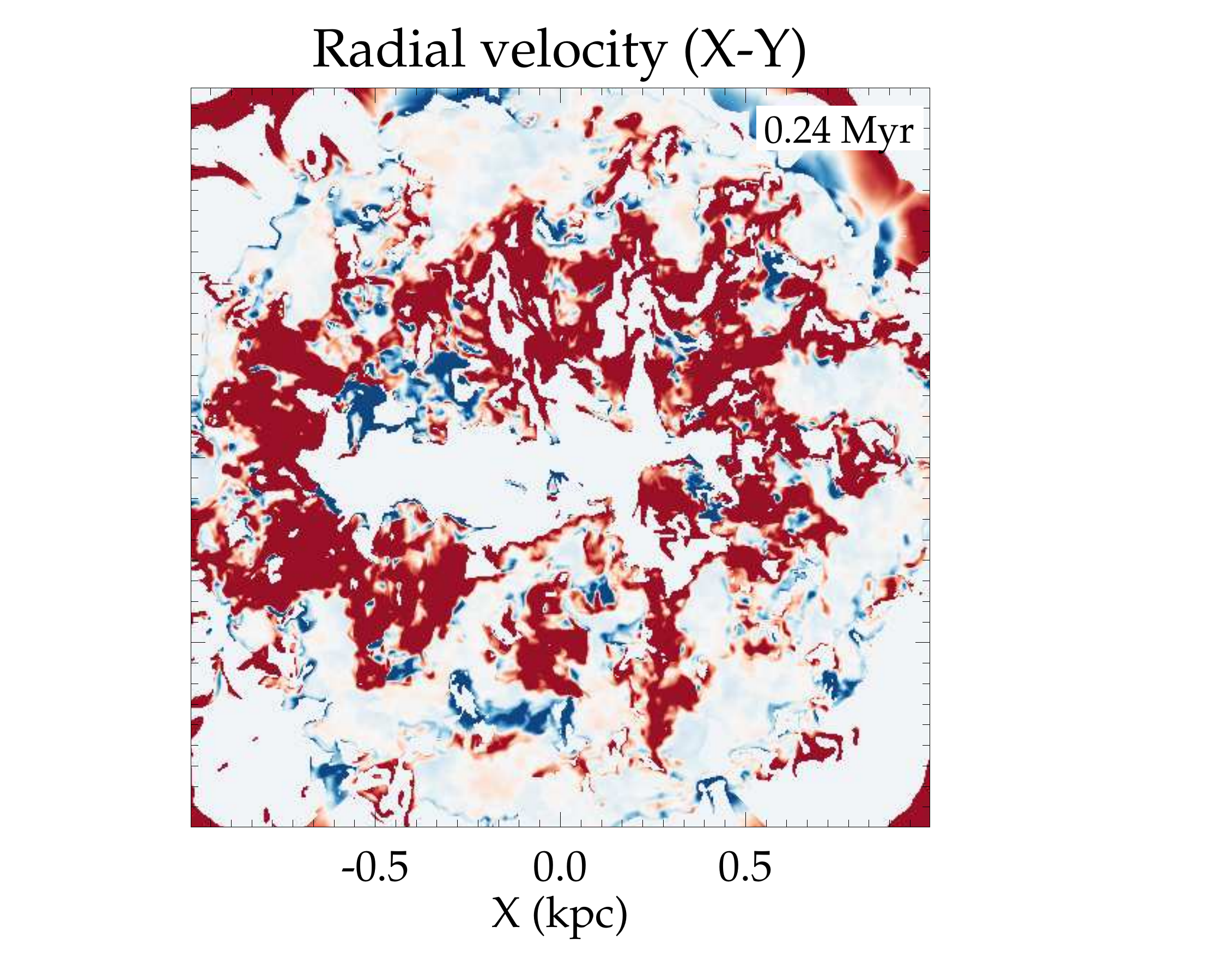} \vspace{-0.1cm}\hspace{-2.5cm}
\includegraphics[width=6.2cm, keepaspectratio]{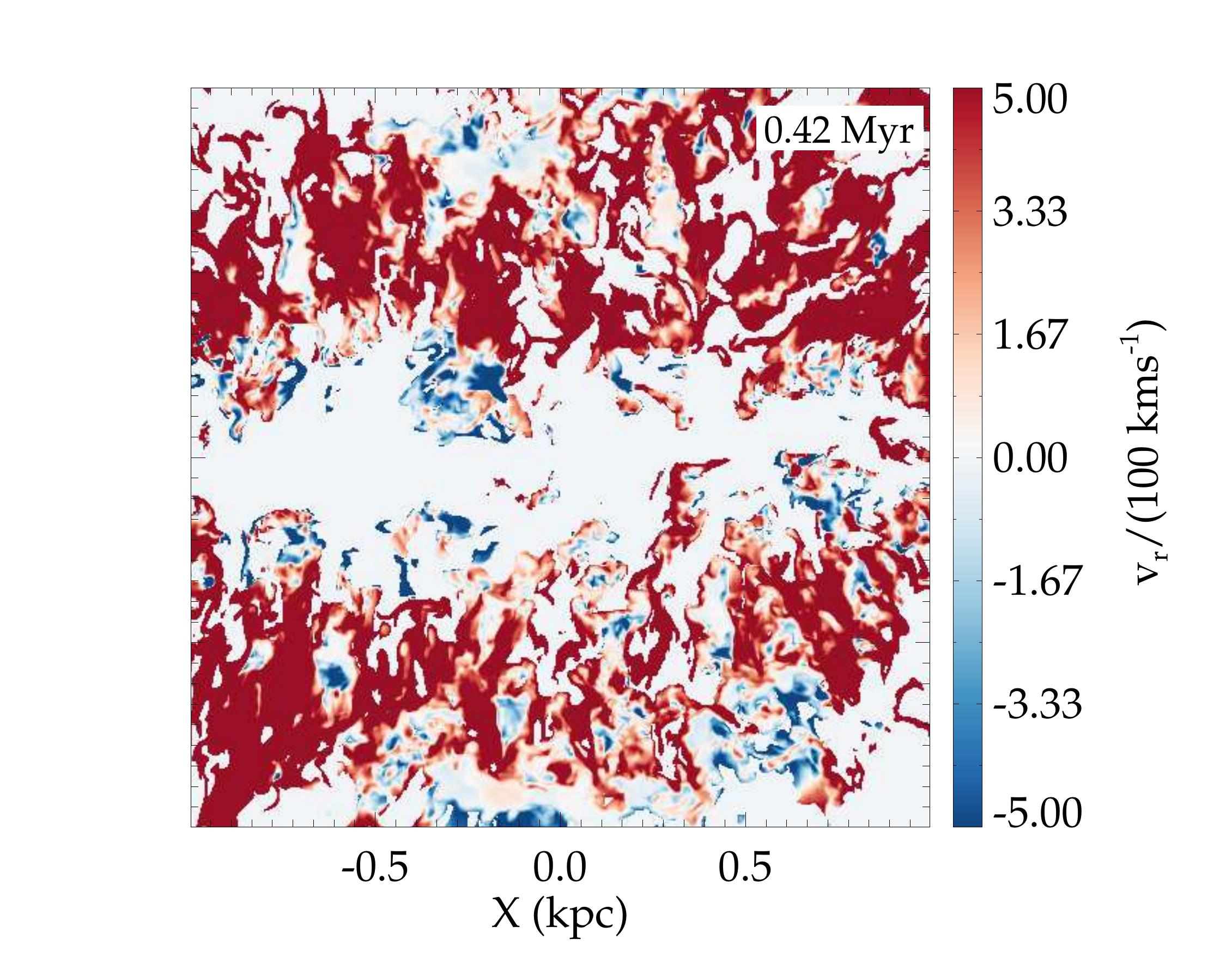} 
\caption{\small Same as Fig.~\ref{fig:jet44xy} but for simulation \PFV{}, for which the jet power is $\Pjet=10^{45}\ergs$. The central column of panels for which $t=0.24\Myr$ corresponds roughly to the current extent of the jets in \icfost{}.}
\label{fig:jet45xy}
\end{figure*}
\begin{figure*}
\centering
\includegraphics[height=5.6cm]{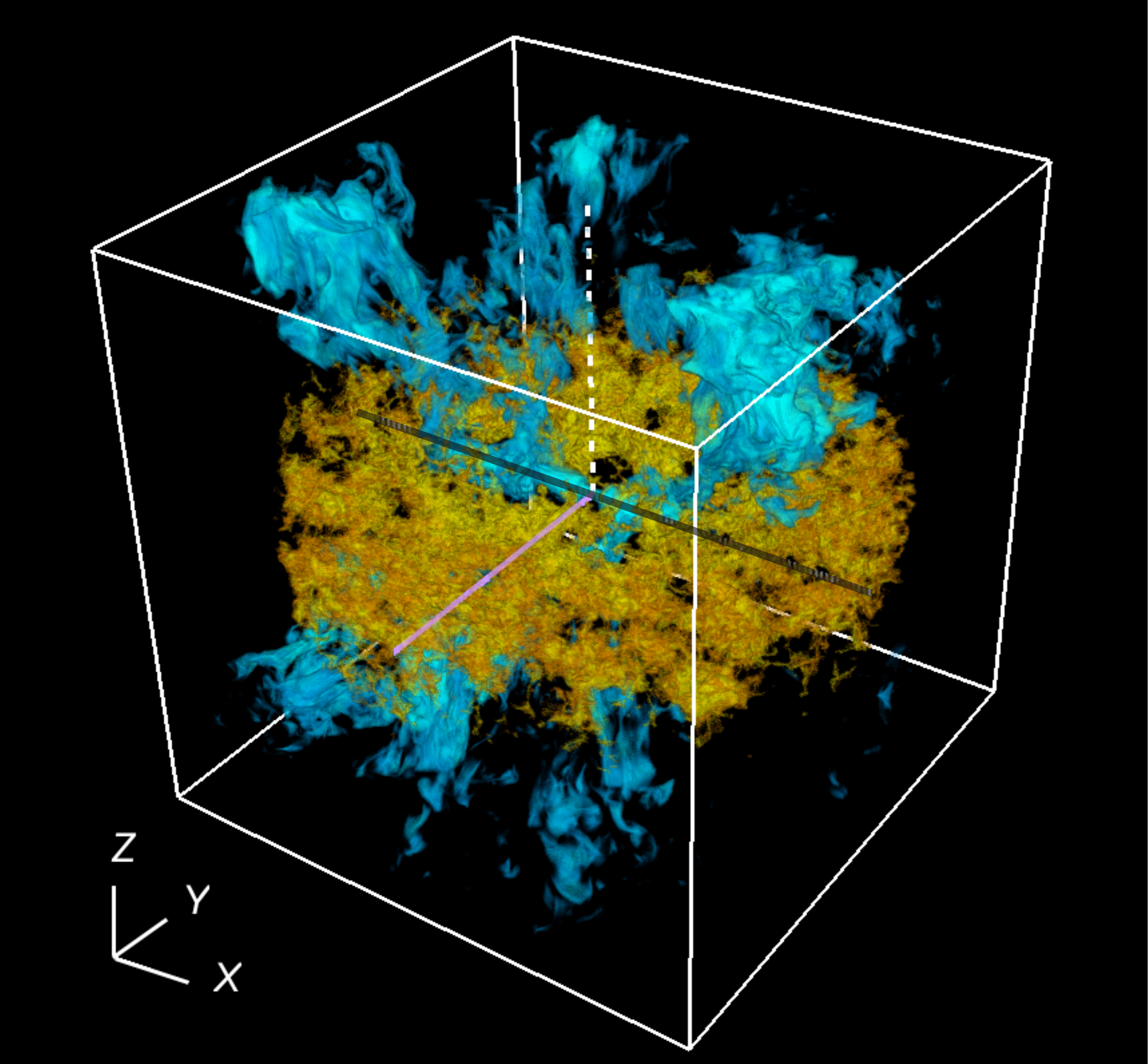}\hspace{-0.1cm}
\includegraphics[height=5.6cm]{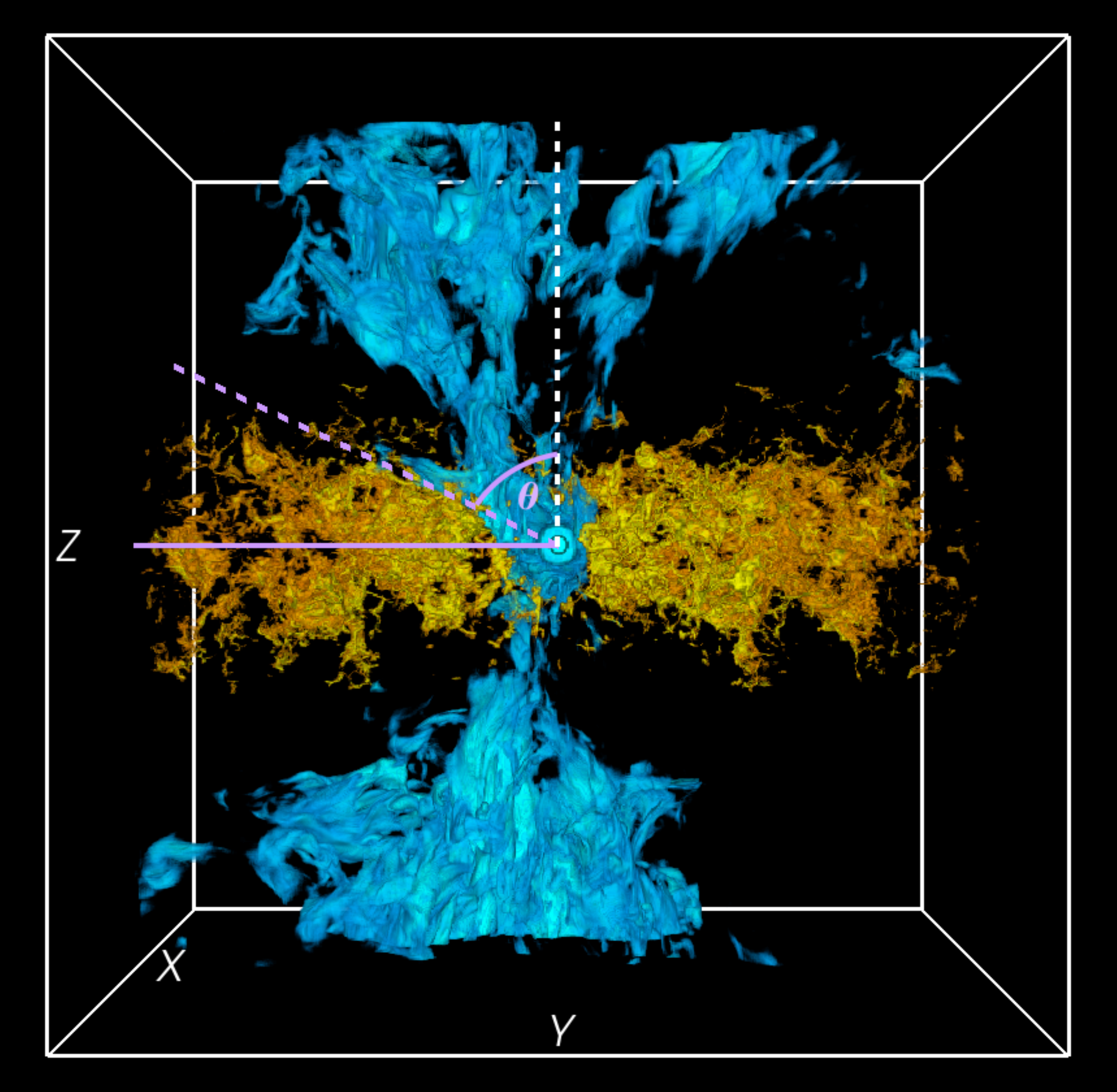}\hspace{-0.1cm}
\includegraphics[height=5.6cm]{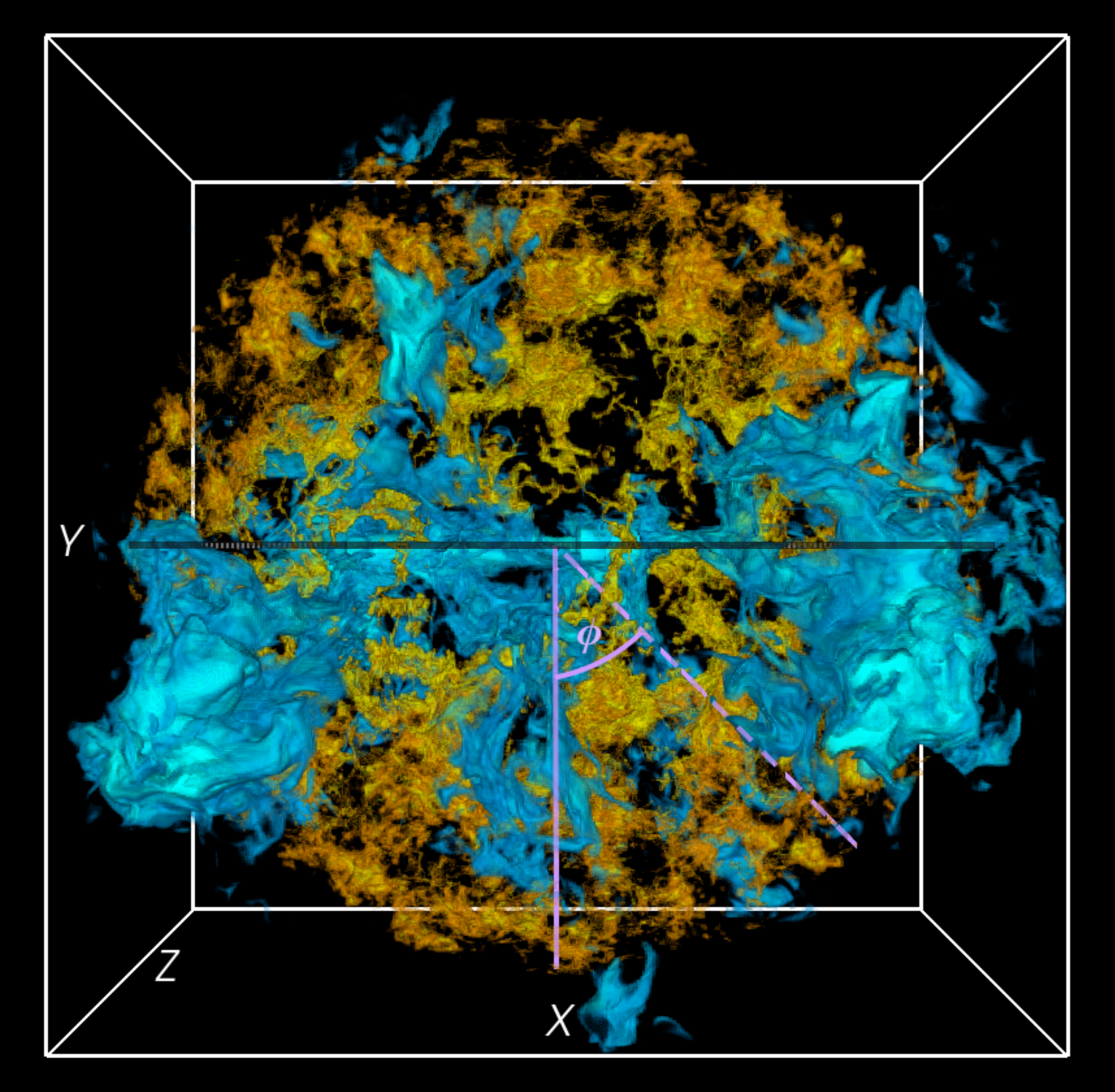}
\caption{\small Volume render of the simulation snapshot at time $t=0.24\Myr$ of simulation \PFV{}. The yellow--orange material represents dense disc gas and the blue material the jet plasma. Diffuse warm and hot gas and gas mixed with jet plasma are not shown as these phases have a large volume filling factor. The grey dotted line traces the jet axis, the white dashed line is the rotational symmetry axis of the disc, and the purple dashed line is the default view normal used in most of the plots in the paper. Left: a perspective from an angle. Center: A side-on view along the jet axis, with the front half of the galaxy clipped away. The purple dashed line shows a line of sight that is tilted by an angle $\theta$ with respect to the disc normal, but is perpendicular to the jet axis. Right: A face-on view onto the disc along the $Z$--axis. The purple dashed line here represent a line of sight along the plane of the disc, but rotated by an angle $\phi$ from the default view normal.}
\label{fig:jet45rt}
\end{figure*}
\begin{figure*}
\centering
\hspace{-2cm}
\includegraphics[width=6.cm, keepaspectratio]{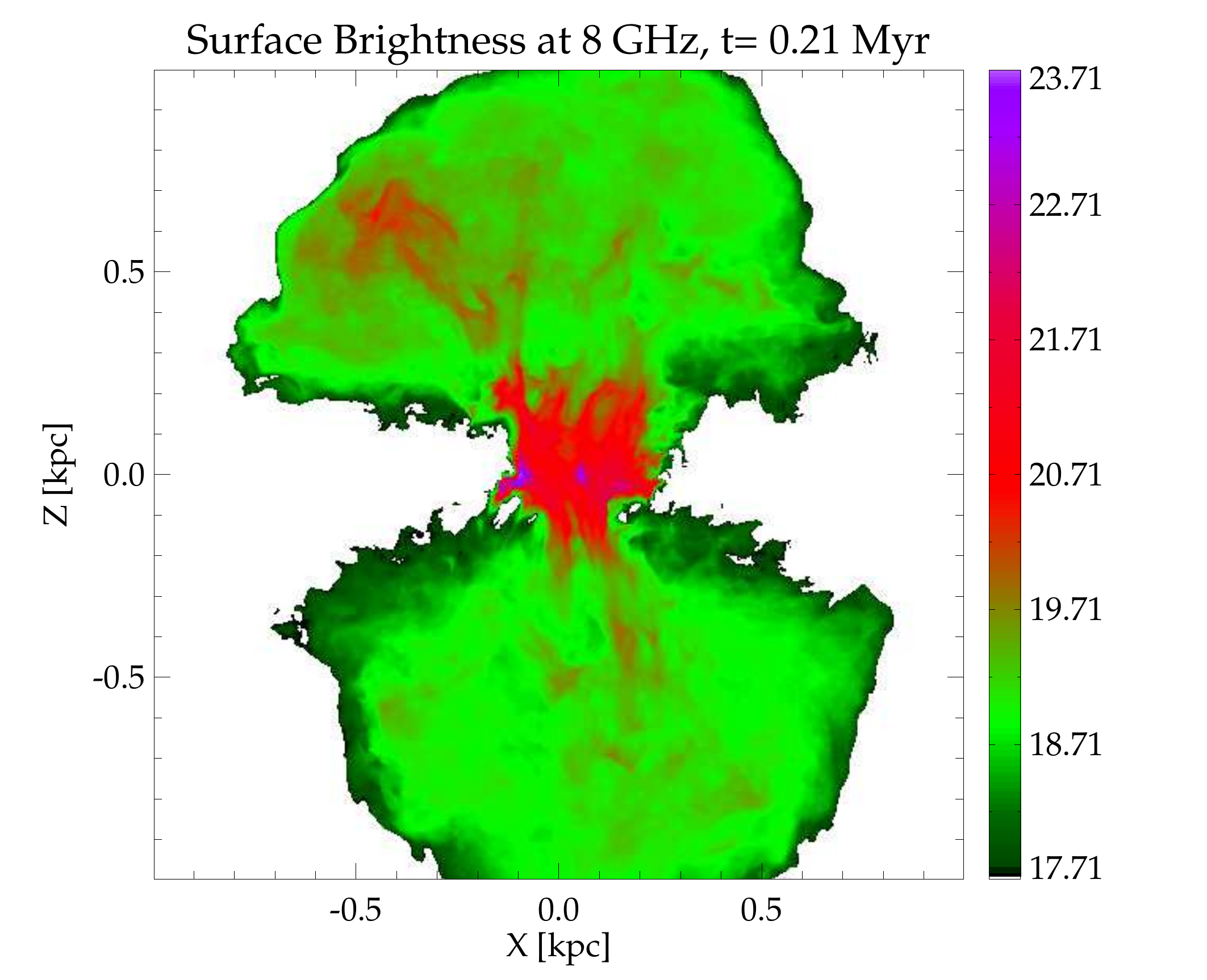}
\includegraphics[width=6.cm, keepaspectratio]{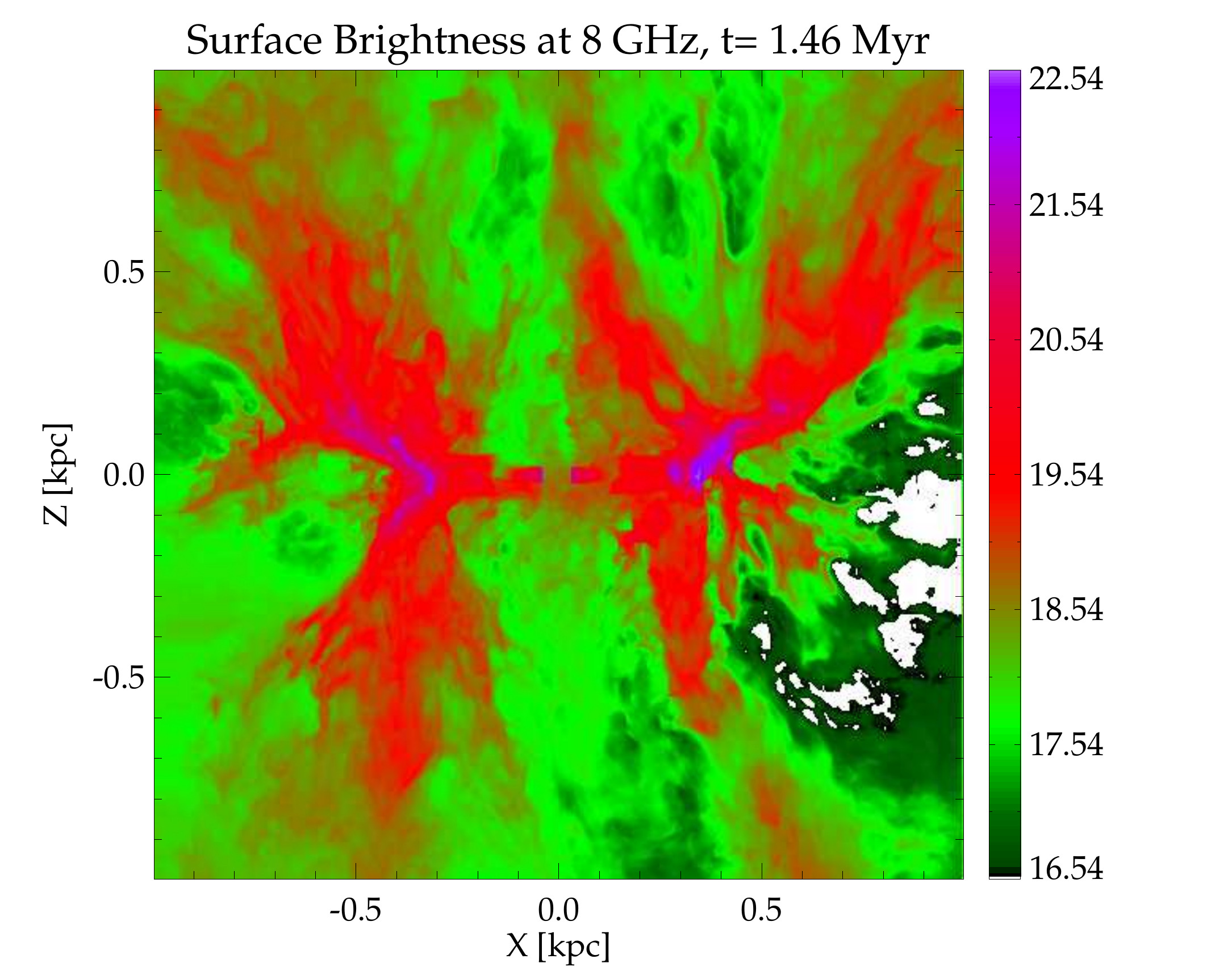} \vspace{-0.1cm}\hspace{-0.1cm}
\includegraphics[width=6.cm, keepaspectratio]{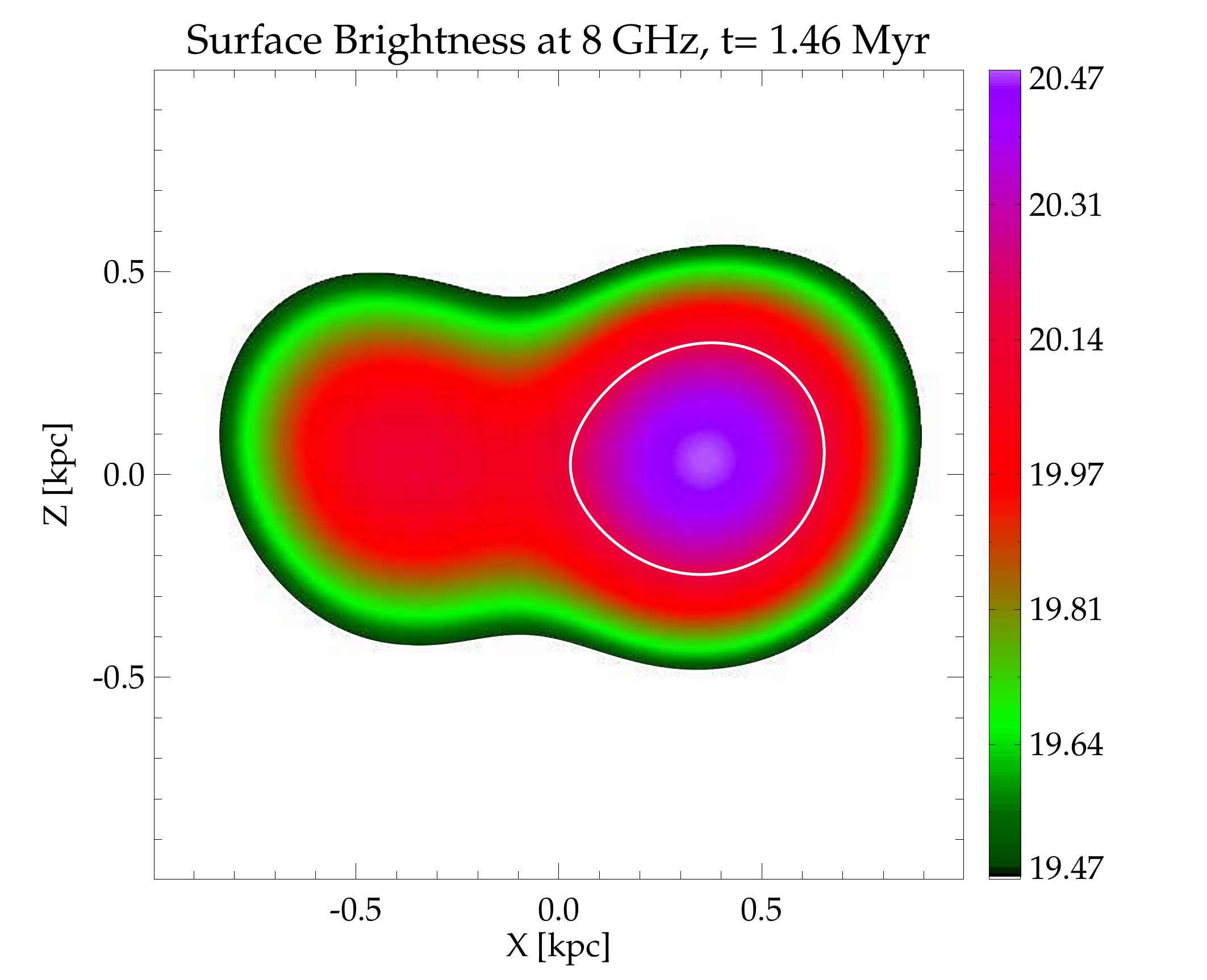} \vspace{-0.1cm}\hspace{-2.5cm}
	\caption{\small Synthetic $8\GHz$ surface brightness from simulation \PFF{}  taking into account free-free absorption. The ray tracing is performed with rays parallel to the $Y$--axis, in the direction of $-y$. Left: Unconvolved image at the resolution of the simulation at $t=0.21\Myr$. Center: Unconvolved image at the resolution of the simulation at $t=1.49\Myr$. Right: The image convolved with a beam of size $234 \mbox{pc} \times 234\mbox{pc}$. The white line denotes the contour for 10\% of peak brightness. The power at 8 GHz of the left and middle panels are $6.6\times10^{21} \WpHz$ and $2.2\times10^{21} \WpHz$. }
\label{fig:44-8ghz}
\end{figure*}

The evolution of the simulation with a jet power of $10^{44}\ergs$ (\PFF) is shown in three snapshots in Figs.~\ref{fig:jet44xz} (side-on) and \ref{fig:jet44xy} (face-on), and the evolution of the simulation with a jet power of $10^{45}\ergs$ (\PFV) is shown in Figs.~\ref{fig:jet45xz} (side-on) and \ref{fig:jet45xy} (face-on). Each figure contains three rows of panels showing mid-plane density slices, mid-plane temperature slices, and mid-plane velocity slices, respectively. For \PFF{}, the snapshot at time $t=1.49\Myr$ corresponds roughly to the observed conditions in \icfost{}, based mainly on the extents of the jet streams, while for \PFV{} it is the snapshot at time $t=0.24\Myr$. A 3D rendering of simulation \PFV{} at $t=0.24\Myr$ from different perspectives is shown in Fig.~\ref{fig:jet45rt}.

The evolution of the jet and ISM in simulations of AGN feedback in a multiphase ISM on kiloparsec galaxy scales is well-documented in the literature \citep{saxton05a, Sutherland2007a, Wagner2012a, mukherjee16a, bieri17a}. As in the simulations described by, e.g., \citet{mukherjee16a}, the interaction between the jet and ISM is very strong because the interaction depth of the jet \citep[the column density of clouds that the jets impact,][]{Wagner2012a} is very high if the jet axis lies in the plane of the gas-rich galactic disc.

The thermal and ram pressures in the jet plasma drive an energy bubble into the ambient ISM. The jet plasma in the bubble, being orders of magnitudes lighter than the dense phases of the ISM, percolates through paths of least resistance between clouds, a phase termed the ``flood-and-channel'' phase \citep{Sutherland2007a}. In the process, the clouds are compressed but they also experience strong hydrodynamic ablation driven through the Kelvin-Helmholtz instability. The ablated material forms cometary tails and partly mixes with the inter-cloud flow, which consists of swept up hot-phase ISM, and jet plasma. 

The clouds and filaments of ablated cloud material are long-lived because of efficient radiative cooling. Radiative cooling does not strongly affect the global evolution of the AGN bubble as the outflow remains largely energy-driven, at least on kpc scales. The dynamics of the warm and cold phases are, however, strongly affected as clouds fragment and their cores and filaments are efficiently accelerated by the surrounding ram pressure and thermal pressure gradients. 

We see in Figs~\ref{fig:jet44xz}, \ref{fig:jet44xy}, \ref{fig:jet45xz}, and \ref{fig:jet45xy} that cloud cores are accelerated to $100$--$200\kms$, while diffuse filaments exceed velocities of $500\kms$. Some cloud cores are displaced and most of the ablated gas is carried along inter-cloud channels to larger disc radii. In these simulations, the jet never fully breaks out of the disc within the duration of the simulations. In the face-on velocity maps of Fig.~\ref{fig:jet44xy} and \ref{fig:jet45xy} we see that the highest gas velocities occur near the locations of the jet head (i.e., the radio hot spots) at $(X, Y) \approx (\pm0.5, 0)\kpc$. The corresponding density maps also show that this is where the strongest cloud compression occurs. We also note that, as seen in the side-on density slices of Figs~\ref{fig:jet44xz} and \ref{fig:jet45xz}, the jet plasma quickly sweeps away the hot halo gas, replacing it with jet plasma and entrained ISM from the disc. The jet also excavates a cavity in the central parts of the galaxy devoid of dense gas. 

The third panels in Fig~\ref{fig:jet44xz} and \ref{fig:jet45xz} show the temperature in the $X-Z$ plane. The jet-ISM interaction forms a multiphase ISM as discussed previously in \citet{mukherjee16a,mukherjee17a}. Shocks driven by the energy bubble heats the initially cold disc to $T\gtrsim 10^5$K, with shocked clouds being driven vertically as discussed earlier. The tenuous cavity filled with relativistic plasma is at a higher temperature ($T\gtrsim 10^8$K).

The main difference between the two simulations presented here (P44 and P45) is that the higher powered jet penetrates through the disc much faster (almost 10 times faster) than the lower-powered jet. As a result, the higher-powered jet ends up excavating a roughly cylindrical cavity about the jet axis, rather than a spherical cavity about the central core. Of course, the resulting velocity dispersion caused by the higher powered jet is much stronger, as expected from the scaling relation of feedback efficiency and jet power \citep{Wagner2011a}. Accordingly, the more powerful jet in simulation \PFV{} is able to clear its path faster and reach an extent of $0.5\kpc$ within $0.24\Myr$, while the lower-powered jet in simulation \PFF{} requires $1.49\Myr$ to reach the same size. The difference in velocity dispersion between the two simulations is discussed more quantitatively in \S~\ref{sec:pv}.

In order to compare the simulations with observations, we generate a number of synthetic radio surface brightness images. Since the simulation is purely hydrodynamic and we do not track the evolution of relativistic electrons, we estimate the synchrotron emissivity by assuming that the energy densities of the magnetic field and relativistic electrons are fixed fractions, respectively $f_{\rm e}$ and $f_{\rm B}$, of the internal energy density.  We adopt nominal values of $f_{\rm e} = f_{\rm B} = 0.1$ but the resulting images can be readily scaled to other values. The surface brightness is calculated by integrating the radiative transfer equation through the volume, taking into account free-free absorption by the dense clouds. The method of generating the synthetic images is presented in more detail in Bicknell et al. ({\it submitted}). 

In Fig.~\ref{fig:44-8ghz}, we show a synthetic synchrotron surface brightness image at an observed frequency of $8\GHz$ generated from our simulation at  $t=704\kyr$, corresponding approximately to the observed extent of the radio jet. The left panel in the figure shows the surface brightness at the resolution of our simulation, and the right panel shows the surface brightness convolved with a beam of size $234 \mbox{pc} \times 234\mbox{pc}$, which corresponds to the resolution of the image at 8 GHz in \citet{morganti98a}. Our synthetic radio image resembles the observed radio image, in particular at the western ``hotspot'', where the radio surface brightness is particularly high due to the strong interaction of the jet with a large dense cloud. Although the distribution of the dense gas in the disk is statistically symmetric, the asymmetric features arise from interactions with the local fluctuations in the density field resulting in the clumpy inhomogeneous ISM.  Note that there is no core component because we do not model the sub-parsec scale nuclear region of the galaxy. 

We emphasize that the radio morphology at the observed resolution belies the complex distribution of the jet plasma in the galaxy, which bears little resemblance to classical radio sources. For example, the jets, as they propagate through the disc, do not develop well-defined lobes that arise when overpressured jet plasma expands against a smooth ambient medium. The regions of bright radio emission are not classical hot-spots in the sense of knots or terminal shocks, but are best described as a ``splatter region'', where the jet head interacts strongly with a dense cloud in its path \citep[see Fig.~5 in][]{Oosterloo2000a}, splitting and splattering the main jet stream in all directions.

Because the jet is so light compared to the clouds, it is easily deflected (or even split), as we see in the white contours of the western jet in the panels in Fig~\ref{fig:jet44xz} for $t=1.49\Myr$ and $t=2.10\Myr$. Throughout the simulation, the jet plasma also escapes perpendicular to the disc, entraining with it cloudlets that are accelerated to several $100\kms$. Although most of the mass of the cold and warm gas remains contained within the disc, the acceleration of ablated gas is also efficient perpendicular to the disc. This effect is a somewhat serendipitous discovery, to which we return in \S~\ref{sec:other-results}.

\subsection{Position-velocity diagrams}\label{sec:pv}
\begin{figure*}
\centering
\includegraphics[height=5.4cm]{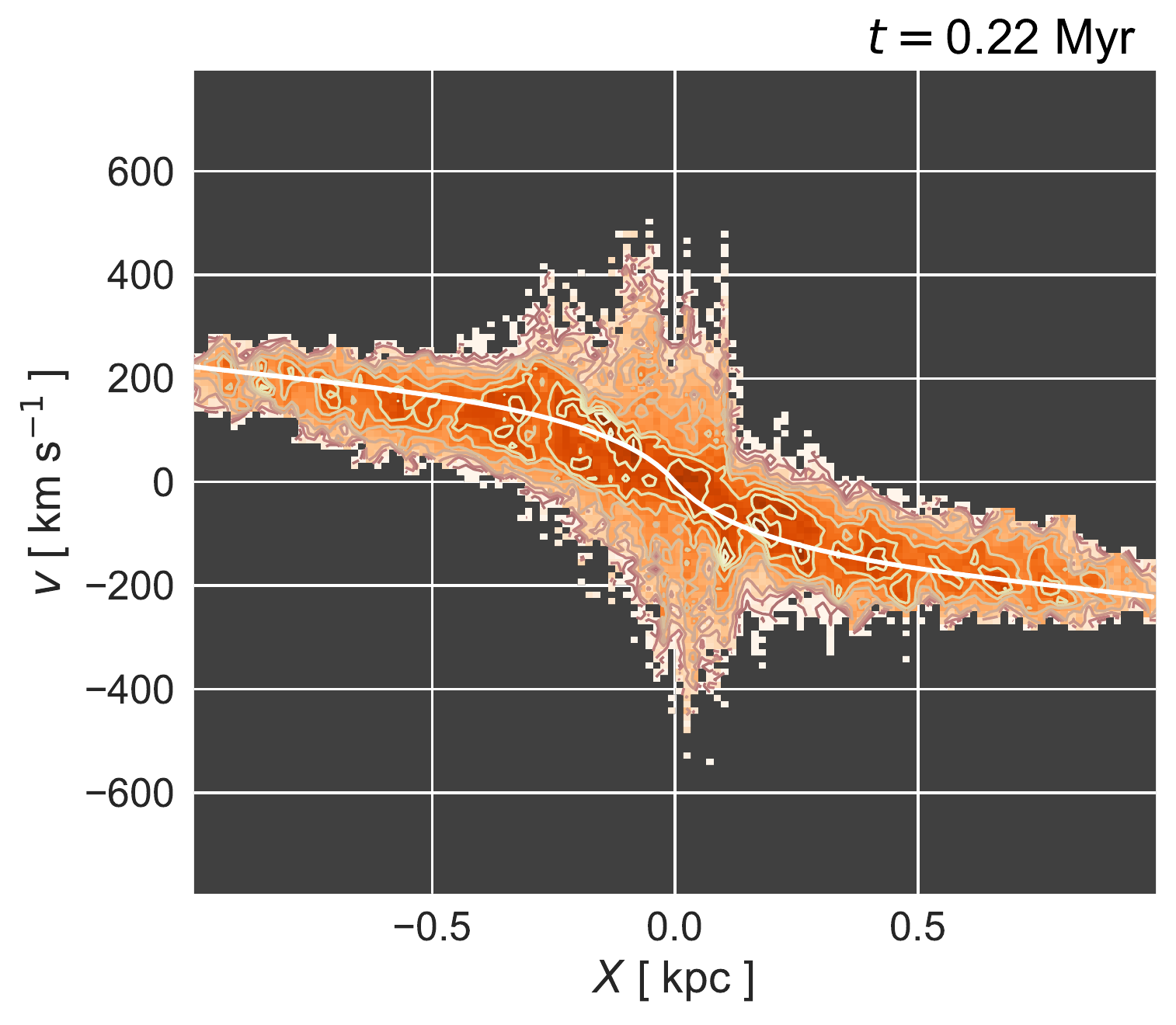}\hspace{-0.2cm}
\includegraphics[height=5.4cm]{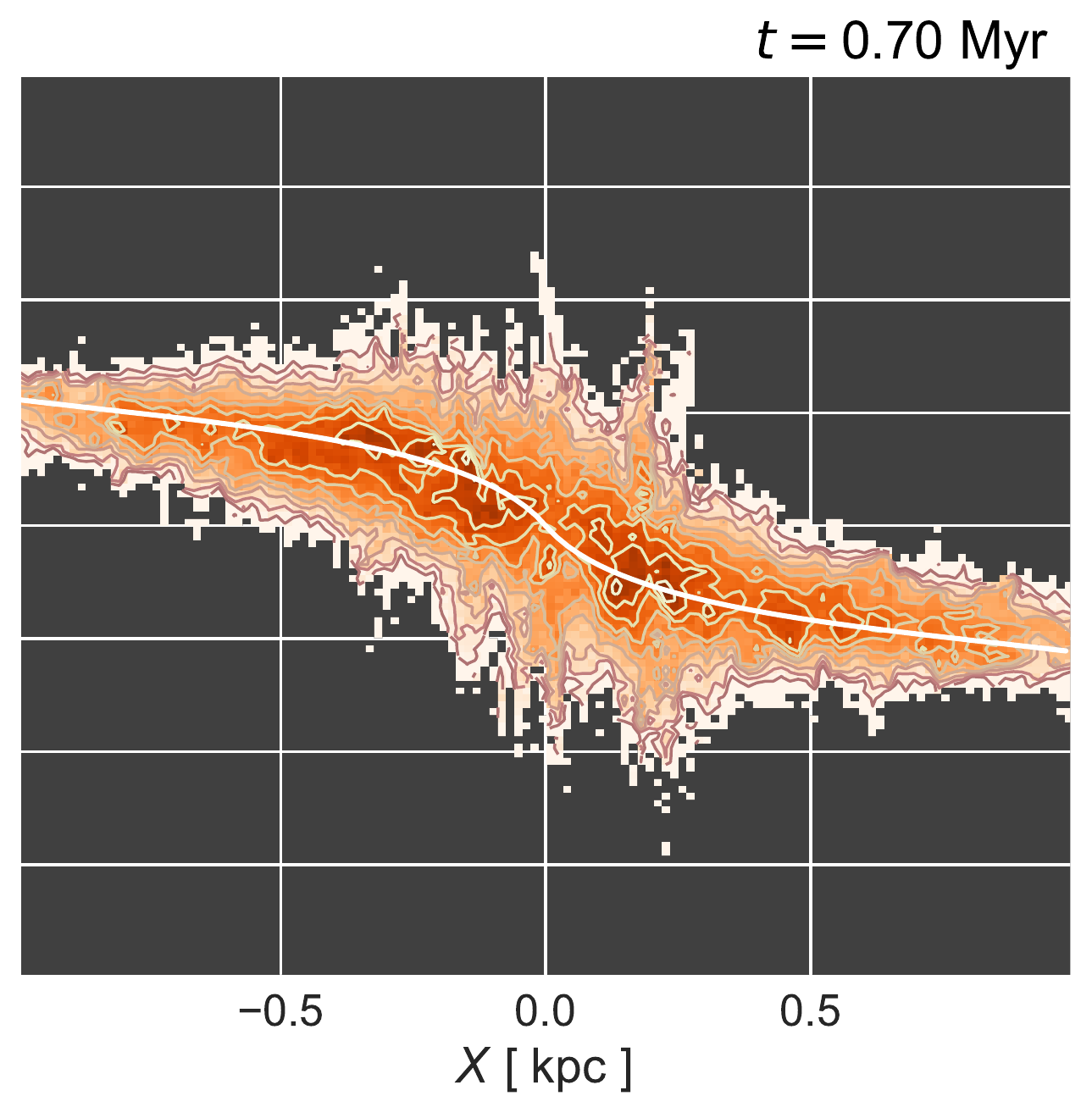}\hspace{-0.2cm}
\includegraphics[height=5.4cm]{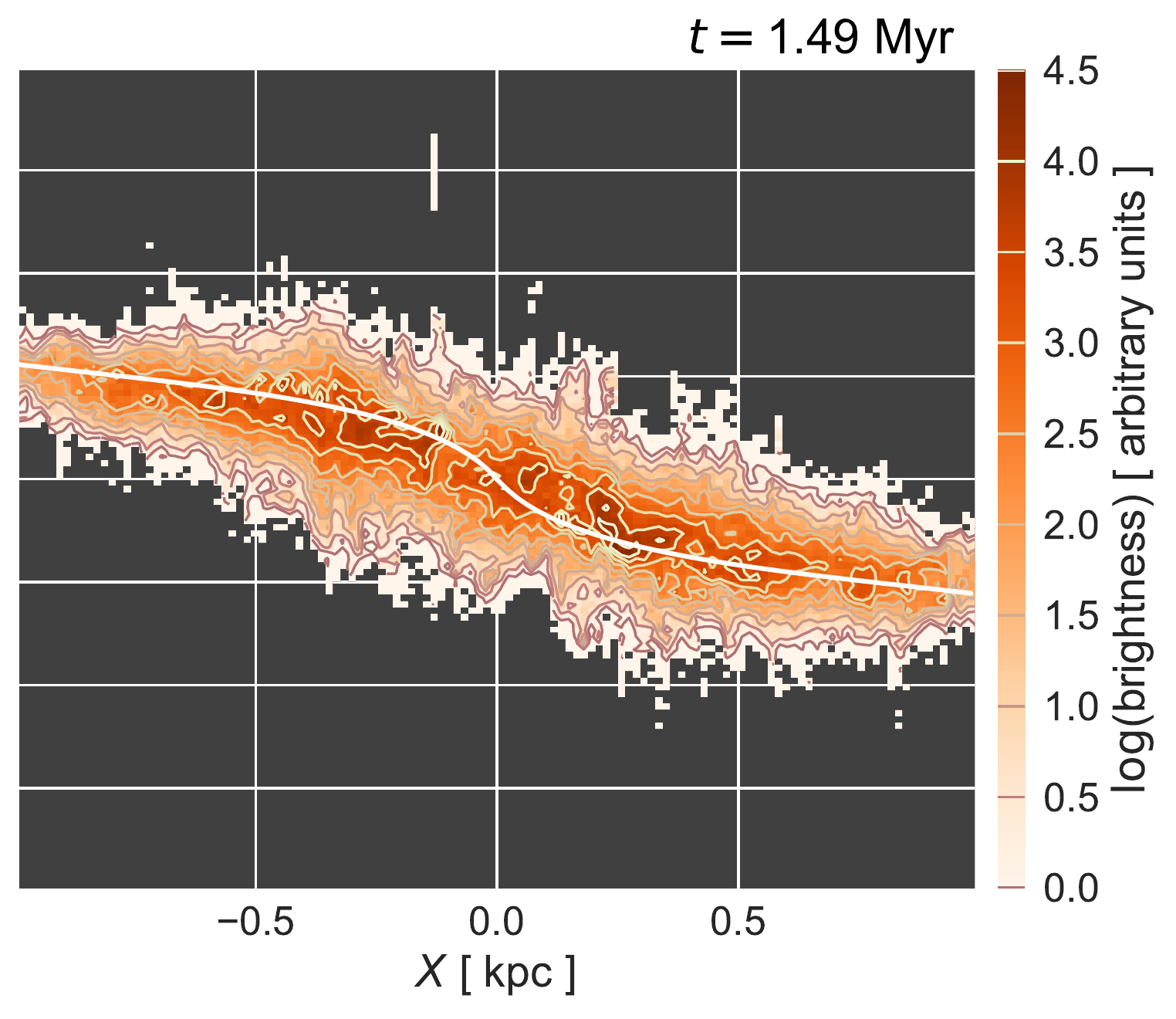}\\
\caption{PV diagrams generated from the \PFF{} simulation for three different snapshots. The right panel corresponds approximately to the current state of \icfost{} with regard to the extent of the jets.}
\label{fig:pv44}
\end{figure*}
\begin{figure*}
\centering
\includegraphics[height=5.4cm]{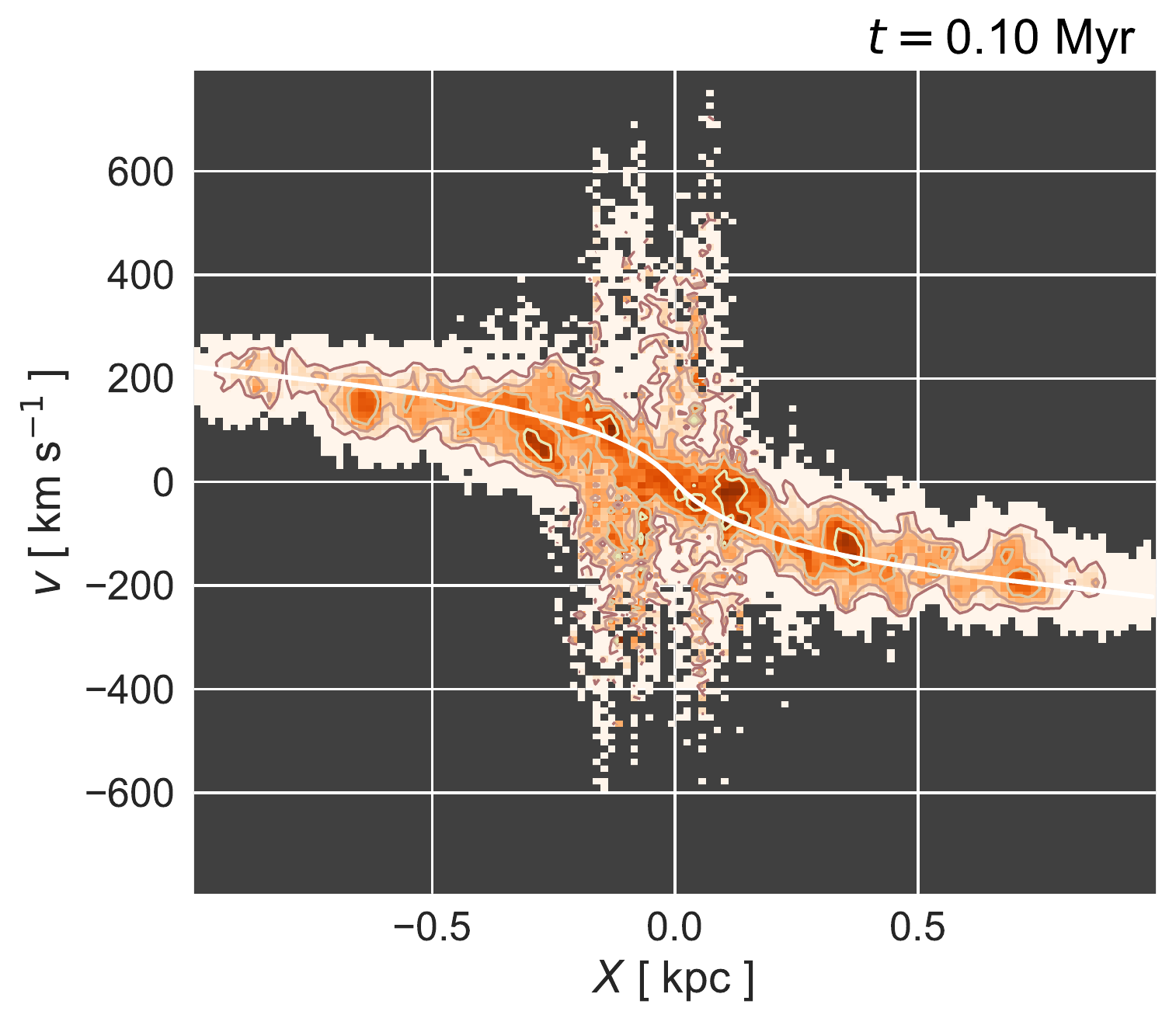}\hspace{-0.2cm}
\includegraphics[height=5.4cm]{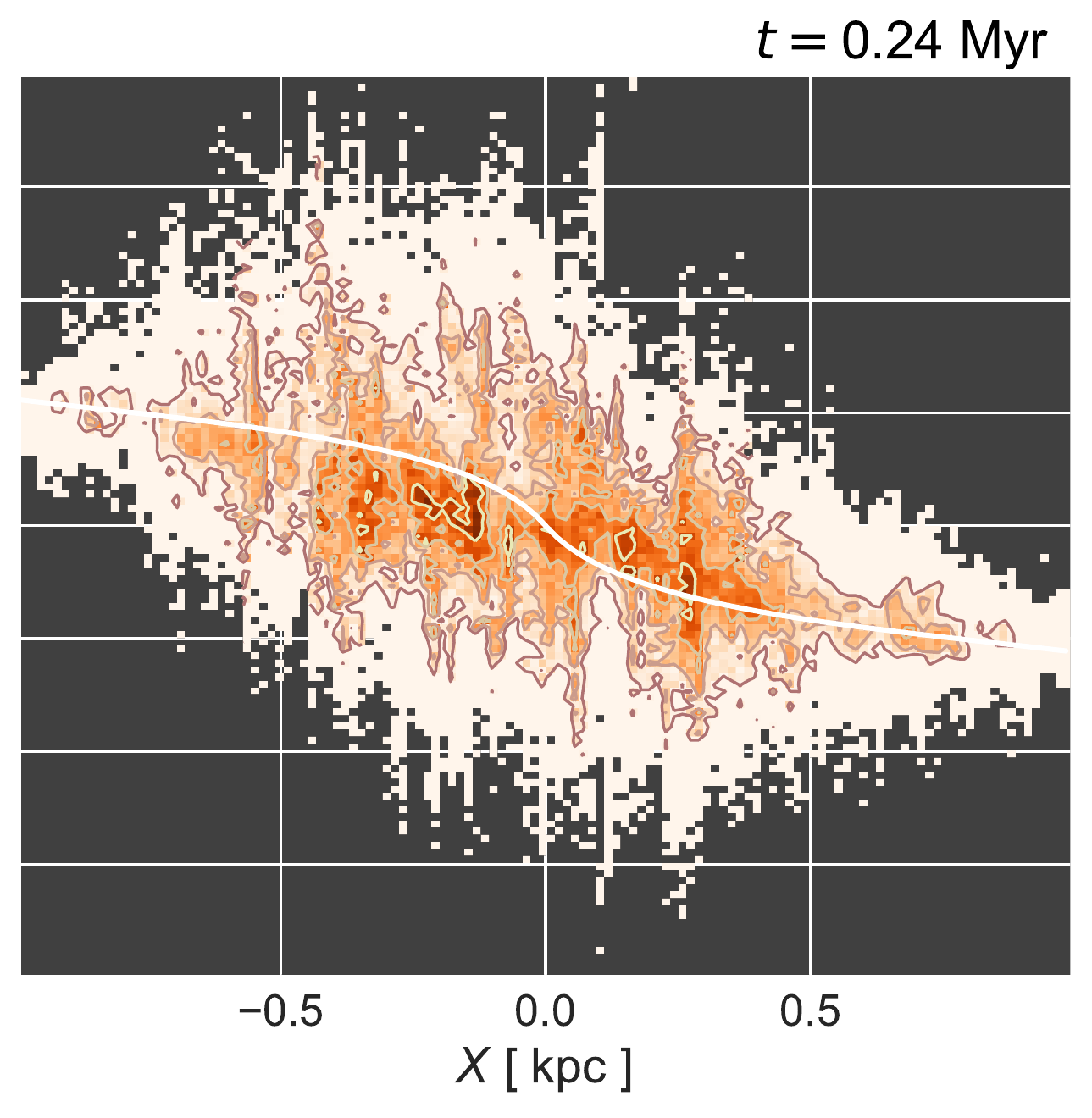}\hspace{-0.2cm}
\includegraphics[height=5.4cm]{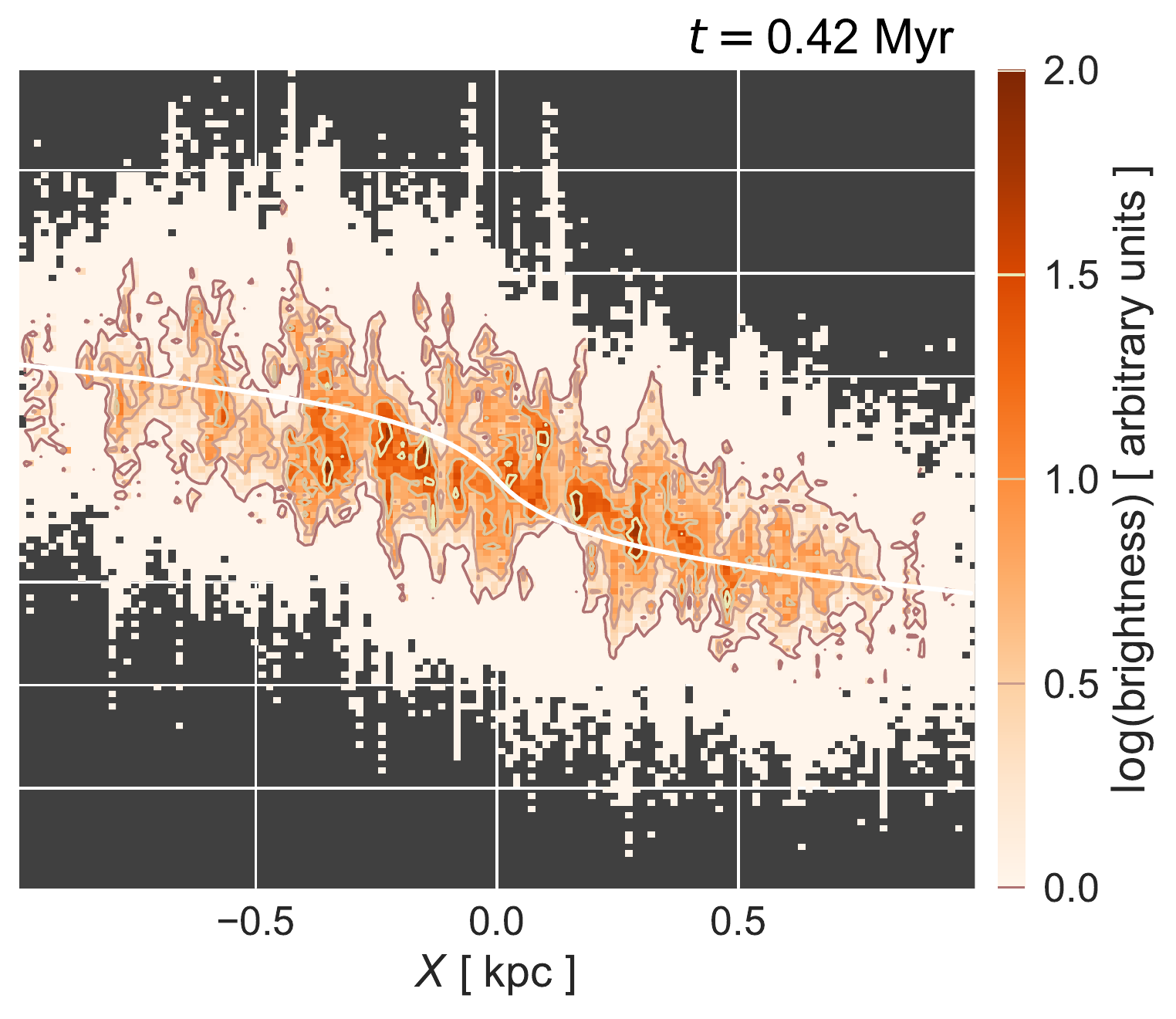}\\
\includegraphics[height=5.4cm]{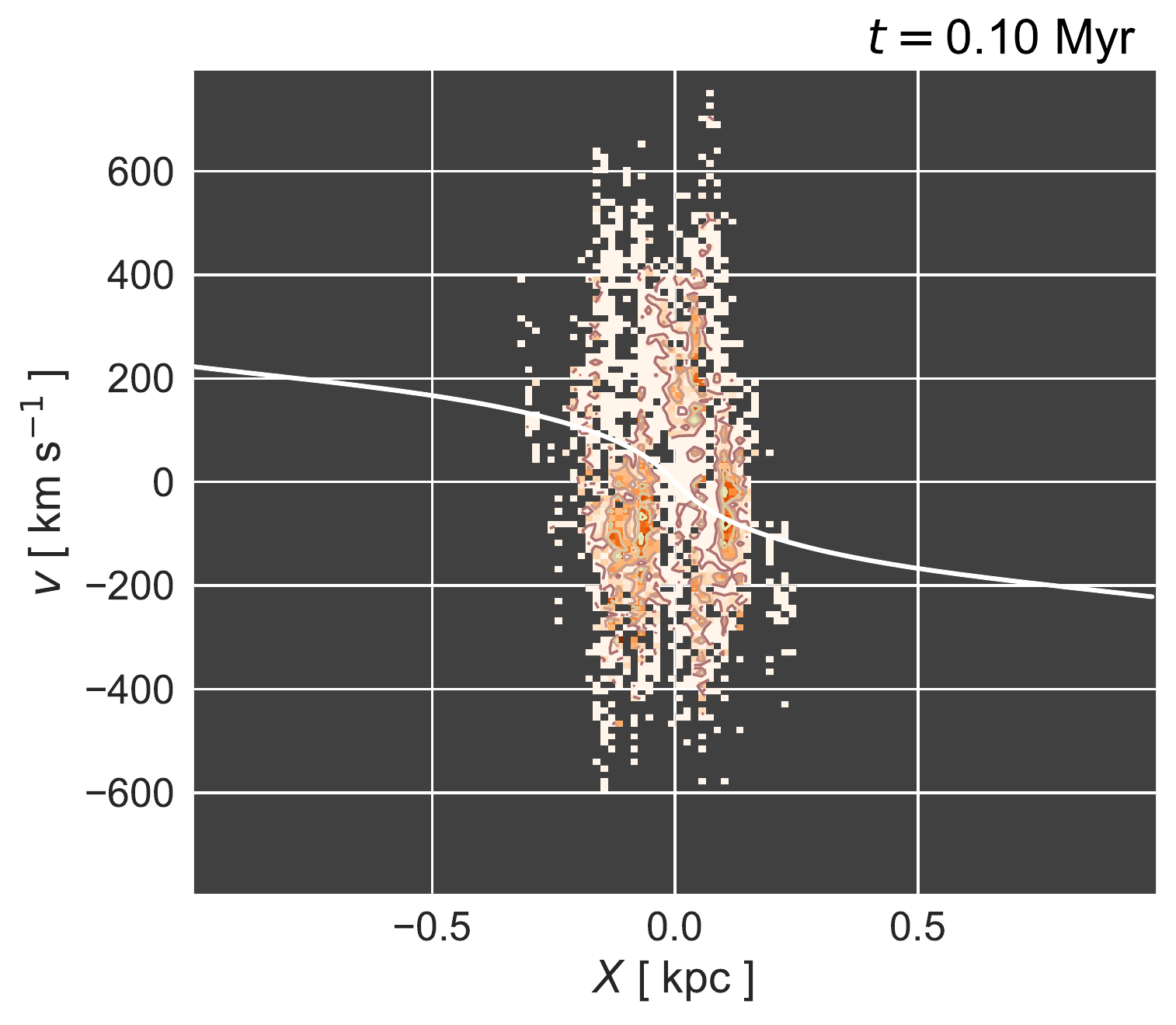}\hspace{-0.2cm}
\includegraphics[height=5.4cm]{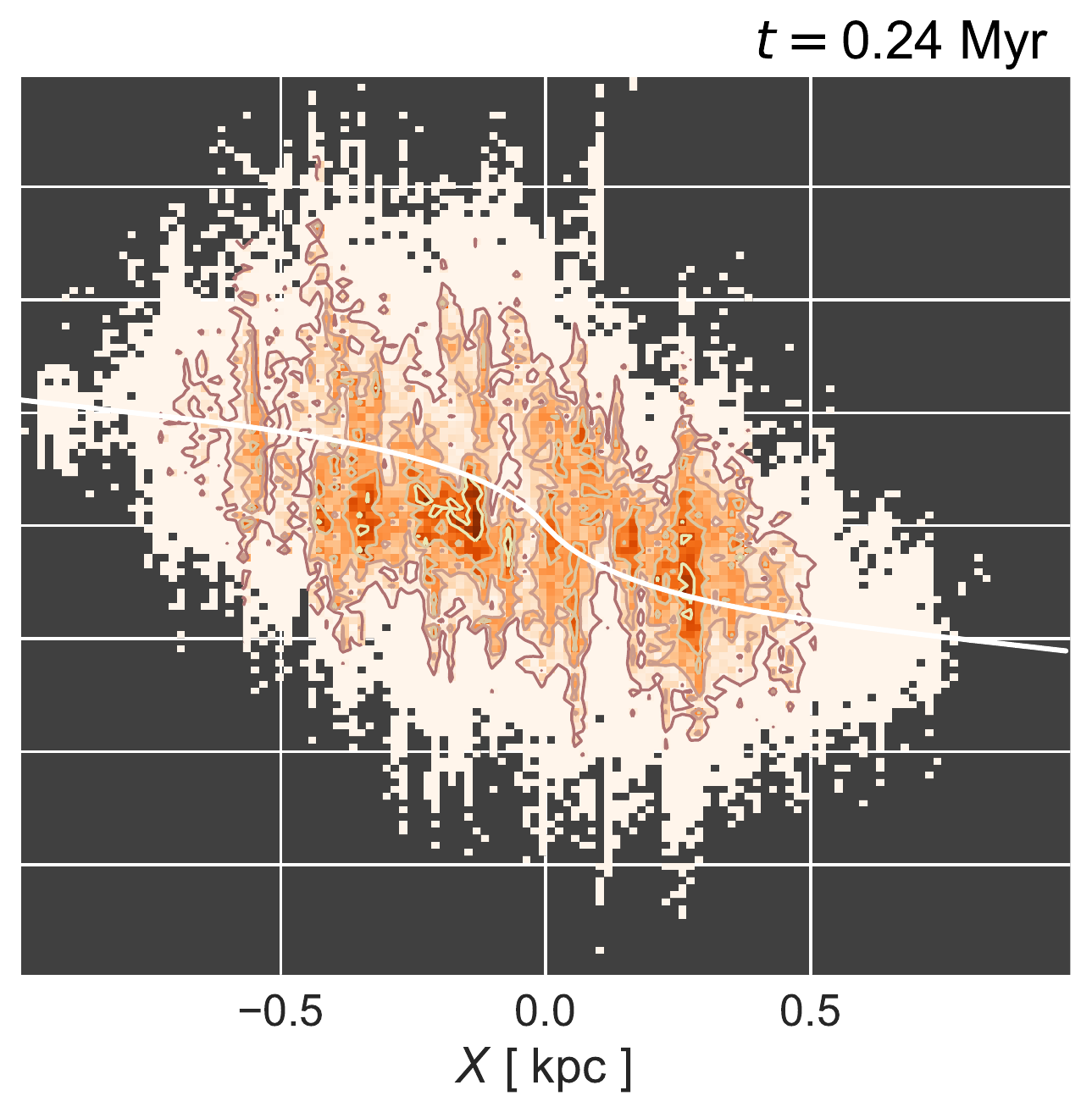}\hspace{-0.2cm}
\includegraphics[height=5.4cm]{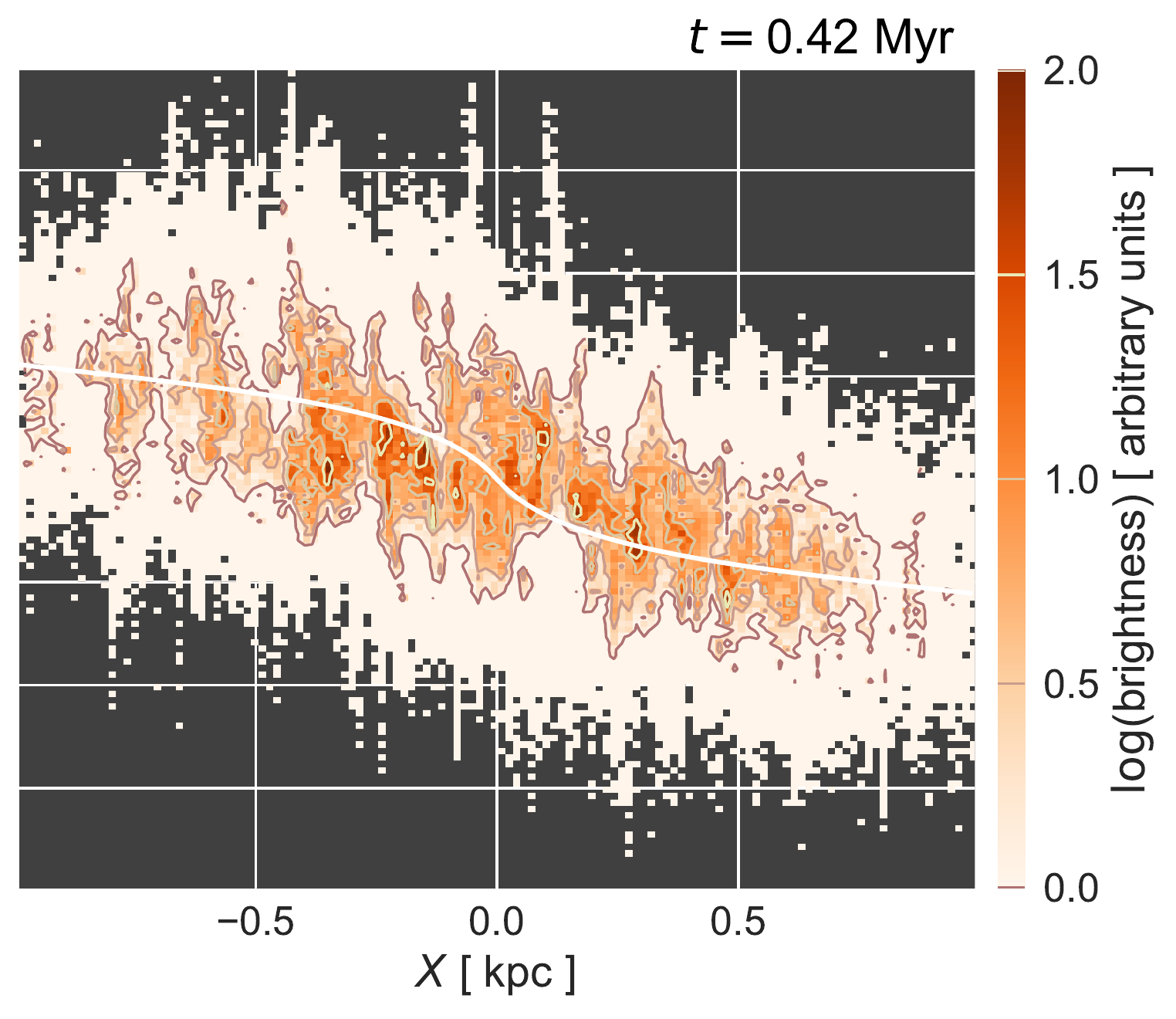}
\caption{PV diagrams for the model in which the jet power is $10^{45}\ergs$, a factor of 10 higher than that for the model shown in Fig.~\ref{fig:pv44}. The top row shows the PV diagrams for all gas according to Eq.~\eqref{eqn:emiss}, and the bottom row shows the PV diagram only for gas that is perturbed by the jet. The dynamic range covered by the colormap and contours is set to 2 dex, although the white area shows the extent of the data down to 4.5 dex in dynamic range to aid comparison with Fig.~\ref{fig:pv44}.}
\label{fig:pv45}
\end{figure*}
\begin{figure*}
\centering
\includegraphics[height=5.4cm]{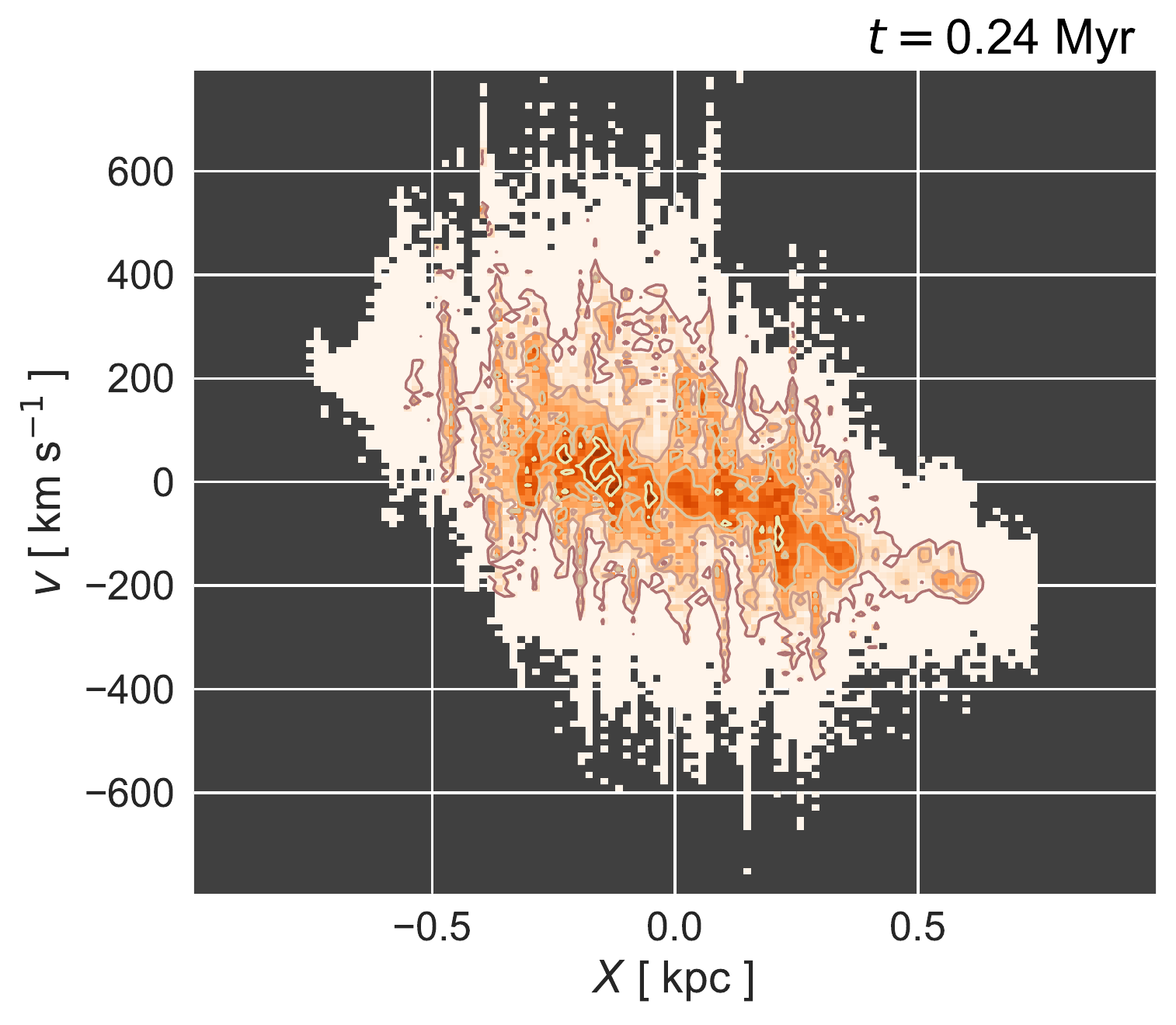}\hspace{-0.2cm}
\includegraphics[height=5.4cm]{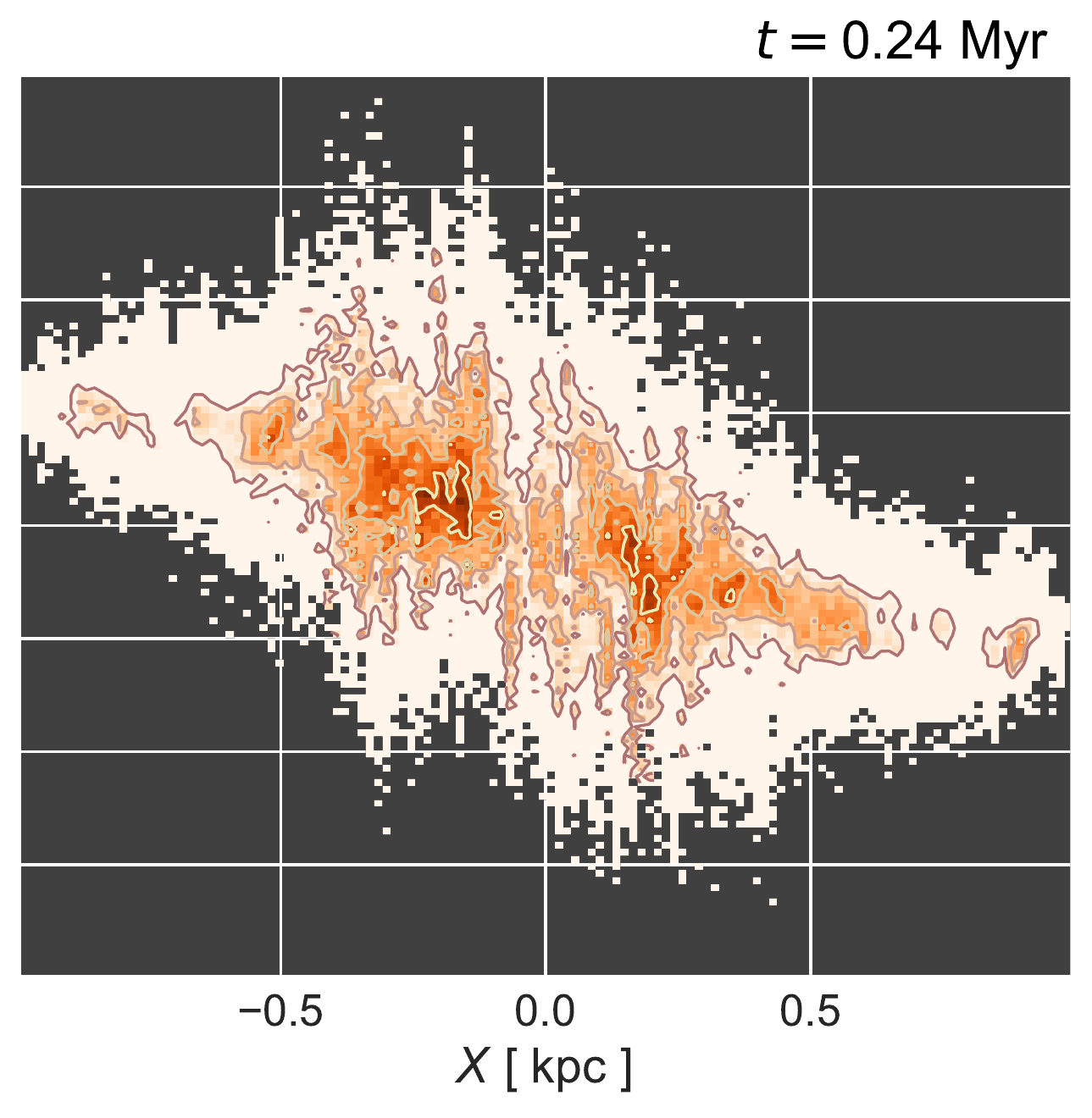}\hspace{-0.2cm}
\includegraphics[height=5.4cm]{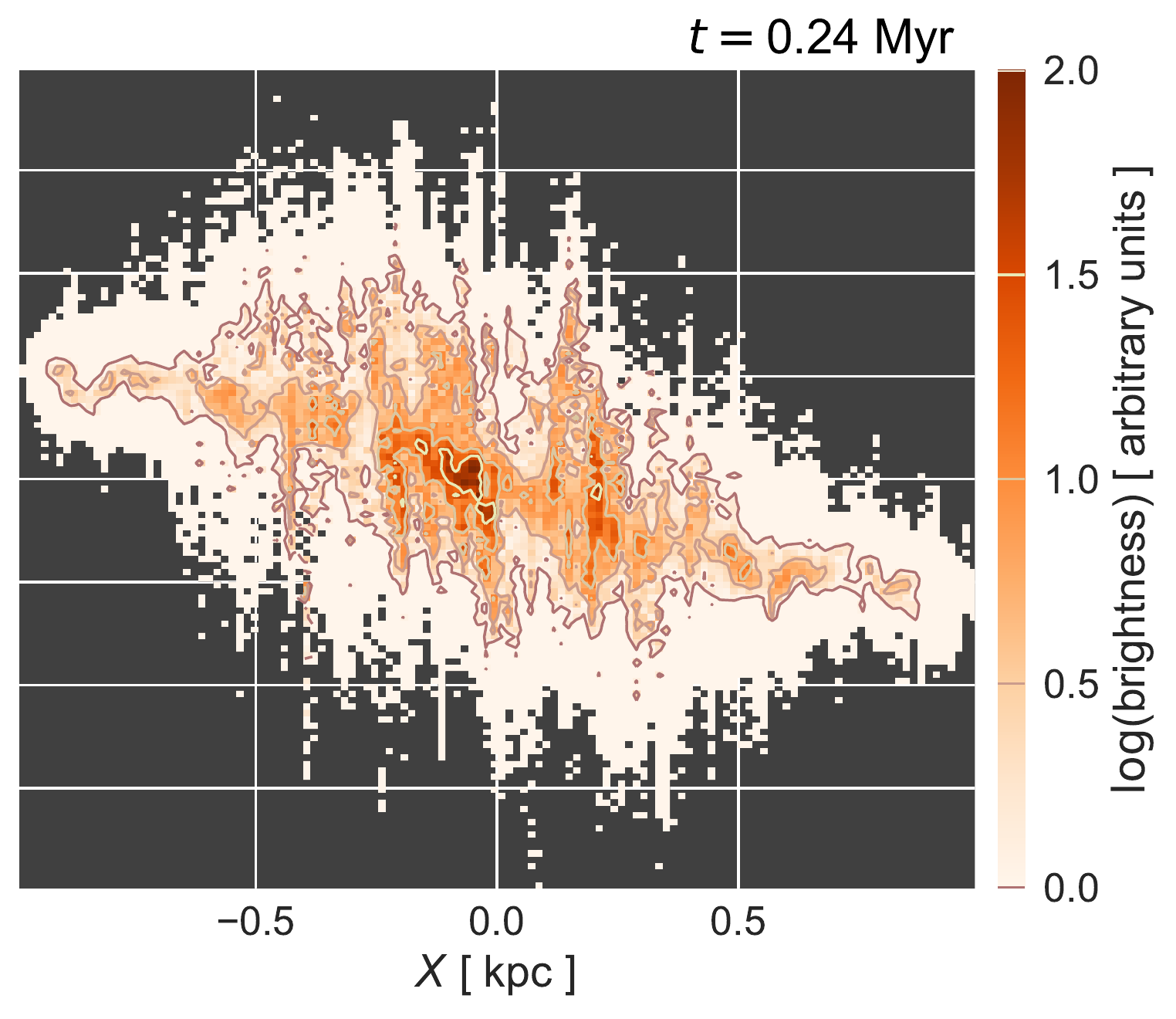}
\caption{PV plots for simulation \PFV{} for the case in which the plane of the disc is inclined by 26$^\circ$ to the line of sight, that is, $\theta=74\deg$ (left) and for the cases in which the line of sight is parallel to the disc but oriented by an azimuthal angle of $\phi=\pm45^\circ$ toward the jet axis (center and right). }
\label{fig:pvang}
\end{figure*}

From the velocity field in the simulations we construct synthetic PV diagrams and compare them to the PV diagrams of \ceeoh{2}{1} published by M15. In our simulations we do not directly track the molecular gas, nor do we follow the chemistry of molecule formation, so that we utilize temperature, cloud tracer, and density thresholds to infer the possible presence of molecular gas and to calculate the emissivity.

Let $\trw$ be the cloud tracer variable ($0 \leq \trw \leq 1$), $\critdens$ the critical density for a molecular transition, and $n$ and $\fmol$ be the particle number density and the molecule fraction by mass in a cell, respectively. We take the emissivity, $\emiss$, to be

\begin{equation}
\emiss =
\begin{cases}
    A \trw^2 n^2 & \text{if}\; n  <   \critdens \; \text{and} \; \fmol > 0 \\
    B  \trw n    & \text{if}\; n \geq \critdens \; \text{and} \; \fmol > 0 \\
    0            & \text{otherwise} \;, 
\end{cases}\label{eqn:emiss}
\end{equation}
where
\begin{equation}
\fmol = 
\begin{cases}
    \trw    & \text{if} \; T < 5000\kelvin \; \text{and} \; n \geq 10\cc \; \text{and} \; \trw \geq 0.98 \\
    0       & \text{otherwise} \;. 
\end{cases} \label{eqn:fmol}
\end{equation}
The values for the temperature, density, and jet tracer thresholds in Eq.~\eqref{eqn:fmol} were chosen to select a sufficient number of dense, cool cells in the simulations that are likely to contain molecular gas. The normalizations $A$ and $B$ are arbitrary, but satisfy $B = A \trw \critdens$. We took the critical density $\critdens=2\times10^4\cc$ for \ceeoh{2}{1}, although the PV diagrams are not very sensitive to the choice of $\critdens$ since only a small fraction of cells in our simulations have densities above the critical density.

The emissivity in each cell is then integrated along the line of sight through a slit of width $444\pc$ across a length of $1\kpc$ along the mid-plane of the galaxy. The alignment and width of the slit are the same as those used by M15. We bin the integrated emission into 128 bins along the slit and 128 bins in velocity space between -$800\kms$ and $800\kms$. This results in a somewhat finer spatial and velocity resolutions than those of the M15 data.

When producing the synthetic PV diagrams, we assume that the gas is optically thin. 
All the disturbed and outflowing gas in \icfost{} is known to be optically thin \citep{Oosterloo2017a, Dasyra2016a}. Furthermore, our simulations show that strong velocity dispersions exist throughout the disc, suggesting that the ISM in the disc may be a mixture of gas that is optically thin\footnote{The optical depth of line emission in a medium with velocity dispersion depends on the velocity gradient as $\tau \propto \frac{1}{|dv/dz|}$ \citep[see e.g.][]{castor70a,neufeld93a} } and optically thick. In any case, our synthetic PV diagrams likely overestimate the surface brightness along the rotation curve relative to the surface brightness of the dispersed gas. Modeling the detailed distribution and radiative transfer of CO transitions for the disc is beyond the scope of this work and we therefore do not attempt to directly compare the range of the modeled and observed CO surface brightness in the PV diagram, focusing instead only on the gas kinematics. However, the fact that the kinematics of the neutral gas \citep{Oosterloo2000a} and the PV diagrams of ionized gas phases \citep{Morganti2007a} are similar to the PV diagram of \ceeoh{2}{1} (M15) indicates that the underlying gas motions are physically related and that our model, therefore, properly captures the jet-ISM interactions that result in the observed velocity dispersions. 

Given the above simplifications and keeping in mind that the dynamical range in density is limited by the spatial resolution of the simulations, the synthetic PV diagrams presented here are only meant to show qualitative trends and qualitative agreement with the ALMA data obtained by M15.

In the simulations we can easily distinguish between gas impacted by the jet and gas in the unperturbed disc using the tracer variable for the jet plasma and selecting cells for which the jet tracer is greater than a small but non-zero value, typically $10^{-8}$. The small value ensures that most cells in the interior of clouds are selected, even if the mixing of the jet plasma and cloud material is dominated by numerical diffusion.

In Fig.~\ref{fig:pv44} and Fig.~\ref{fig:pv45} we show PV diagrams for model \PFF{} and \PFV{}, respectively. The solid white curve traces the projected circular velocity (Eq.~\ref{eqn:vcirc}) weighted by the mean density profile of the turbulent disc given by Eq.~\eqref{eqn:disc}. It approximates the mean rotation of the disc gas represented in the PV plots at $t=0$ (see also Fig.~\ref{fig:ic}). The brightness is in arbitrary units, but limited to a dynamical range of 4.5 dex, although for the PV diagrams of simulation \PFV{}, we limit the colored contours to 2 dex, because the velocity dispersions attained by the more powerful jet are higher. In fact, the velocity dispersions in simulation \PFV{} are almost a factor of 2 higher compared to those in simulation \PFF{}, as expected from the scaling $\Pjet \propto \sigma^{1/4}$ found by \citet{Wagner2011a}. 

The top row in Fig.~\ref{fig:pv45} shows the PV diagrams for all gas for which $\fmol > 0$, and the bottom row shows the PV diagrams only for gas perturbed by the jet. A visual comparison of the two rows of PV diagrams underlines how strong the effect of the jet on the dynamics of the clouds throughout the entire disc is. 

We recall that there are five notable features in the observed PV diagrams of \icfost{} presented in M15: 1) Broad velocity dispersions of order a few $100\kms$ across the spatial region overlapping with the jet; 2) Spikiness 3) Asymmetry in the velocity dispersions; 4) Strong dispersion in the forbidden quadrants (i.e. motions in the counter-rotational direction); 5) Straightening of the rotation profile compared to that predicted by the gravitational potential. 

To a greater or lesser extent, we recover all these features in the synthetic PV diagrams generated from the simulations. The broadening of the velocity dispersion and the spikiness of the broadening is particularly clear, especially for simulation \PFV{}, but also for simulation \PFF{} while $t<0.7\Myr$. In the splattering regions of strong jet-ISM interactions, the broadening is by as much as $600\kms$; the clouds here are dispersed nearly isotropically, giving rise to emission from the forbidden quadrants. At $t=1.49\Myr$, however, the jet in simulation \PFF{} does not appear capable of producing the observed velocity dispersion. Simulation \PFV{}, on the other hand, exhibits strong velocity dispersion even at $t=0.24\Myr$ when the jet has evolved roughly to the extents observed in the $8\GHz$ radio image. The asymmetry in the broadening is not as pronounced as in the data in M15 for either simulation, which may indicate an even more irregularly clumped ISM than we have assumed. 

On the other hand, the straightening of the rotation profile along the central 0.3 pc region seen in the PV diagram, which reduces the steepening of the gradient compared to the solid white line, is clearly seen in both simulations. At $t=1.49\Myr$ and at $t=0.24\Myr$, simulations \PFF{} and \PFV{} show shallower rotation profiles, compared to the rotation profile at time $t=0$. Simulation \PFV{} displays a considerably stronger straightening of the rotation profile, compared to simulation \PFF{}. In simulation \PFV{}, the dense gas in the disc also appears to be more strongly broken up, as evidenced by the gaps between regions delineated by the highest brightness contour levels.

The degree to which the rotation profile is straightened due to jet-ISM interactions depends on how much of the inner dense clouds is displaced to larger radii. For simulation \PFV{} at $t=0.24\Myr$, the gradient along the brightest parts is almost constant, meaning that most of the gas is redistributed to large disc radii. However, the process may be more complicated. Bulk cloud displacement does occur, but in the inner few kpc, most cloud material is ablated and clouds are dispersed in their entirety. We must also always bear in mind that we may not be capturing cloud survival properly -- magnetic fields and turbulence may play a role, but we are also limited by numerical diffusion at the grid resolution and unresolved turbulent dissipation. We cannot, therefore, make a strong statement as to whether simulation \PFF{} or simulation \PFV{} represents a closer match to the observed data as far as the straightening of the rotation profile is concerned.

M15 outlined a model that explained the dispersions seen in the PV diagrams through radial mass transport by the pressure of the jet-driven, expanding radio bubbles. This model, although confined to a simple geometry, was able to explain the dispersion in the gas and the straightening of the rotation curve. Our simulations indicate that the physics in their model is correct, and in addition capture the stochastic nature of jet--cloud interactions that lead to the asymmetric and jagged features in the PV diagrams.

There is some freedom in the choice for the direction of line-of-sight with respect to the plane of the disc and the jet axis. We have tested various viewing angles and the PV diagrams are fairly similar for line-of-sight inclinations of less than $\pm45^\circ$ with respect to either the plane of the disc or the jet axis. Note that the pairs of viewing angles $(\theta, \phi)$ and $(-\theta, \phi)$ for any $\theta$ and $\phi$ are symmetric as far as the average density and velocity field are concerned. However, pairs of viewing angles $(\theta, \phi)$ and $(\theta, -\phi)$ differ intrinsically because the orientation of the jet flux relative to the velocity field of the rotating disc gas along lines of sight offset from the center are different. 

In Fig.~\ref{fig:pvang}, we show the PV plot for simulation \PFV{} for the case in which the line of sight is inclined by $\theta=74^\circ$ to the disc normal and for the cases in which the line of sight is oriented by an azimuthal angle of 45$^\circ$ with respect to the normal to the jet axis along the plane of the disc ($\phi=\pm 45^\circ$). In all three cases the main features of the PV diagram for simulation \PFV{} along $(\theta, \phi) = (0, 0)$ (Fig.~\ref{fig:pv45}, middle panel) are retained, but individual regions of strong dispersion vary. For example, when the viewing angle is tilted along $\phi$ so that the component of the jet flux parallel to the line of sight act in the opposite direction to the component of the rotating disc gas parallel to the line of sight, the emission from the forbidden quadrants becomes enhanced (right panel in Fig.~\ref{fig:pvang}). Broadly speaking, however, the features of the synthetic PV diagrams are insensitive to the choice of line of sight within a few tens of degrees, and we attribute this to the isotropic nature of the gas dispersions caused by the jets.

\subsection{Other results}\label{sec:other-results}
\begin{figure}
\centering
\includegraphics[width=7cm, keepaspectratio]{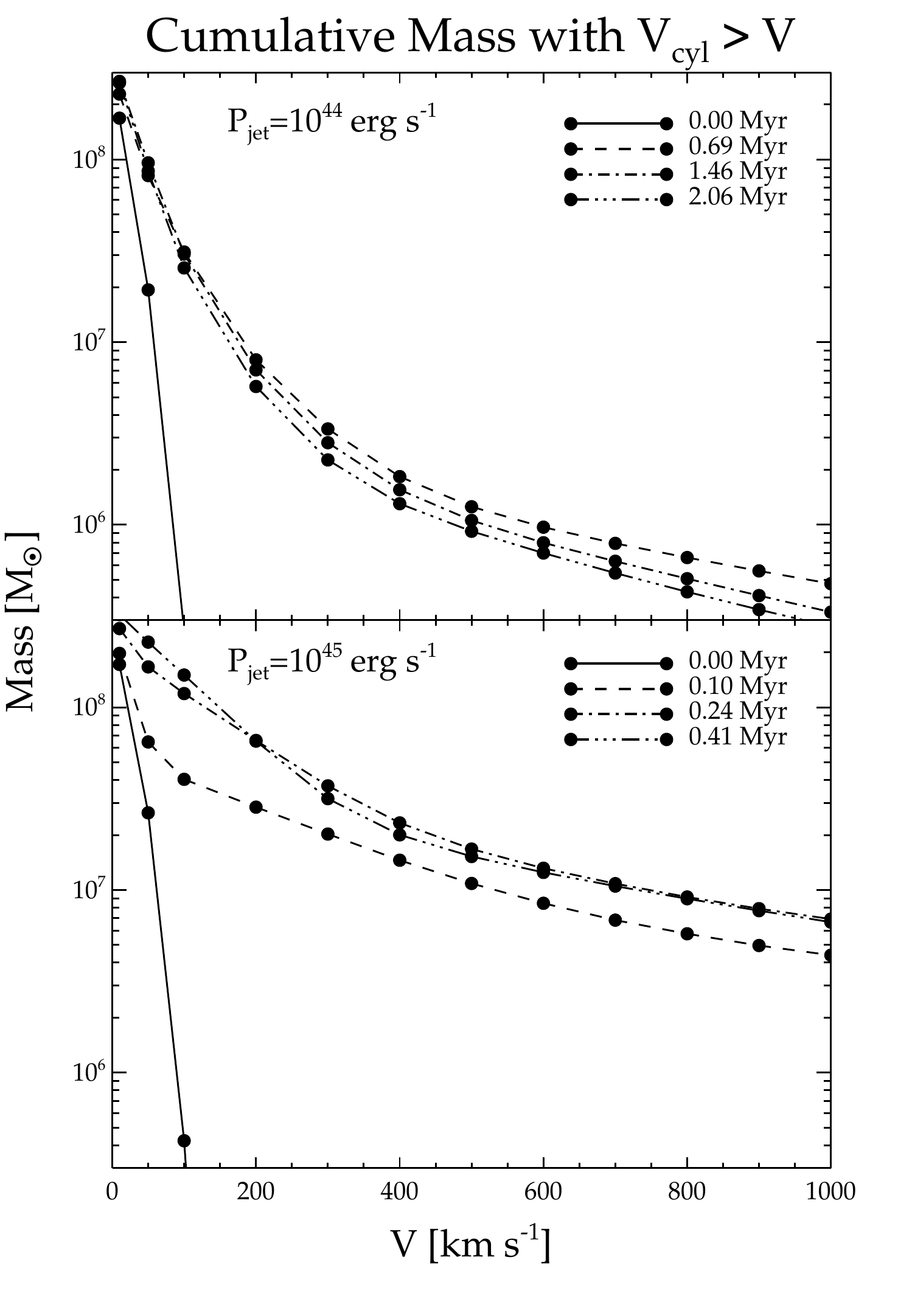}
\caption{Mass of gas with cylindrical velocity ($\vcyl$) greater than a given value, at different times. After the jet has evolved significantly, $8\times10^6\Msun$ of gas has an outflow velocity of $\gtrsim200\kms$ for simulation \PFF{}. For simulation \PFV{} the total mass of gas with $\vcyl \gtrsim 200 \kms$ is $\sim 7\times10^7\Msun$. }
\label{fig:massfrac_vrad}
\end{figure}
\begin{figure}
\centering
\includegraphics[width=7cm, keepaspectratio]{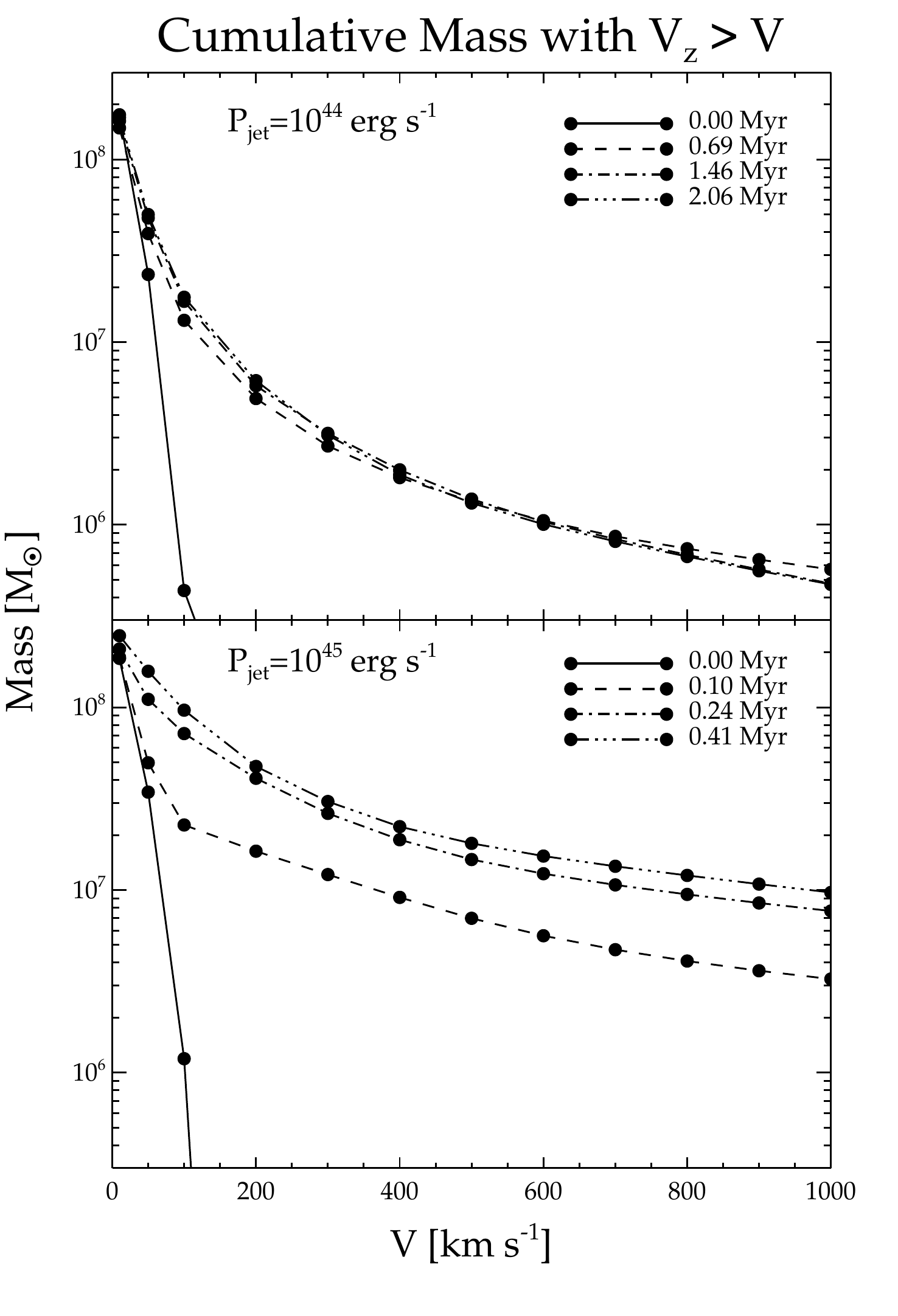}
\caption{Mass of gas with vertical velocity ($\vel_z$) greater than a given value, at different times. After the jet has evolved significantly, $6\times10^6\Msun$ of gas has an outflow velocity of $\gtrsim200\kms$ for simulation \PFF{}. For simulation \PFV{} the total mass of gas with $\vel_z \gtrsim 200 \kms$ is $\sim 5\times10^7\Msun$. The outflowing masses in the plane of the disc are similar to the outflowing masses in the vertical direction (see Fig.~\ref{fig:massfrac_vz})}
\label{fig:massfrac_vz}
\end{figure}
\begin{figure*}
\centering
\includegraphics[height=5.4cm]{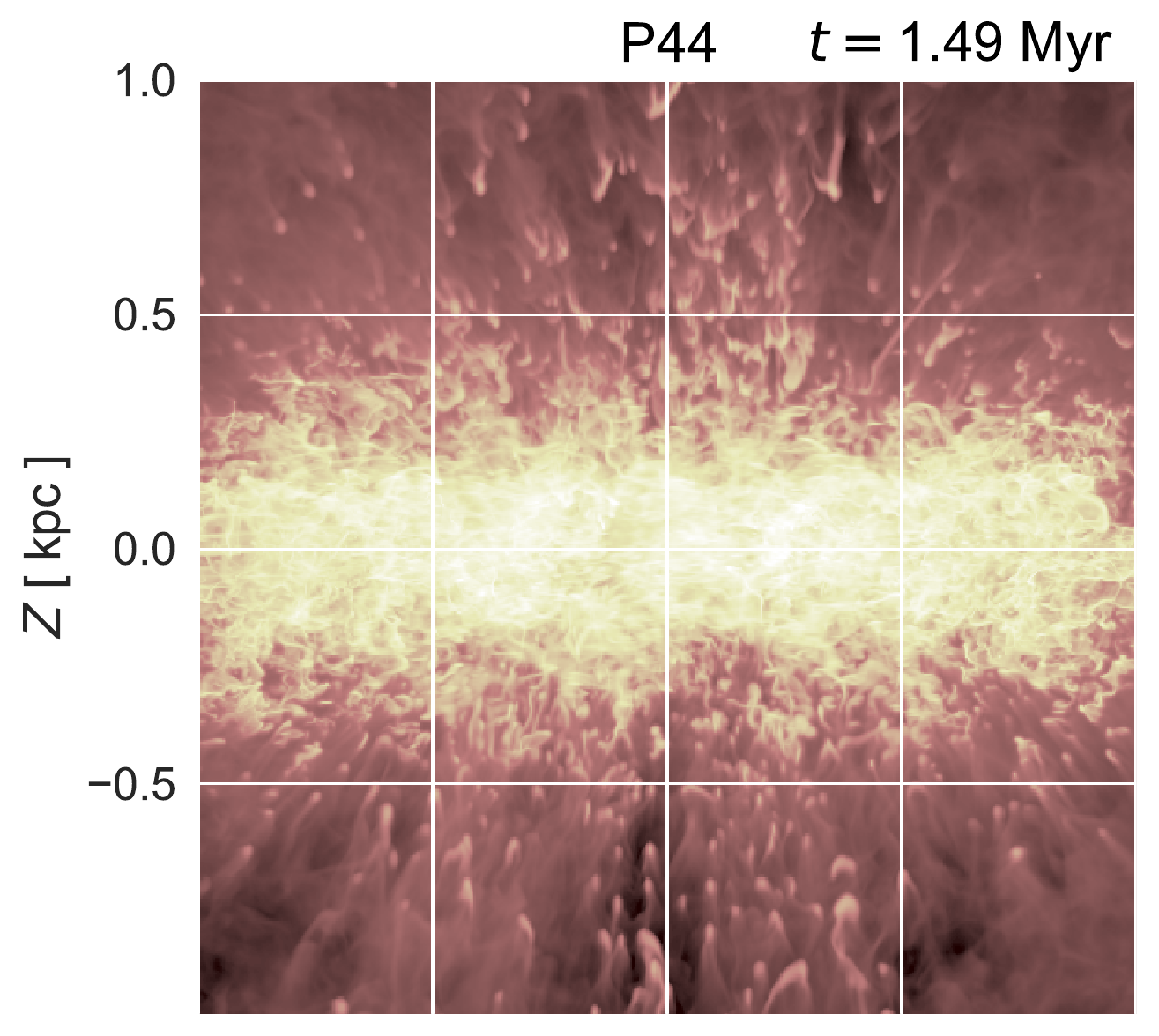}\hspace{-0.05cm}
\includegraphics[height=5.4cm]{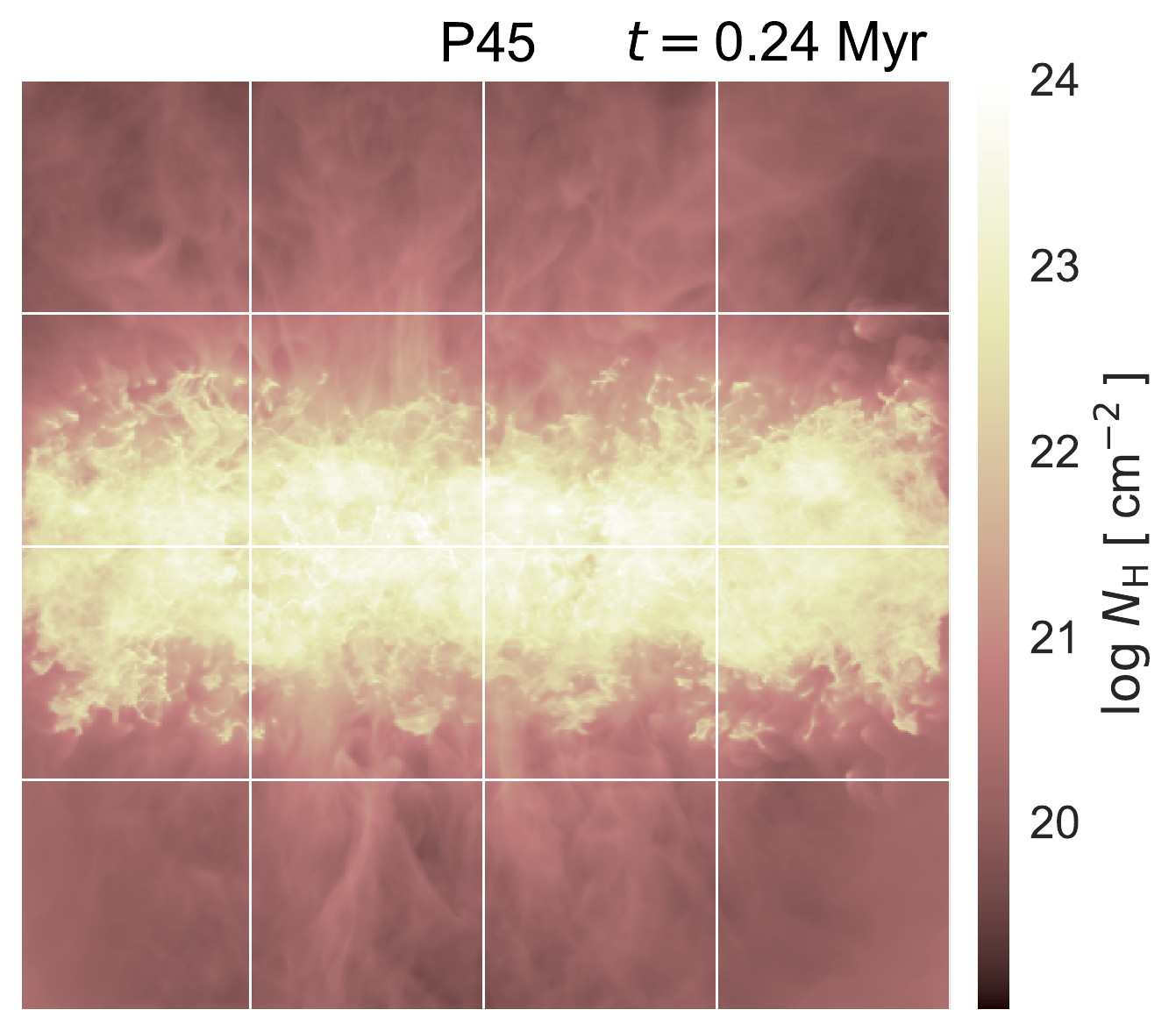}\\\hspace{-0.05cm}
\includegraphics[height=5.88cm]{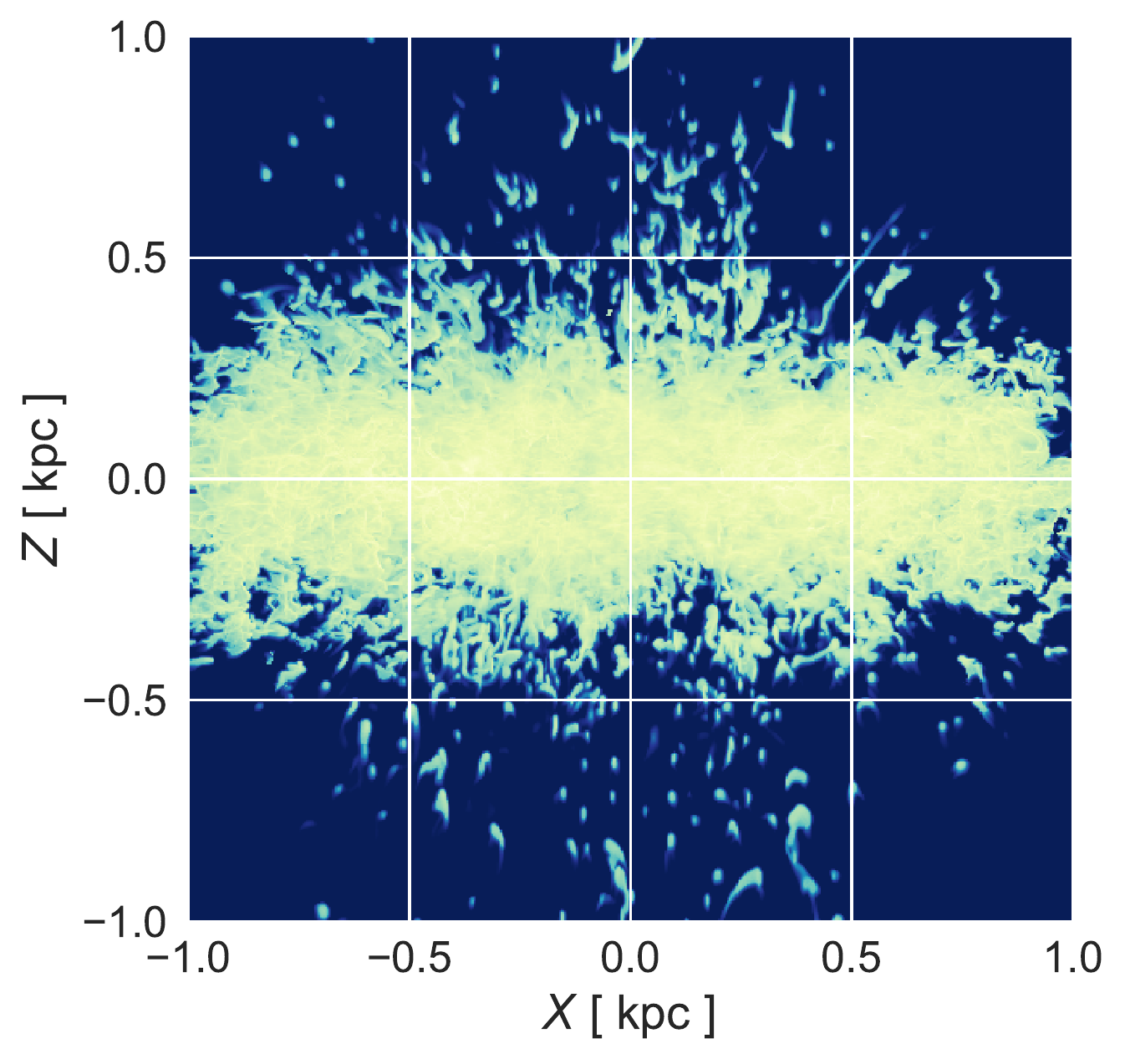}\hspace{-0.22cm}
\includegraphics[height=5.88cm]{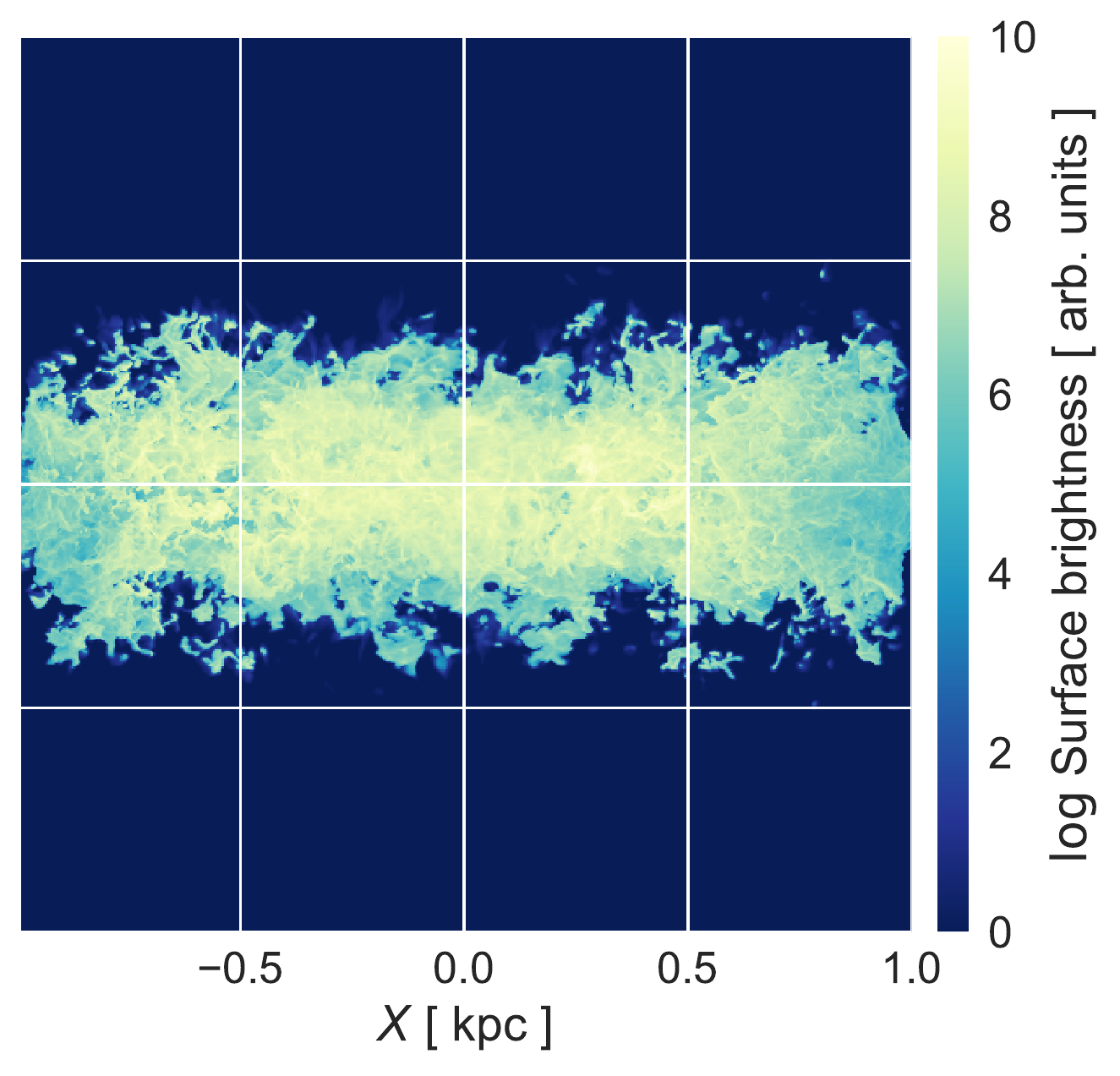}
\caption{Comparison of side-on column density and \halpha{} surface brightness between simulations \PFF{} and \PFV{}. Top panels are the column density, bottom panels are \halpha{} surface brightness. The left panels are simulation \PFF{}, and the right are \PFV{}. Simulation \PFF{} shows cometary clumps lifted off the disc of the galaxy due to energy deposited in the disc by the jet during its long confinement time. The much shorter confinement time of the jet in simulation \PFV{} does not lead to such an effect and the vertical outflow is mainly hot gas beyond $10^5\kelvin$.}
\label{fig:nhha}
\end{figure*}
\begin{figure*}
\centering
\includegraphics[height=5.4cm]{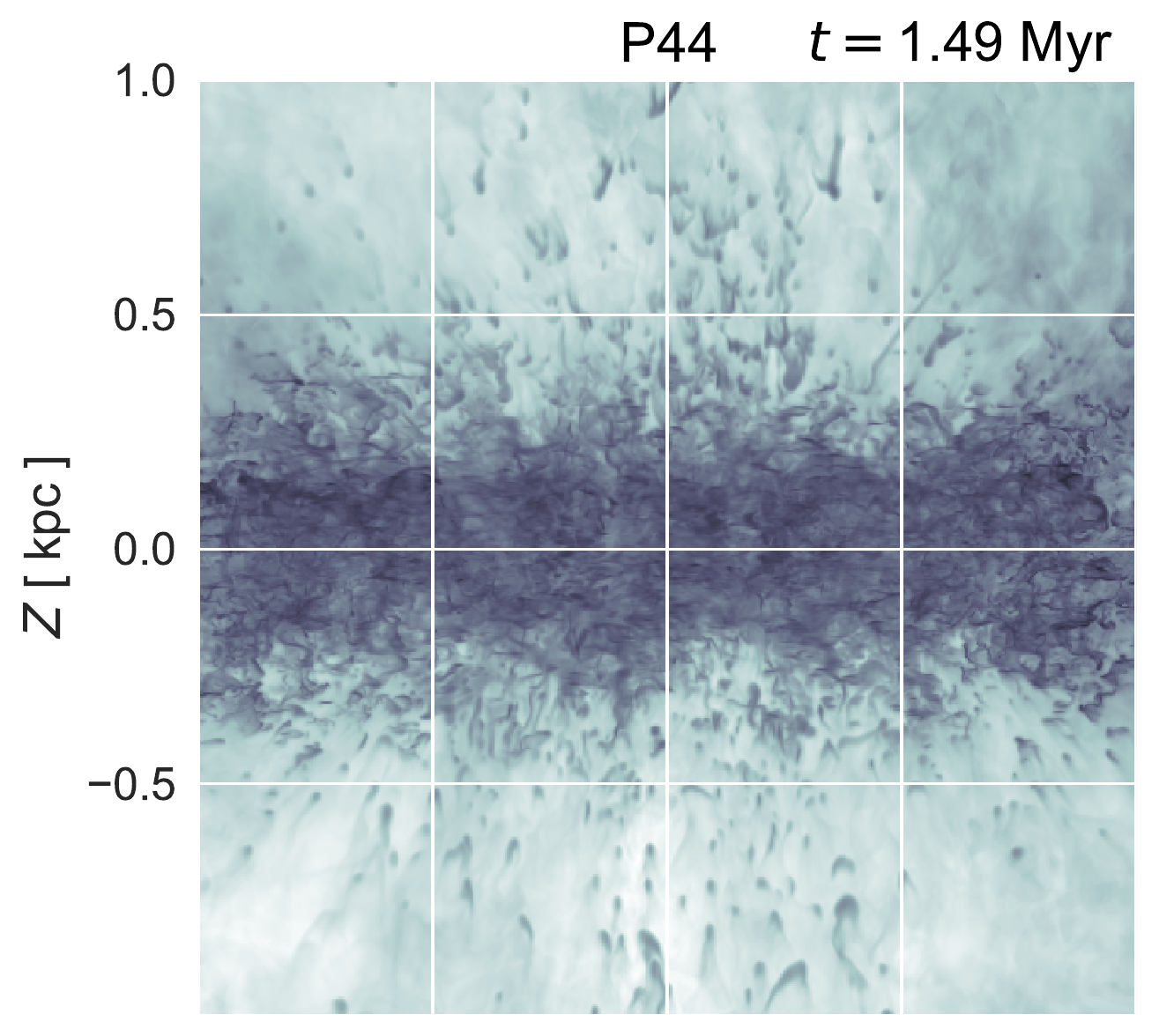}\hspace{-0.05cm}
\includegraphics[height=5.4cm]{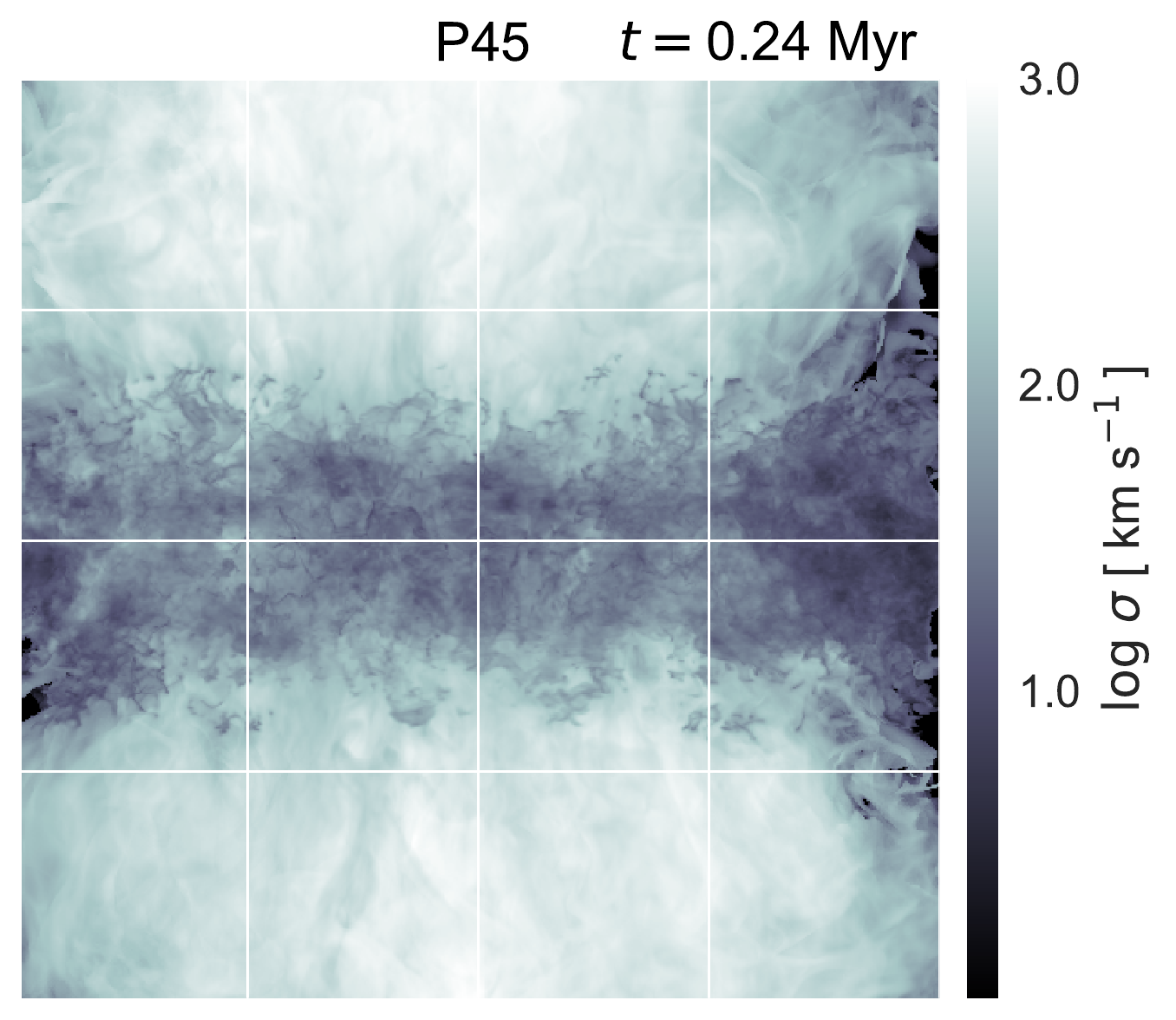}\\\hspace{-0.25cm}
\includegraphics[height=5.88cm]{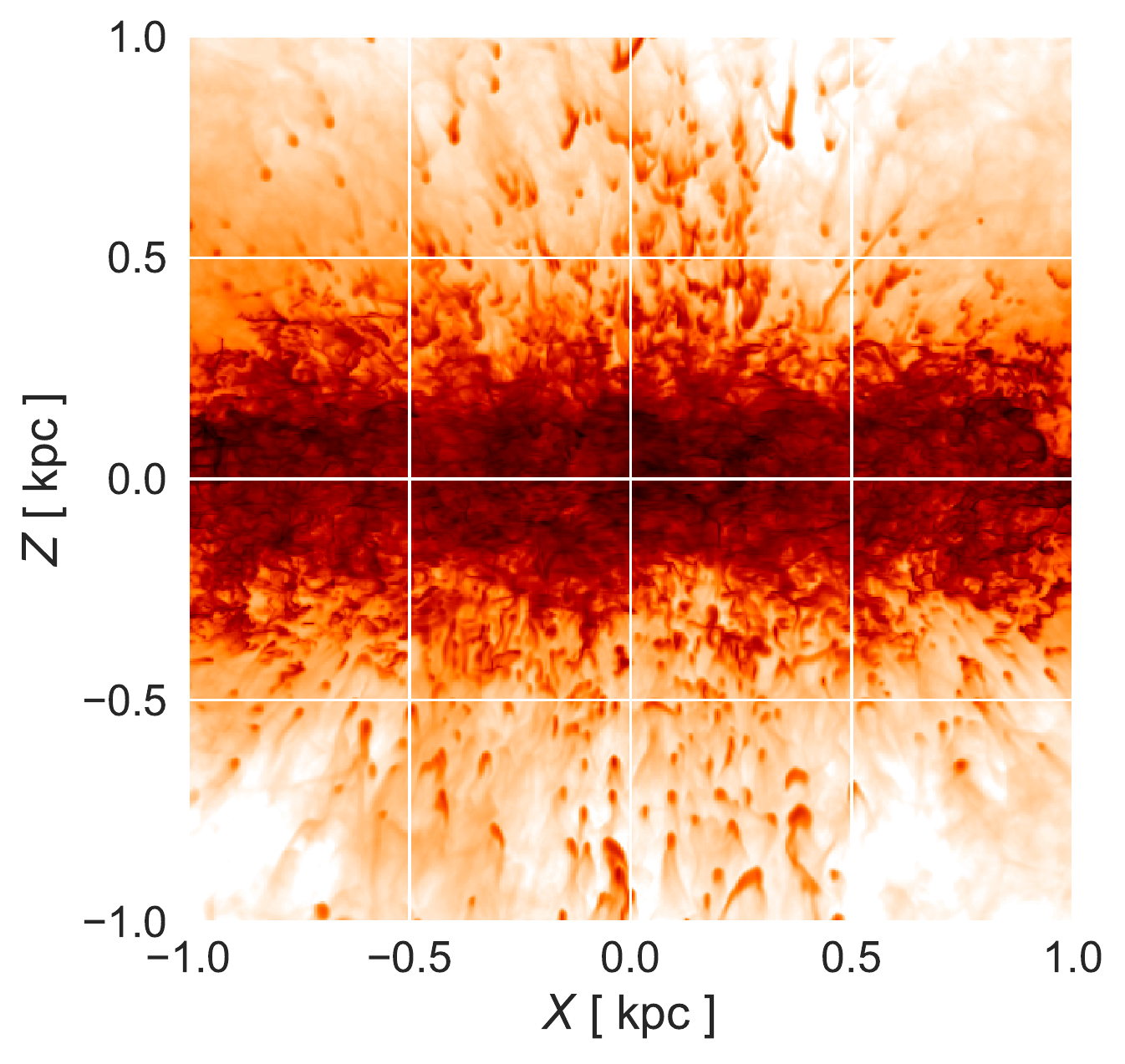}\hspace{-0.22cm}
\includegraphics[height=5.88cm]{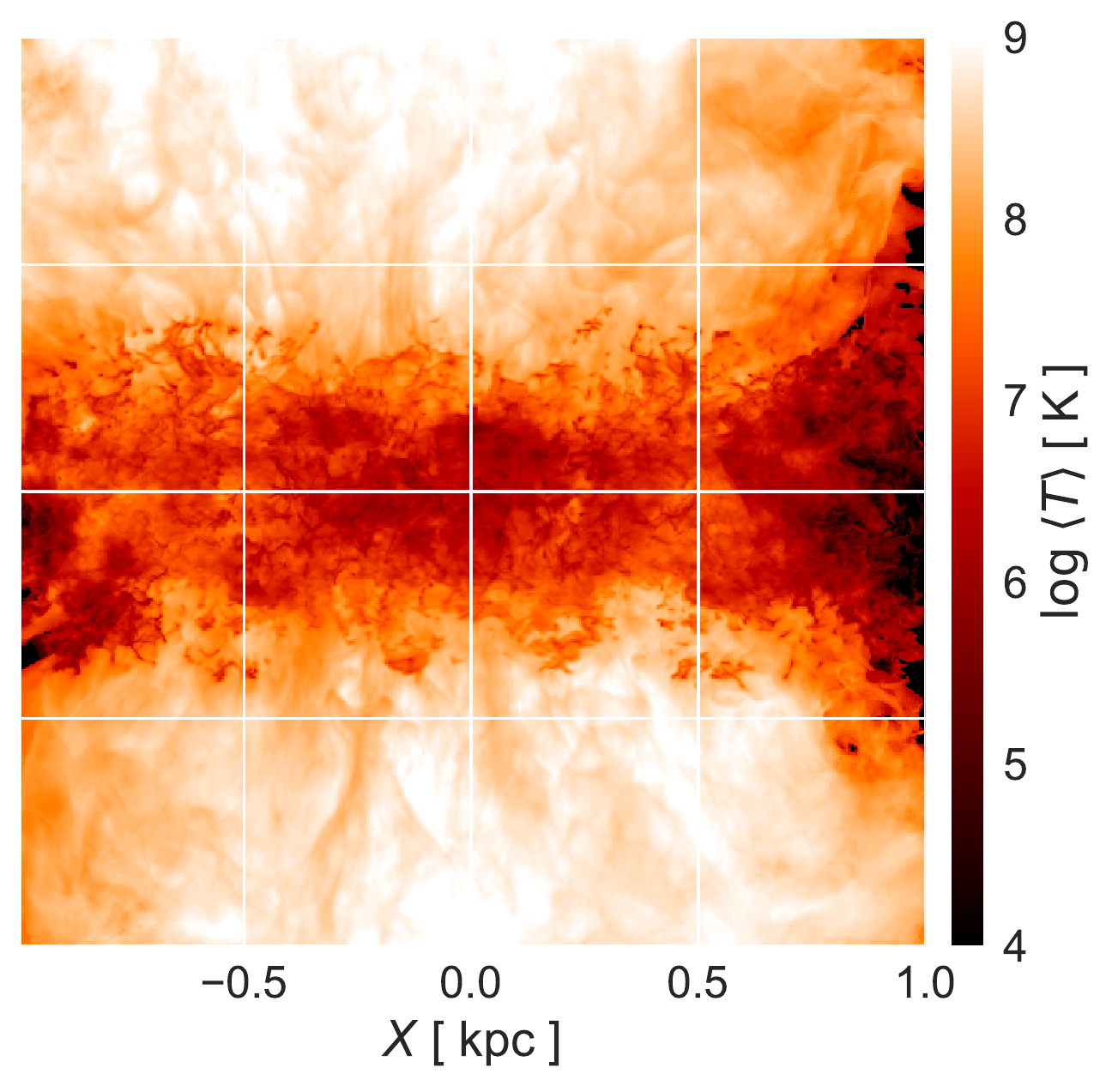}
\caption{Comparison of side-on line-of-sight velocity dispersion and mean temperature along the line of sight between simulations \PFF{} and \PFV{}. Top panels are the velocity dispersion, bottom panels are mean temperature. The left panels are from simulation \PFF{}, and the right are for simulation \PFV{}. The velocity dispersion and mean temperature are density-weighted and a dense phase tracer threshold of $\trw>0.98$ is applied. Simulation \PFF{} shows cometary clumps lifted off the disc of the galaxy due to energy deposited in the disc by the jets during their long confinement time. The much shorter confinement time of the jets in simulation \PFV{} does not lead to such an effect and the vertical outflow is mainly hot gas beyond $10^6\kelvin$.}
\label{fig:sgte}
\end{figure*}
\begin{figure*}
\centering
\includegraphics[height=5.88cm]{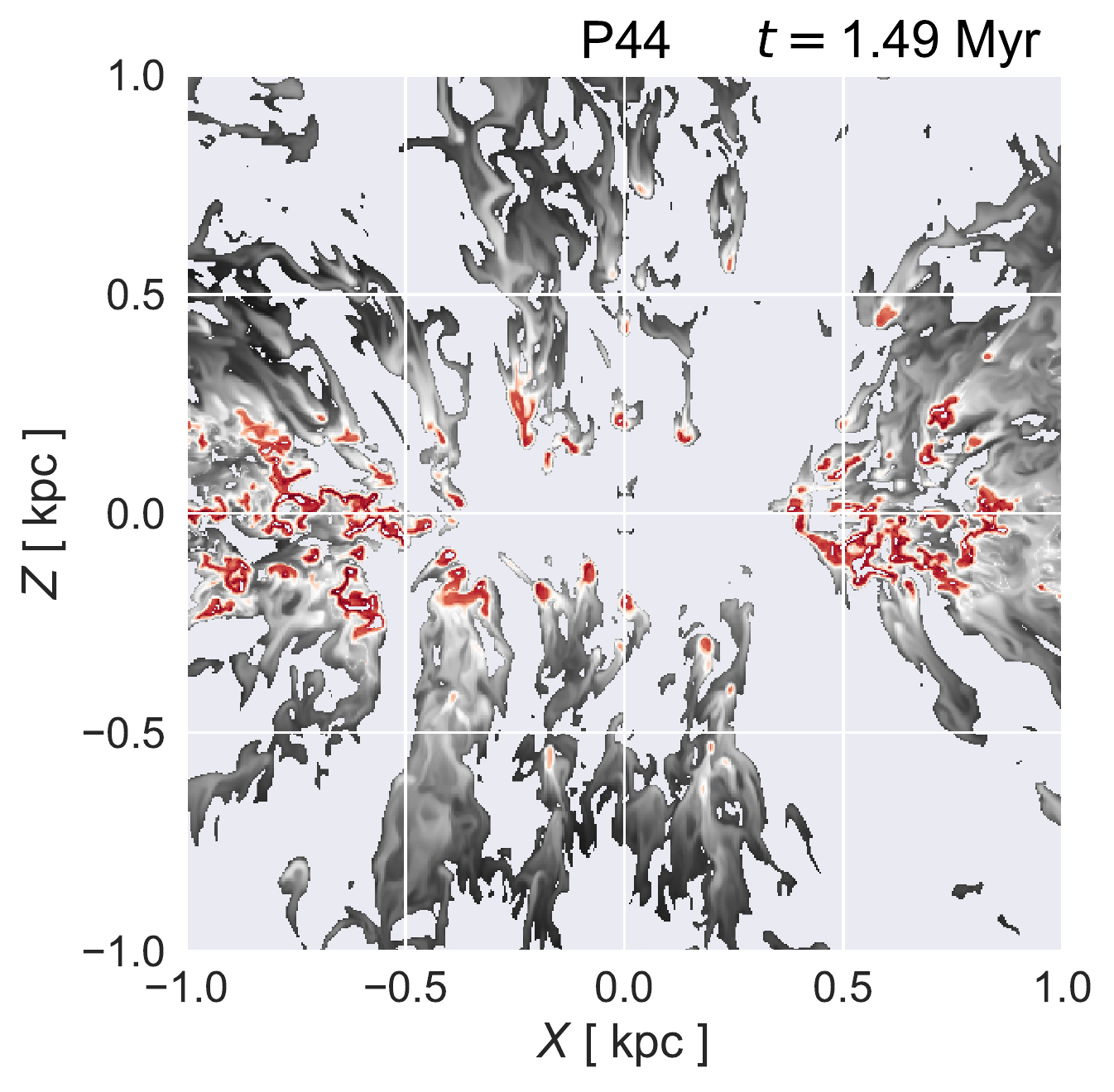}\hspace{-0.22cm}
\includegraphics[height=5.88cm]{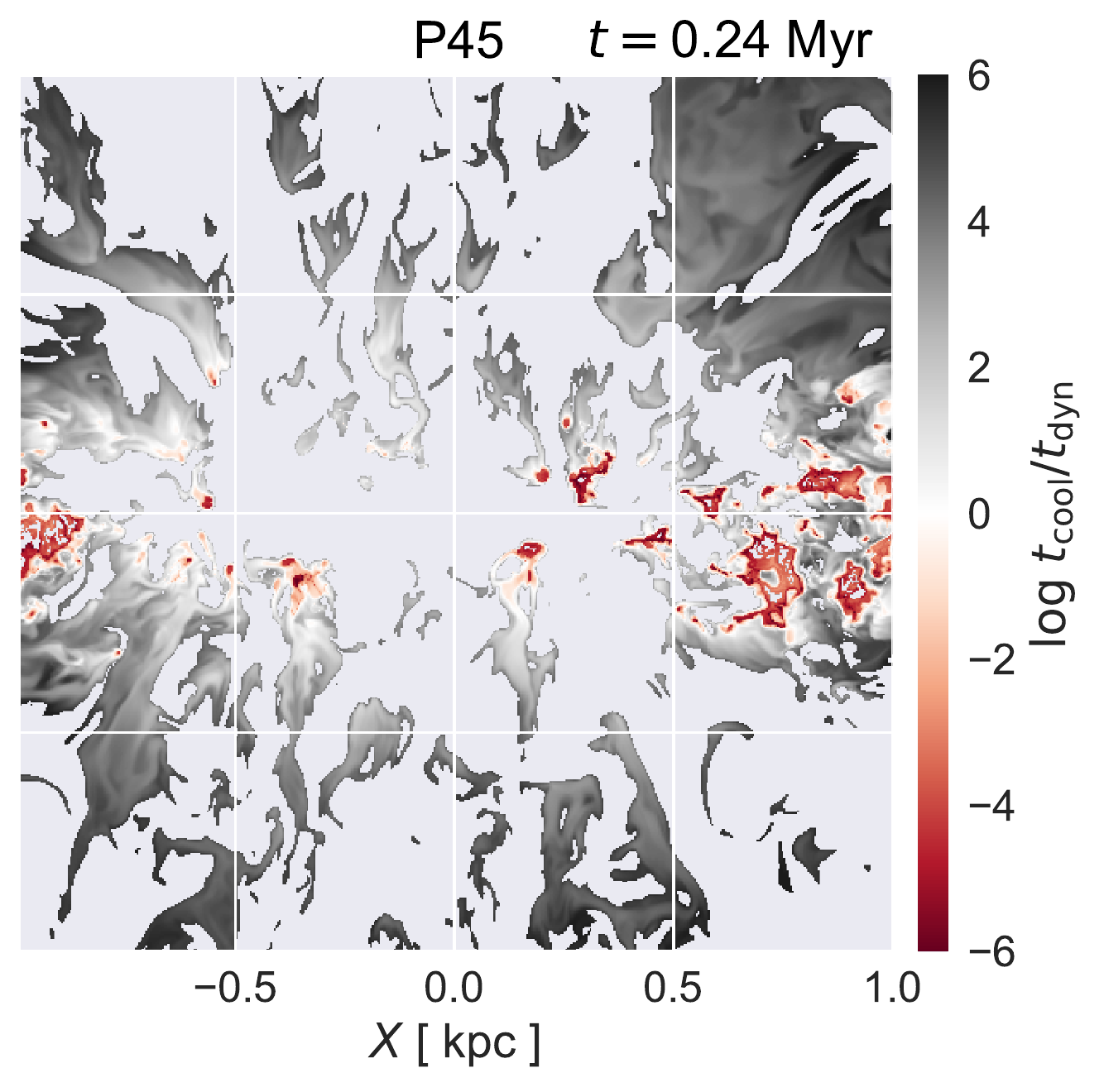}
\caption{Mid-plane slices perpendicular to the $Y$--axis of the ratio of cooling time to dynamical time for simulations \PFF{} (left) and \PFV{} (right). The dynamical time is defined as $1\kpc / \vrad$, where $\vrad$ is the radial velocity. Dense, slow-moving cloud surfaces and clump cores have short cooling times relative to their dynamical time, while the cooling times of gas filaments ablated from clouds exceed their dynamical time.}
\label{fig:tcool}
\end{figure*}

We measure the mass in the outflowing gas in our simulations, distinguishing between gas flowing along the plane of the disc, and gas flowing perpendicularly away from the galactic plane. In Fig.~\ref{fig:massfrac_vrad}, we show the evolution of the mass of gas with cylindrical radial velocity $\vcyl > \vel$ as a function of $\vel$, at different times for both simulations. 

In Fig.~\ref{fig:massfrac_vz}, we show a similar plot of the mass in gas with vertical velocity $\vel_z > \vel$ as a function of $\vel$. For simulation \PFF{}, gas of mass of $8\times10^6\Msun$ is outflowing along the disc plane with a velocity greater than $200\kms$ while in the case of simulation \PFV{}, it is $7\times10^7\Msun$. The outflowing gas masses perpendicular to the disc are similar. This is likely due to the fact that the jet plasma propagates quasi-isotropically within the confines of the disc, and, as a consequence, the mechanical advantage is equally effective in the $z$--direction as it is directed radially in the plane of the disc. In addition, dense gas that is accelerated in the $z$--direction experiences less hindrance by other clouds located ahead. 
The outflowing gas masses translate to total mass outflow rates of $\sim 8\Msunyr$ and $70\Msunyr$, for simulations \PFF{} and \PFV{}, respectively, assuming a characteristic dynamical timescale of $\tdyn\approx1\Myr$. \citet{Oosterloo2017a} find total outflowing gas masses of at least $5\times10^6\Msun$, and a mass outflow rate of around $12\Msunyr$. Thus, in simulation \PFF{}, we find a mass outflow rate similar to that found observationally by \citet{Oosterloo2017a}, but the mass outflow rate for simulation \PFV{} is about an order of magnitude larger. 

In \S~\ref{sec:evolution} we mentioned that the jet plasma is also escaping perpendicular to the disc, entraining with it small clumps and filaments of disc material. While the $8\GHz$ radio images do not reveal the jet plasma escaping perpendicular to the disc, the $1.4\GHz$ continuum image presented in Fig.~2 of \citet{Morganti1998a} shows radio emission aligned with the minor axis of the galaxy. This radio structure may be the result of a starburst wind, but we propose the alternative explanation that it is radio plasma escaping perpendicular to the disc, as seen in our simulations. In fact, the galaxy-wide star formation rate of \icfost{} based on a far-infrared luminosity of $\LFIR\approx1.6\times10^{10}\Lsun$ is approximately $5\Msunyr$ \citep{Satyapal2005a} with an associated $1.4\GHz$ luminosity of $\Popfghz\approx8\times10^{21}\WpHz$ \citep{Murphy2011a}. It appears unlikely that the radio structures of almost $10\kpc$ in extent are bubbles associated with a star-formation activity of less than $5\Msunyr$. A direct comparison between the observed $1.4\GHz$ morphology and a synthetic radio map of the extended jet plasma would, however, require a simulation box size at least 5 times larger than that we have employed.

In Fig.~\ref{fig:nhha} we show in the top two panels the column densities along $Y$ for the simulations \PFF{} and \PFV{}. In the second row of the same figure, we show the \halpha{} surface brightness viewed along $Y$ in the optically thin limit. The \halpha{} emissivity is obtained directly from the cooling function. The snapshots are those for which the extents of the jets are in agreement with the currently observed extents of the jets. The striking difference between the two simulations is the presence of clumps lifted off the disc in simulation \PFF{}, whereas simulation \PFV{} shows only diffuse filaments entrained in the jet plasma outflow perpendicular to the disc.  

The difference between the two cases is a direct consequence of the power of the jets and the resulting confinement time. In simulation \PFV{}, the higher-power jet heats and strips the outer layers of clouds on a shorter time scale, dispersing the inner region of the ISM. The regions at larger disc radii remain more intact. In simulation \PFV{}, on the other hand, the jet plasma takes more time to percolate through the ISM, shocking and fragmenting the clouds. The jet plasma escaping perpendicular from the disc has more time to entrain and accelerate cloudlets in bulk. The acceleration of cloudlets is also aided by the vertical thermal pressure gradient. A similar result of low-power jets affecting a wider extent of the ISM has been shown in our previous papers \citep{mukherjee16a,mukherjee17a}. 

This interpretation of the difference in evolution is supported by maps of the line-of-sight velocity dispersion and mean temperature, shown in Fig.~\ref{fig:sgte}. For these plots, we weighted the velocity and temperature by density, and applied a cloud tracer threshold of $\trw > 0.98$. We clearly see that simulation \PFV{} results in a higher velocity dispersion and higher mean temperature within in the disc than simulation \PFF{}, and produces a higher-velocity, hotter, more diffuse vertical outflow than simulation \PFF{}. 

Also, as noted earlier in Fig.~\ref{fig:massfrac_vz}, the more powerful jet drives out more gas than the weaker jet. However, that gas is mostly in the hot diffuse form not visible in, e.g., \halpha{}. Thus the morphology of shocked gas observed from emission lines, and the phase of the outflowing gas (cold, warm or hot) can help distinguish between different models of feedback.

Finally, we consider the possibility that molecules form in the dispersed or outflowing gas. In particular, if diffuse outflowing gas can condense again to cool dense filaments that form molecules, the velocity dispersions seen in the molecular phase may be achievable with a lower jet power. Since we do not follow the formation of molecules in our simulations, we  test the plausibility of this scenario by comparing the cooling time to the dynamical time of the dispersed or outflowing gas. We show mid-plane slices for simulations \PFF{} and \PFV{} of the ratio of cooling time to dynamical time in Fig.~\ref{fig:tcool}. For the dynamical time, we take $\tdyn=1\kpc / \vrad$ and the cooling time is $\tcool = \varepsilon / \Lambda(\rho, T)$, with a warm phase tracer cut of $\trw>0.98$. 

In both simulations, the cooling time in dense, slow-moving cloud surfaces and clump cores is shorter than the the dynamical time, but the cooling times of gas filaments ablated from clouds exceed their dynamical time. The comparatively long cooling time in fast, diffuse filaments may indicate that molecule formation in these is unlikely, although a simulation including the chemistry of molecule formation, such as that performed by \citet{Richings2017a}, is required to make a more conclusive statement.

\section{Summary and Discussion}\label{sec:summary}
We have performed 3D relativistic hydrodynamic simulations of the jet-ISM interactions in \icfost{}, a radio galaxy in which the jet is aligned with the plane of the kpc-scale disc, and in which strong velocity dispersions are observed in all gas phases. We have set up a fractal, turbulent, rotating disc with parameters informed by observations, and have presented two simulations with different jet powers: $10^{44} \ergs$ (P44) and $10^{45} \ergs$ (P45). 

The jets in the simulation evolve very differently to the classical ``dentist's drill'' \citep{scheuer82a}. In the simulations, we recover the evolution of the flood-and-channel phase typical of jet-ISM interactions, in which clouds are shocked and gas is ablated from the clouds in the ISM. In this geometry, with the jet axis lying in the plane of the gaseous disc, the jet is strongly confined and jet plasma vents perpendicular to the disc.

The jet plasma that vents perpendicular to the disc, entrains gas ablated from the clouds. In the case of simulation \PFF{}, in which the jet is confined 6 times longer than in the case of simulation \PFV{}, clumps of dense gas are lifted off the disc; these may  possibly be visible in emission from warm ionised medium (e.g.\halpha{} as we show in Fig.~\ref{fig:nhha}).

In order to directly compare the gas kinematics resulting from jet-ISM interactions with gas kinematics inferred from recent ALMA observations by M15, we have constructed synthetic PV diagrams from the density, temperature, velocity, and cloud tracer fields in our simulations. We compare our synthetic PV diagrams to the PV diagram presented in M15 for \ceeoh{2}{1} and find that our simulations broadly reproduce the key features in the observational data, including an increase in velocity dispersion to a few $100\kms$ in the regions within approximately $0.6\kpc$, in which the jet has interacted strongly with the dense clouds in the ISM. In this region, the CO surface brightness is enhanced and coincident with the $8\GHz$ radio image.

The asymmetry in the PV diagrams between the eastern and western sides of the galaxy, as well as the jagged features in the PV plots are also reproduced. These features are largely explained by the clumpy, inhomogeneous nature of the ISM, through which the jet requires a substantial amount of time to penetrate. The forbidden quadrants in the PV diagram show a strong signal because the jet, especially near the jet head, disperses the gas nearly isotropically, resulting in gas being accelerated against the direction of the disc rotation.

The peaks in the radio surface brightness correspond to these locations, where the main jet stream impacts a massive dense cloud in its path and splitting into secondary jet streams while compressing, fragmenting, and dispersing the cloud. Such a ``splatter region'' is particularly prominent, both in the observations and in our simulations, in the western part of the galaxy, where a significant  velocity dispersion is observed in the molecular gas as well as in the ionized and neutral phases.

We have presented two models with jet powers $10^{44}\ergs$ and $10^{45}\ergs$, respectively. As expected, a higher jet power  leads to stronger velocity dispersion in a much shorter time ($1.49\Myr$ for simulation \PFF{}, and $0.24\Myr$ for simulation \PFV{}). Our results indicate that the ISM in our simulation is a reasonable representation of the ISM in \icfost{}. $\Pjet=10^{44}\ergs$ is a lower limit to the jet power because jets of even lower power would likely not result in the observed velocity dispersion, unless molecules are efficiently produced in the diffuse outflowing gas over the short dynamical time of the outflow. A simple estimate of the dynamical time versus cooling time for the outflowing gas does not favour this scenario. The simulations by \citet{Richings2017a} also suggest that timescales larger than a few $10^5$ years are required to obtain a significant amount of molecular gas in an AGN-driven outflow, although their simulations employed a smooth ISM.

The values of the jet power of \icfost{} quoted in the literature, including by M15, are of the order of $10^{43}\ergs$--$10^{44}\ergs$. They are obtained through scaling relations between radio power and cavity power derived for classical evolved radio jets doing work against smooth halos of galaxies or clusters \citep{cavagnolo10a}. Such scaling relations are known to be  unreliable \citep{Godfrey2016a}, and would not apply to jets propagating into the clumpy ISM of a galactic disc, as is the case for \icfost{}. 

From our simulations, we find the radio power at 1.4 GHz for the P44 simulation at time $\sim 1.49$ Myr to be $\sim 5.9 \times10^{21} \WpHz$ and that for the P45 simulation at $\sim 0.24$ Myr to be $\sim 4 \times 10^{23} \WpHz$. The above estimates assume a nominal value of $f_e=f_B=0.1$, as discussed Sec.~\ref{sec:results}. For an electron-positron plasma, the values of these fractions may be 0.5 each, which would increase the simulated radio powers by a factor of 92 (discussed in more detail in Bicknell et al. {\it submitted}).  The observed radio power at 1.4 GHz was estimated to be $3\times 10^{23} \WpHz$ \citep{Morganti2015a}. Thus the actual mechanical power of the jet is likely to lie in between the two values considered in this work $10^{44}\ergs$ and $10^{45}\ergs$. This is an order of magnitude higher than that estimated from the simple scaling relations obtained as an empirical fit to the  estimates of radio power and energy in the radio cavities  inferred from radio and X-ray data respectively \citep{birzan08a,cavagnolo10a}. However, although the mechanical power of the jet in \icfost{} implied by our simulations differ from the mean relations derived in \citet{cavagnolo10a}, the values are  consistent with the scatter in the measurements of the jet power \citep[e.g. Fig.~1 of][]{cavagnolo10a} at values of radio powers similar to that of \icfost{}.

The kinematics of the gas affected by the jets as represented by the PV diagrams in M15 and \citet{Oosterloo2017a} favour a jet power closer to $10^{45}\ergs$, but the mass outflow rates favour a jet power closer to $10^{44}\ergs$. Currently, we are not able to put stronger constraints on the jet power, but future studies that include the kinematics of the emission line gas, and an estimate of the jet age may help in this regard. \emph{Our work has shown, however, that the power of jets and the nature of the intervening ISM with which it interacts  can be constrained by careful modeling of the perturbed velocity fields that the jet-ISM interactions generate.}

\section*{Acknowledgements}
This research was supported by the Australian Research Council through the Discovery Project, The Key Role of Black Holes in Galaxy Evolution, DP140103341. This research/project utilised resources and services from the National Computational Infrastructure (NCI), which is funded by the Australian Government. AYW has been supported in part by ERC Project No. 267117 (DARK) hosted by Universit\'e Pierre et Marie Curie (UPMC) - Paris 6, PI J.~Silk. RM gratefully acknowledges support from the European Research Council under the European Union's Seventh Framework Programme (FP/2007-2013) /ERC Advanced Grant RADIOLIFE-320745. This work has greatly benefited from discussions with Kalliopi Dasyra, David Neufeld, Pierre Guillard, and Joseph Silk. We also thank the referee for the helpful suggestions.




\bibliographystyle{mnras}
\bibliography{bibl} 




\appendix



\section{Propagation time of the jet}\label{sec:tjet}

We approximate the dynamical time of the jet propagating through the disc with the advance speed of the jet head calculated using Equations 20 and 21 in \citet{Safouris2008a} \citep[see also Equations 23 and 24 in][]{Sutherland2007a}. Let $\zeta$ be the ratio between jet density and the density of the ambient medium into which the jet is propagating. The time for the jet head to traverse a distance $L$ is then:
\begin{equation}
\tjet = \frac{L}{c} \left(1 + \frac{1}{\Gamma}\sqrt{\frac{\chi}{\left(1 + \chi\right) \zeta}}\right) \left(1 - \frac{1}{\Gamma^2}\right)^{-\frac{1}{2}} \;.
\label{eqn:tjet}
\end{equation}
Strictly, the equations mentioned above were derived for the advance speed of the terminal hot spot for jets propagating into a homogeneous medium. However, by using the average density of the multiphase interstellar medium in our simulations, we obtain a reasonable estimate for the jet head advance speed. Using $L=0.5\kpc$, $\zeta=10^{-5}$, $\chi = 0.48$, and $\Gamma=4$, we obtain $\tjet\approx0.4\Myr$. This estimate agrees within a factor of a few with the duration of our simulation up to the point where the jets reach a distance of $0.5\kpc$ from the center.


\bsp	
\label{lastpage}
\end{document}